# Megascopic Quantum Phenomena

## A Critical Study of Physical Interpretations

**Michal Svrček**

CMOA Czech Branch, Carlsbad, Czech Republic

---

**Abstract**  A historical study of metaphysical interpretations is presented. A megascopic revalidation is offered providing responses and resolutions of current inconsistencies and existing contradictions in present-day quantum theory. As the core of this study we present an independent proof of the Goldstone theorem for a quantum field formulation of molecules and solids. Along with phonons two types of new quasiparticles appear: rotons and translons. In full analogy with Lorentz covariance, combining space and time coordinates, a new covariance is necessary, binding together the internal and external degrees of freedom, without explicitly separating the centre-of-mass, which normally applies in both classical and quantum formulations. The generally accepted view regarding the lack of a simple correspondence between the Goldstone modes and broken symmetries, has significant consequences: an ambiguous BCS theory as well as a subsequent Higgs mechanism. The application of the archetype of the classical spontaneous symmetry breaking, i.e. the Mexican hat, as compared to standard quantum relations, i.e. the Jahn-Teller effect, superconductivity or the Higgs mechanism, becomes a disparity. In short, symmetry broken states have a microscopic causal origin, but transitions between them have a teleological component. The different treatments of the problem of the centre of gravity in quantum mechanics and in field theories imply a second type of Bohr complementarity on the many-body level opening the door for megascopic representations of all basic microscopic quantum axioms with further readings for teleonomic megascopic quantum phenomena, which have no microscopic rationale: isomeric transitions, Jahn-Teller effect, chemical reactions, Einstein-de Haas effect, superconductivity-superfluidity, and brittle fracture. We demonstrate how the megascopic extension of the microscopic theory deals with the various paradoxes, such as the arrow of time, the Jordan-von Weizsäcker problem, to explain classicality, Santilli's "no reduction theorem", the Schrödinger cat paradox and Wigner's friend, and the Kauzmann paradox of residual entropy. Only the Copenhagen interpretation seems to be able to incorporate telicity via megascopic mirroring of each of its basic axioms, and in that way does confront the abovementioned paradoxes.

## 1. Introduction

Quantum theory was at first formulated and construed within the framework of the Copenhagen School. This interpretation was a result of long years of intensive exploration of the microworld answering one of the greatest scientific paradoxes of all times: on the one hand maintaining classical concepts in the description of experimental results, and on the other hand to build a novel perception of the microworld, the latter not the world of reality as we know from Newton's equations of the macroworld, but rather the abstract world of Aristotelian potentiality. The primary consequences of this conception were indeterminism and complementarity, which were introduced for the first time in science. Since classical physics has influenced philosophy for centuries, leading to such extreme views as mechanical materialism, it was indeed difficult for many scientists at first to accept the Copenhagen interpretation in its entirety. There were numerous unsuccessful attempts to explain microprocesses in a deterministic way by means of hidden parameters. The sharpest discussions proceeded between Bohr and Einstein, which culminated in the most ingenious objection of Einstein, known as the EPR paradox [1]. Later Bell discussed this paradox and proved what is known today as Bell's theorem [2]. Based on this insight, subsequent experimental tests fully confirmed the validity of all quantum mechanical predictions also in concert with the celebrated Copenhagen interpretation.

The initial work of Bohr and later realized as the Copenhagen interpretation basically concerned the hydrogen atom, but how to continue? What about molecules, crystals, liquids, and other forms of macroscopic matter? Do we understand them as well, too? Detailed many body treatments were developed during the last half century, recommending how to deal with complex systems composed of elementary particles. These many body treatments have one common feature: they respect one classical rule, i.e. knowing the physical law of motion of the smallest parts, we can unambiguously predict the motion of the whole. Although the majority of scientists believe this, one may still ask: is this really true? We have of course no reason to doubt this belief, yet what is in full accordance with our deep-rooted experience, we might still find such examples or paradoxes that convince us otherwise. Unfortunately such examples and paradoxes do exist, although they usually are considered only to be of peripheral interest. Nevertheless this will be the topic of the presented article.

One may ask whether the Copenhagen interpretation is bad, as Einstein and all supporters of hidden parameter schemes tried to prove, or whether the elementary particle equations are at fault? Of course, no! On the contrary, the Copenhagen interpretation was formed in a very accurate, tangible and original way. However the simple characteristics of the hydrogen atom, did not reveal the whole truth about a many-body quantum system alongside yielding an inappropriate relationship between the micro-, the macro- and the mega-world. As will be demonstrated in this paper, there is also the controversial problem regarding the occurrence of the alleged quantum jumps, which have no origin in the original Copenhagen microscopic formulation. Hence this article will promote a so-called 'second quantum floor' describing megascopic quantum phenomena, in contrast to the 'first floor', defined by the standard description of microscopic phenomena within the Copenhagen interpretation. We will in addition expound, which heretofore being unexplained or even misinterpreted, upon the phenomena that belong to this 'second floor'.

## 2. The clamped-nuclei paradox

Let us start with the simplest formulation of the many-body problem based on the Schrödinger equation. Solving it exactly is in practice impossible, and therefore its simplified solution known as the Born-Oppenheimer (B-O) approximation [3] is widely used. Quite recently Sutcliffe & Woolley [4] offered an extensive historical justification of the development of quantum theory, in particular the quantum chemical solution of Schrödinger's equation, discussing the applicability as well as the limitations of Potential Energy Surfaces (PES) in quantum chemistry. As conclusion they wrote: „This qualitative modification of the internal Hamiltonian, the extra choice of fixed nuclear positions in the 'electronic' Hamiltonian, is ad hoc in the same sense that Bohr's quantum theory of the atom is an ad hoc modification of classical mechanics. An essential feature of the answer is put in by hand. We know that both modifications have been tremendously useful and our point is not that something else must be done in practical calculations on molecules. The point is how the successful description of molecules involving the clamped-nuclei modification at some stage can best be understood in terms of quantum mechanics. In the case of the Bohr atom the resolution of the inconsistency in mechanics applied to the microscopic realm was achieved quite quickly with the formulation of quantum mechanics; in the molecular case, no such resolution is at present known."

In other words, we have enough numerical verifications available, but what we need, is the elucidation why the clamped-nuclei modification works so well. Their statement is an imperative challenge to everybody. In passing we should not forget that two scientific sub-disciplines, quantum chemistry and solid state physics, are mostly dependent on the B-O approximation, making them two sand-castles. This begs the question: What can we expect: the confirmation of the B-O approximation that satisfies our desire for perfection – or a disclosure of something new that goes beyond our contemporary knowledge of physics and chemistry? We will contend in this paper that the latter is closer to the truth.

What about the objection that we can avoid all the above mentioned problems by solving the Schrödinger equation exactly. I quote from Monkhorst's article [5] an analogous critique of the above mentioned PES concept: „The BO approximation is strictly valid only near the multidimensional PES minimum. On the other hand, the BH (Born-Huang) treatment inspired the thinking that the entire PES is usually valid to describe the dynamics of a molecule, including dissociation into various fragments, i.e., chemical reactions. However, this view is only acceptable if the PES-associated wave functions interact only weakly for all nuclear configurations. It is usually quite impossible to verify this, and most consumers of the PES concept assume it. Their only ''defense'' is the adiabatic nature of a molecular wave function, where the electronic wave functions adjust ''instantaneously'' to the evolving nuclear configuration. In fact, this view is central to the teachings of quantum chemistry."

Monkhorst also describes his coupled-cluster method for the solution of the system of electrons and nuclei on the same footing, completely evading the B-O approximation: „Therefore, when the need arises to address the limitations of the PES, it seems valuable to remove it entirely from the toolbox of the molecular scientist. One option is the molecular coupled-cluster (MCC) method I formulated 10 years ago [6]. This method takes a very atomic view of a molecule: instead of fixing the nuclei as in the BO approximation, the electrons and nuclei are both described quantum-dynamically within a centrosymmetric shell structure. The coupled-cluster method, duly generalized, is used to describe the correlation among all particles."

Being aware of the practical computational limitations of this exact method, he adds: „Even though the MCC method seems attractive, and (I hope) computationally tractable once implemented, it can be only practical for small molecules. I cannot see how the PES concept, and its attendant molecular structure ideas, will be superseded with this method for large molecules. Its qualitative appeal, its semiquantitative success, and its deep roots in the chemists' minds will keep the PES as real as it is now."

The reasons given above is most likely why most of chemists prefer calculations based on the B-O approximation. As far as the error caused by this calculation, on the adiabatic level the Born-Huang (B-H) ansatz [7] is applied, and as Kutzelnigg proved [8], the B-H ansatz fully compensates for this error. Note however that Kutzelnigg's proof indicates only the numerical equivalence, and solely on the adiabatic level. There are no objective equivalences between the exact solution of the Schrödinger equation and solutions based on the whole of the B-O approximation and B-H ansatz at all, even on the adiabatic level.

Fortunately, a small number of scientists were inspired by Monkhorst's idea, and performed exact calculations of some of the simplest molecules. Cafiero & Adamowitz wrote [9]: „The model of the molecule… is quite similar to an atom, as has been noted by Monkhorst [5]. We have the analogue of the nucleus with the heavy particle at the center of the internal coordinate system, and we have the analogues of electrons in the internal particles. The main difference between this model and an atom is that the internal particles in an atom are all electrons and in the molecular atom or atomic molecule the internal particles may be both electrons and nuclei (or, as we should more correctly say, pseudo-particles resembling the electron and the nuclei). Formally this difference manifests itself in the effective masses of the pseudo-particles and in the way the permutational symmetry is implemented in the wave function."

And further looking at the conclusion from their calculations: „Since in the nonadiabatic treatment we have included the nuclei in the wave function, we must determine the molecular structure by calculating expectation values of the distances between the nuclei. Since the operators representing the internuclear distances… do not commute with the Hamiltonian…, we cannot measure these distances exactly for stationary states of the system. Furthermore, since the ground state is spherically symmetric, we cannot gain additional information from the wave function such as the expectation values of the x, y, z coordinates of any of the particles, since these will average out to zero in all cases. Lastly, the expectation values for the distances between any two particles in each subset of identical particles (including the distance to the particle at the origin, if that particle belongs to the subset) will be equal. This is a widely known fact for electrons, i.e., due to antisymmetry, or indistinguishability, all of the electrons are on average at the same distance from each other, as well as from each nucleus in the molecule. That this same fact applies to nuclei is, perhaps, less known."

These first calculations of their kind gave us a remarkable insight into the actual differences between the exact and the clamped-nuclei approaches. The exact solution communicates the concept of the "molecular atom or atomic molecule", where we can only see one undivided isolated composite object, while we are unable to say anything more about the individual molecules involved in such a system. The hierarchy elementary particles → atoms → molecules doesn't exist here as we know it from the B-O approximation; instead it is replaced with the reduced hierarchy elementary particles → molecular atoms or atomic molecules. The exact solution leads always to a full symmetry picture of the molecule, and the indistinguishability principle, valid for nuclei as well, doesn't allow any "symmetry breaking", such as e.g. an isomerism. As Sutcliffe & Woolley state: „The Isolated Molecule model doesn't capture isomerism, nor optical activity." It is indeed a pity that exact solutions are so complex that till now they have only been tested on very small systems, i.e. up to ten particles, electrons and nuclei included. It would be very interesting to test them on something bigger at least on the simplest Jahn-Teller [10] systems. It seems that the indistinguishability of the nuclei and their shell nature, no J-T symmetry breakings are possible at all, cf. the case of isomerism, and that we will obtain fully symmetrical solutions for the ground states. Although standard theories of the J-T effect [11,12] go beyond the B-O approximation and are nonadiabatic, they are nevertheless all of clamped-nuclei type.

Thus we can see that even exact solutions of Schrödinger's equation do not yield the full picture of all quantum chemical phenomena. Yet we cannot avoid the clamped-nuclei concept in the B-O approximation, leading to a serious paradox: how can an approximation produce results that by no means follow from the exact solution? What is the nature of the B-O model when it is not viewed as an approximation? Where is the true origin of the clamped-nuclei concept, allowing the description of individual molecules? This is indeed a quantum paradox, since we don't know any classical analogies. How do we reach the B-O approximation? Perhaps by simply transgressing the law of nature due to the hierarchical way of quantization - first the electrons are quantized and the nuclei remain classical, and then they are quantized a posteriori. But quantum mechanics demands simultaneous quantization, regardless how heavy the nuclei are compared to the electrons. Sutcliffe & Woolley ask: „The interesting question is how to get from the quantum theory of an Isolated Molecule to a quantum theory of an individual molecule by rational mathematics." The problem is thus, on "what" quantum theory the mentioned rational mathematics do apply. If we understand quantum theory as quantum mechanics based on the original Copenhagen interpretation, employing the Schrödinger equation for particles in the many-body systems, we find no answer. The input of electrons and nuclei in the mechanical Schrödinger equation exhibits a full symmetry regarding both types of particles. As we will see in this paper, many-body systems are running in dual mode, and we need to search the answer not under the mechanical pattern, but under the field concept where simultaneous quantization can be performed.

The concepts "isolated system" and "individual system" are well defined in the classical world, but only the first one has a similar meaning in the microworld, whereas the second one needs a radical redefinition on the quantum level. However, in speaking about elementary particles as "individual systems", there is no problem. The problems start when we try to apply this concept to composite systems like molecules or crystals.

In regard to the phenomena mentioned above, the explanation calls for the concept of clamped-nuclei, but, as just said, this concept, from the viewpoint of an exact quantum mechanical many-body formulation, does not work. The Copenhagen interpretation must be incomplete in this respect.

## 3. The paradox of time irreversibility

The fundamental problem of both classical and quantum physics is still the question of the arrow of time. Why does time have a direction?

A half century ago, Santilli, then a nuclear physicist, stated, what he called a no-reduction theorem [13]: „A macroscopic irreversible system cannot be consistently reduced to a finite number of elementary constituents all in reversible conditions and, vice versa, a finite number of elementary particles all in reversible conditions cannot consistently characterize a macroscopic irreversible system." He further said, claiming that he has discussed this problem with Heisenberg and Dirac and that none of the quantum theory founders knew the answer: „The above theorem establishes that irreversibility originates at the most ultimate structure of nature." Till now the scientific community does not seem to take this paradox seriously, nevertheless Santilli continues: „During the 20th century it was generally believed that the irreversibility over time of our macroscopic environment was "illusory" (sic) because, when macroscopic events are reduced to their elementary particle constituents, irreversibility "disappears" (sic) and one recovers nice elementary particles in the reversible conditions necessary for the applicability of special relativity, quantum mechanics and quantum chemistry." Although he was correct with the formulation of the "no reduction theorem", he is certainly wrong with his later interpretation. His approach is to introduce the so-called "Lie-admissible algebras" for the construction of irreversible equations of the microworld, seemingly giving up the concept of reversible quantum equations. As a consequence the macroscopic limit of his irreversible equations will lead also to classical irreversible expressions, e.g. replacing the standard Newtonian laws, unravelling puzzles arising from his earlier theorem once and for all. In doing so he does not object to microscopic irreversibility even when no irreversible equations are needed. This appears justified since any intervention during measurement is an irreversible act, causing the reduction of the wave function of the system, ultimately in agreement with microscopic irreversibility. Yet he destroys the philosophical logic of a consistent quantum philosophical microscopic view simply for the limited purpose to explain macroscopic irreversibility. A better alternative would be to accept the incompleteness of quantum theory as stated by the Copenhagen interpretation, rather than rewriting physics in its entirety.

Looking into the history of quantum mechanics, Bohr initially described the Heisenberg principle of uncertainty as a purely epistemic problem. In this context there is no problem to accept the "objective reality" in the macroscopic limit. But today we know that the principle of uncertainty is also ontological, and there is unfortunately no classical limit relating to some "objective reality", even if, on the macroscopic scale, "objective potentiality" still persists. This raises the question: how to relate the various forms of the macroscopic level of "objective potentiality" into one reality. The problem is that classical physics describes events at the scale of continuous time. On the contrary, quantum equations cannot describe any event, since they arise only after outside intervention, i.e. after a measurement which can lead to irreversible processes. However, there are also many irreversible processes, e.g. chemical reactions that occur independently on any observer, i.e. without any specific measurement process. Consequently, if there is no macroscopic ontological limit of quantum laws that applies to the microworld, then there must exist yet another type of events, that occurs on the macroscopic level, which are responsible for the time irreversibility, but have no roots in the original Copenhagen School. This is of fundamental importance, since if and when we find this, until now unknown type of events, we will recover the concept of unitary transformations as pertaining to reversible equations on both the microscopic as well as the macroscopic levels.

Classical physics recognizes reversible processes as the primary ones along with the full reversibility on the time axis, whereas irreversible processes violates this symmetry with the arrow of time oriented only in the "forward" direction. This stance also carries over in microscopic quantum theory. Santilli recognized the ascending order of irreversibility, but nevertheless made an inaccurate inference in the assumption of irreversibility conceiving reversibility as a constricted limit. There are essentially no irreversibility in the primary equations of physics, which leads us to explore deep-seated formulations that might reveal precise relations between reversibility/irreversibility allowing solutions that provide the proper characteristics depending on the case at hand. Therefore the classical concept of the "arrow of time" becomes inadequate in the quantum world and without such readings the Copenhagen interpretation remains incomplete.

## 4. Bohm's five prophetic statements

Bohm is largely known for his attempts to invigorate the theory of hidden parameters, originally formulated by de Broglie as the contentious pilot-wave theory, adding the well-known conception of the famed quantum potential [14]. However, this appears to lead nowhere. The theory of hidden parameters cannot be helpful for a solution of the incompleteness problem associated with the Copenhagen interpretation. Nevertheless, as is less known, thirty years after he published his new version of non-local hidden parameters, Bohm arrived at five important critical conclusions regarding the aforementioned lack of completeness. He himself never found the resolution, since his own Bohmian Mechanics indeed suffers from the same shortcomings as the Copenhagen interpretation, believing that it will be answered in the future by some new modifications of the hidden variable idea. As can be seen, it is easier to be critical rather than finding the correct solution.

These five critical points of Bohm are of extraordinary relevance, because neither the Copenhagen interpretation nor any other rendering is able to cope with them, to take their ideas in to consideration fulfilling their requirements. I will cite them all and give my reflections and detailed remarks below. The first four are taken from his book Wholeness and the Implicate Order [15], the last fifth from his dialogue with the biologist Sheldrake [16].

The first quote runs:

1. „Nevertheless, in most of the work that is being done in physics today the notions of formative and final cause are not regarded as having primary significance. Rather, law is still generally conceived as a self-determined system of efficient causes, operating in an ultimate set of material constituents of the universe (e.g. elementary particles subject to forces of interaction between them). These constituents are not regarded as formed in an overall process, and thus they are not considered to be anything like organs adapted to their place and function in the whole (i.e. to the ends which they would serve in this whole). Rather, they tend to be conceived as separately existent mechanical elements of a fixed nature."

Here Bohm discerns two categories of causes - efficient and final, according to the ancient classification introduced by Aristotle. Nature is full of phenomena based on final causes, but unfortunately the science of physics (classical, relativistic or quantum) is unable to incorporate them and represent them in some form of meaningful equations. What is even worse, physicists often do not recognize final causes at all treating the relevant phenomena as they would be induced by efficient causes. A nice example is the spontaneous symmetry breaking. Since it is believed that nothing in nature is spontaneous, physicists not seeing any explicit efficient cause, they call such symmetry breakings "spontaneous" imparting adverse formulations of such phenomena that are based on this type of symmetry violation. We will discuss this problem in more detail below.

The second statement reads:

2. „When it comes to the informal language and mode of thought in physics, which infuses the imagination and provokes the sense of what is real and substantial, most physicists still speak and think, with an utter conviction of truth, in terms of the traditional atomistic notion that the universe is constituted of elementary particles which are 'basic building blocks' out of which everything is made. In other sciences, such as biology, the strength of this conviction is even greater, because among workers in these fields there is little awareness of the revolutionary character of development in modern physics. For example, modern molecular biologists generally believe that the whole of life and mind can ultimately be understood in more or less mechanical terms, through some kind of extension of the work that has been done on the structure and function of DNA molecules. A similar trend has already begun to dominate in psychology."

The quantum physics community, as understood today by a vast majority of physicists, constitutes an unluckily fulfilment of the materialistic ideas of Democritus. As concerns ancient Democritus' atomism, the concepts of quantum physics differs little or not at all from classical mechanics. In fact the great men of classical physics (Newton, Leibniz, Laplace and others) sincerely believed that if someone gives them the laws of motion for all existing constituents of matter, they could predict the motion of the whole universe. No matter how much the laws of motion of elementary particles in quantum physics are profoundly dissimilar, both mathematically and philosophically, to the laws of classical physics, yet by introducing complementarity and uncertainty no one dares to contest any unambiguousness in the formulation of many-body quantum equations. As a result many physicists, particularly cosmologists, believe in the existence of a wave function describing the whole of the universe, one of the first and main anticipation in the Everett relative state interpretation [41]. Bohm's conclusion is absolutely correct, and we will prove here that the conventional many-body formulation in quantum physics is ambiguous as well, and that microscopic quantum laws are not the ultimate and unique platform for understanding all natural phenomena.

The third citation becomes:

3. „So fragmentation is in essence a confusion around the question of difference and sameness (or one-ness), but the clear perception of these categories is necessary in every phase of life. To be confused about what is different and what is not, is to be confused about everything… The question of fragmentation and wholeness is a subtle and difficult one, more subtle and difficult than those which lead to fundamentally new discoveries in science. To ask how to end fragmentation and to expect an answer in a few minutes makes even less sense than to ask how to develop a theory as new as Einstein's was when he was working on it, and to expect to be told what to do in terms of some programme, expressed in terms of formulae or recipes."

This is indeed a very important objection. We will demonstrate a relation between Bohm's dilemma of wholeness versus fragmentation and the problem of isolated versus individual as formulated by Sutcliffe & Woolley. The microscopic quantum world simply does not match our concepts of individual or fragmented, albeit these words have clear and well-defined meanings on the classical level. And further it is most astonishing to compare how the authors mentioned characterize this problem and evaluate its complexity. Bohm emphasized that the challenge is more difficult than the problem of relativity confronting Einstein, while Sutcliffe & Woolley emphasized that it is harder to solve than the problem of quantum theory for Bohr.

The fourth excerpt reads:

4. „Here, we may hope to get some clues by considering problems in a domain where current theories do not yield generally satisfactory results, i.e. one connected with very high energies and very short distances. With regard to such problems, we first note that the present relativistic quantum field theory meets severe difficulties which raise serious doubts as to its internal self-consistency. There are the difficulties arising in connection with the divergences (infinite results) obtained in calculations of the effects of interactions of various kinds of particles and fields. It is true that for the special case of electromagnetic interactions such divergences can be avoided to a certain extent by means of the so-called 'renormalization' techniques. It is by no means clear, however, that these techniques can be placed on a secure logical mathematical basis. Moreover, for the problem of mesonic and other interactions, the renormalization method does not work well even when considered as a purely technical manipulation of mathematical symbols, apart from the question of its logical justification. While it has not been proved conclusively, as yet, that the infinities described above are essential characteristics of the theory, there is already a considerable amount of evidence in favour of such a conclusion."

This is entirely a quantum field problem. Quantum mechanical patterns, derived from the classical Hamilton-Jacobi formalism, doesn't seem to lead to any similar difficulties. The question is rather, whether an analogical quantization of the classical electro-magnetic field Langrangian will provide us with the full quantum field pattern, especially when it is used in closed systems such as nuclei in quantum chromodynamics. Note that original field patterns of quantum electrodynamics represent only the scattering process of a few free particles in vacuum. We will here analyze these inconsistencies in applying such a limited pattern to bound-state massive systems like molecules and crystals.

The final Bohm quotation contains the exchange with Sheldrake:

5. Bohm: 'How is time to be understood?' Now, in terms of the totality beyond time, the totality in which all is implicate, what unfolds or comes into being in any present moment is simply a projection of the whole. That is, some aspect of the whole is unfolded into that moment and that moment is just that aspect. Likewise, the next moment is simply another aspect of the whole. And the interesting point is that each moment resembles its predecessors but also differs from them. I explain this using the technical terms 'injection' and 'projection'.

Each moment is a projection of the whole, as we said. But that moment is then injected or introjected back into the whole. The next moment would then involve, in part, a re-projection of that injection, and so on in-definitely."

Sheldrake: „But don't you get time in physics when you have a collapse of the wave function?"

Bohm: „Yes, but that's outside the framework of quantum physics today. That collapse is not treated by any law at all, which means that the past is, as it were, wiped out altogether… You see, the present quantum mechanics does not have any concept of movement or process or continuity in time; it really deals with one moment only, one observation, and the probability that one observation will be followed by another one…"

In short the question is whether those "injections", promoted by Bohm, would be able to solve the problem of time in quantum mechanics? And further what are those "injections" and what do they represent, and how could we formulate them by rigorous mathematics? They should first of all represent quantum jumps in a similar way as "projections". We also know that "projections" are the results of our conscious observations of the quantum systems, but if "injections" are quantum jumps as well, how do we observe and measure them? Is it possible, that some complementation or extension of the Copenhagen interpretation could finally lead to an ontological justification of those "injections", yet without having any epistemic access to them? In other words, is there some, until now unknown, kind of quantum jumps in nature, which can neither be explained on the basis of present-day quantum theory, nor directly measured, and perhaps with no microscopic origin at all? As already said, this will be the permeating topic of our paper, where in particular we will show how Bohm's "injections" will be closely related to the time reversibility/irreversibility problem of Santilli.

## 5. The paradox of Schrödinger's cat

This is the well-known paradox from the time when Einstein and Schrödinger exchanged arguments regarding the indeterministic features of the newly conceived quantum theory. Nevertheless today, against the background of the mentioned incompleteness of the Copenhagen interpretation, many alternatives are being developed with Schrödinger's cat playing an important testing role. J. D. Norton, he well-known historian of physics, highlights the significance in the following terms [17]: „This paradox of the Schrödinger's cat is the most vivid expression of a lingering problem in the foundations of quantum theory. In the last two decades especially, there has been a huge amount of work devoted to finding variations to standard quantum theory or just new ways to think about the same theory that avoid this problem. There is no consensus on which approach is the correct one or even if some sort of repair is needed."

Norton describes the history of the Schrödinger's cat problem, which followed shortly after the appearance of the famous EPR paradox [1] in March 1935: „In the aftermath of this paper, Einstein and Schrödinger exchanged letters in which they aired their common concerns about quantum theory. In that correspondence, Einstein put to Schrödinger what we now see is an early version of the cat paradox. He outlined a "crude macroscopic example" in a letter to Schrödinger of August 8, 1935: "The system is a substance in chemically unstable equilibrium, perhaps a pile of gunpowder that, by means of intrinsic forces, can spontaneously combust, and where the average life span of the whole setup is a year. In principle this can quite easily be represented quantum-mechanically. In the beginning the $\psi$-function characterizes a reasonably well-defined macroscopic state. But, according to your equation, after the course of a year this is no longer the case at all. Rather, the $\psi$-function then describes a sort of blend of not-yet and of already-exploded systems. Through no art of interpretation can this $\psi$-function be turned into an adequate description of a real state of affairs; [for] in reality there is just no intermediary between exploded and non-exploded."[18]"

As can readily be seen, Santilli's paradox previously mentioned, is a remake of the old gunpowder pile account due to Einstein, as both paradoxes deal with the factual irreversibility of chemical reactions. As we know that Einstein was not successful with his famed EPR objection, this time he did formulate a problem still unresolved.

Norton continues [17]: „Erwin Schrödinger published his "cat" thought experiment in a lengthy paper in the November 29, 1935, issue of the journal Die Naturwissenschaften. Here's the entirety of his original account: "One can even make quite ludicrous examples. A cat is enclosed in a steel chamber, together with the following infernal machine (which one must secure against the cat's direct reach): in the tube of a Geiger counter there is a tiny amount of a radioactive material, so small that although one of its atoms might decay in the course of an hour, it is just a probable that that none will. If the decay occurs, the counter tube fires and, by means of a relay, sets a little hammer into motion that shatters a small bottle of prussic acid. When the entire system has been left alone for an hour, one would say that the cat is still alive provided no atom has decayed in the meantime. The first atomic decay would have poisoned it. The $\psi$-function of the total system would yield an expression for all this in which, in equal measure, the living and the dead cat (sit venia verbo ["pardon the expression"]) blended or smeared out. The characteristic of these examples that an indefiniteness originally limited to atomic dimensions gets transformed into gross macroscopic indefiniteness, which can then be reduced by direct observation. This prevents us from continuing naively to give credence to a "fuzzy model" as a picture of reality. In itself this contains nothing unclear or contradictory. There is a difference between a blurred or unsharply taken photograph and a shot of clouds and mist.""

Due to the Schrödinger's cat narrative, the scientific community were divided: some believed that living and dead cats are really in the state of a quantum superposition until the moment of a conscious measurement, others claimed that the cat had to be either alive or dead yet before the box was opened. Carpenter & Anderson further analysed the laboratory thought experiment with the following conclusion [19]: „The implications arising from the "Schrödinger's cat" thought experiment have led some authors to argue that observation of a measurement by a conscious observer is required to collapse quantum wave-functions. Here we combine Schrödinger's experimental paradigm with a system for splitting the information about the quantum state between two observers, thereby allowing distinct outcomes to be recorded without either observer knowing the state of the measured quantum event. Our results imply that to collapse a quantum wave-function, measurement alone, rather than conscious observation of a measurement, is sufficient."

The last objection made by Einstein against the Copenhagen interpretation was published in his Reply to Criticisms in 1949 [18], which runs in analogy with Schrödinger's cat argument: „As far as I know, it was E. Schrödinger who first called attention to a modification of this consideration, which shows an interpretation of this type to be impracticable. Rather than considering a system which comprises only a radioactive atom (and its process of transformation), one considers a system which includes also the means for ascertaining the radioactive transformation — for example, a Geiger-counter with automatic registration-mechanism. Let this latter include a registration-strip, moved by a clockwork, upon which a mark is made by tripping the counter. True, from the point of view of quantum mechanics this total system is very complex and its configuration space is of very high dimension. But there is in principle no objection to treating this entire system from the standpoint of quantum mechanics. Here too the theory determines the probability of each configuration of all its co-ordinates for every time instant. If one considers all configurations of the coordinates, for a time large compared with the average decay time of the radioactive atom, there will be (at most) one such registration-mark on the paper strip. To each coordinate configuration corresponds a definite position of the mark on the paper strip. But, inasmuch as the theory yields only the relative probability of the thinkable co-ordinate-configurations, it also offers only relative probabilities for the positions of the mark on the paper strip, but no definite location for this mark."

Yet again, cf. the case of gunpowder pile, Einstein's critique was justifiable, and hence there exists until now no resolution to this paradox. Nevertheless, for Einstein the only way out was the acceptance of the statistical interpretation of quantum mechanics, which today is not the accepted majority standpoint. „One arrives at very implausible theoretical conceptions, if one attempts to maintain the thesis that the statistical quantum theory is in principle capable of producing a complete description of an individual physical system. On the other hand, those difficulties of theoretical interpretation disappear, if one views the quantum-mechanical description as the description of ensembles of systems." [18]

As the Schrödinger's cat has inspired many new interpretations of quantum theory, Schreiber in his Master of Science thesis, made an extensive comparison of them and wrote a very nice parody "The Nine Lives of Schrödinger's Cat" [20] where he describes the fate of the poor cat in the nine most popular alternative interpretations. In his conclusion, he says: „Discussion: The orthodox interpretation works well in practice but is ambiguous and therefore unacceptable. Bohr's epistemological views are also somewhat vague and require classical physics to be assumed. For those who are satisfied with epistemology, the decoherent histories approach admirably ties down any ambiguity or vagueness. It may be used to recover classical physics. Those who would like to see an ontological model of the world must look elsewhere. Neither many-world nor many-minds provide such models. The idea that the mind causes collapse does but the model is problematic. Bohm's interpretation does seem to work although it is rather awkward, especially in the context of quantum field theory. The best hope at the moment for an ontological theory is that the world's state collapses according to some fixed set of consistent histories. However, no clear criterion has emerged for the appropriate set of histories. Finally, there are approaches which involve not interpretation of quantum mechanics but modification. These approaches have not been considered here. The most successful of these is the stochastic scheme of 'GRW' (this scheme was put in a very nice form by Bell) although it is not without its problems. Several other attempts have involved introducing non-linear terms in the Schrödinger equation but none of these seems to work."

Schreiber then ends with his final anticipation - what should be done, but it never was: „Final word: Quantum mechanics is a theory lacking an ontological picture of the world. The search for such an ontology has been long, hard and appallingly haphazard. It is time that the entire programme was defined and analysed in a systematic and uniform mathematical way. Strange as this idea may seem, I am convinced that it is possible. When this programme is completed, theoretical physicists will finally be able to put the cat out and take a well-earned rest."

For Norton as well as for Schreiber, none of the present known interpretations is fully acceptable [17]: „My own feeling is that none of these responses is satisfactory… However, if there are new physical laws that would resolve the measurement problem, we can be pretty sure that they are quite exotic and not produced by a small adjustment in our existing theories. For, if these small adjustments are there to be found, eight decades of work by many of the brightest minds in quantum physics has failed to find them. Nonetheless, this "new, as yet unknown, physics" response is the one that all the other responses have to beat. Ask each of them, is this response more plausible than the "new physics" response? In each case, I answer no."

The two aforementioned conclusions suggest that we have to take on an apparent challenge searching for a radical new ontology and a new physics, based on deductive reasoning and rigorous mathematics. As we will show in this paper, this "new physics" imparts neither some new artificial topological spaces beyond the Hilbert space, nor some speculative multidimensional superstring proposals. All we need, should be to find nature's hidden laws, which have not yet been discovered and not yet revealed.

## 6. What is the Copenhagen interpretation?

The Copenhagen interpretation reflects a standard reference of our philosophical and physical views of quantum theory, yet there are remarkable differences of perspectives regarding the perception and understandings between the actual founders of quantum physics. Recognizing that the Copenhagen interpretation may not be complete, it will be valuable to remind the reader of the most important stances before starting to make any attempts to construct a more general theory.

Schlosshauer and Camilleri [21] takes up the unresolved disagreements between Bohr and Heisenberg: „In order to meaningfully speak of observation in quantum mechanics, Bohr concluded, "one must therefore cut out a partial system somewhere from the world, and one must make 'statements' or 'observations' just about this partial system" [22]… Bohr's epistemological demand for a "cut between the observed system on the one hand and the observer and his apparatus on the other hand" also became a key theme of Heisenberg's thinking. However, for Heisenberg, the object–instrument divide was coincident with the quantum–classical divide… It is important to realize that Heisenberg's views on the cut were somewhat at odds with Bohr's own view of the problem. Indeed, in an exchange of correspondence in 1935 Heisenberg and Bohr argued the point without resolution. Some twenty years later Heisenberg would report that "Bohr has emphasized that it is more realistic to state that the division into the object and rest of the world is not arbitrary" and that the object is determined by the very nature of the experiment [23]. In a letter to Heelan in 1975, Heisenberg also explained that he and Bohr had never really resolved their disagreement. Heisenberg remained convinced "that a cut could be moved around to some extent while Bohr preferred to think that the position is uniquely defined in every experiment" [24]."

Heisenberg's argument for the arbitrary position of the cut reads as follows: „The dividing line between the system to be observed and the measuring apparatus is immediately defined by the nature of the problem but it obviously signifies no discontinuity of the physical process. For this reason there must, within certain limits, exist complete freedom in choosing the position of the dividing line [25]."

Heisenberg's quantum-classical divide is also supported by von Weizsäcker, who tries to relate irreversibility and classicality: „Having thus accepted the falsity of classical physics, taken literally, we must ask how it can be explained as an essentially good approximation [when describing objects at the macrolevel]… This amounts to asking what physical condition must be imposed on a quantum-theoretical system in order that it should show the features which we describe as "classical."… I am unable to prove mathematically that the condition of irreversibility would suffice to define a classical approximation, but I feel confident it is a necessary condition [26]."

The position of an arbitrary cut was also shared by von Neumann [27]. He suggested that the entangled state of the object and the instrument collapses to a determinate state whenever a measurement takes place. So what finally causes such a collapse seems to be the mind of the observer. Other physicists also argued the latter boundary, i.e. the observer's consciousness, such as London [28], Peierls [29], Wigner [30], etc.

The philosopher Jan Faye, discusses, in his Stanford Encyclopedia article [31], the ambiguities, significances and implications of the Copenhagen interpretation, in particular Howard's and Henderson's careful analyses: „Don Howard [32] goes as far as concluding that "until Heisenberg coined the term in 1955, there was no unitary Copenhagen interpretation of quantum mechanics." The term apparently occurs for the first time in Heisenberg [33]. In addition, Howard also argues that it was Heisenberg's exposition of complementarity, and not Bohr's, with its emphasis on a privileged role for the observer and observer-induced wave packet collapse that became identical with that interpretation. Says he: "Whatever Heisenberg's motivation, his invention of a unitary Copenhagen view on interpretation, at the center of which was his own, distinctively subjectivist view of the role of the observer, quickly found an audience". This audience included people like Bohm, Feyerabend, Hanson, and Popper who used Heisenberg's presentation of complementarity as the target for their criticism of the orthodox view."

Furthermore he says: „Recently, Henderson [34] has come to a similar conclusion. He makes a distinction between different versions of Copenhagen interpretations based on statements from some of the main characters. On one side of the spectrum there is Bohr who did not think of quantum measurement in terms of a collapse of the wave function; in the middle we find Heisenberg talking about the collapse as an objective physical process but thinking that this couldn't be analyzed any further because of its indeterministic nature, and at the opposite side Johann von Neumann and Eugene Wigner argued that the human mind has a direct influence on the reduction of the wave packet. Unfortunately, von Neumann's dualistic view has become part of the Copenhagen mythology by people opposing this interpretation."

Wigner seems to be the last contributor of the philosophical conception of quantum theory, known today as the orthodox Copenhagen interpretation. Let us take a closer look on a more detailed study of Wigner's writings, such as the preprint of Primas and Esfeld [35], where Wigner's original definition of the measurement process reads:

„[The reduction of the wave packet] takes place whenever the result of an observation enters the consciousness of the observer – or, to be even more painfully precise, my consciousness, since I am the only observer, all other people being only subjects of my observations… The measurement is not completed until its result enters our consciousness. This last step occurs when a correlation is established between the state of the last measuring apparatus and something which directly affects our consciousness. This last step is, at the present state of our knowledge, shrouded in mystery and no explanation has been given for it so far in terms of quantum mechanics, or in terms of any other theory." [30]

Accordingly Primas & Esfeld exhibit Wigner's attempt to avoid the position of excessive subjectivity allowing the person's mind to be in a state of superposition, figuratively known under the name of Wigner's friend.

Moreover they continue: „Wigner's main argument for his view is that it is "difficult to accept the possibility that a person's mind is in a superposition of two states… We ourselves never have felt we were in such superpositions." [36] In this context, Wigner discusses the famous "paradox of Wigner's friend": If one observer describes another observer who observes something like a von Neumann chain and if the first observer describes this whole process in the terminology of quantum mechanics, he will end up with ascribing a superposition of different states of consciousness to the second observer. Wigner avoids this paradox by maintaining that quantum theory does not apply to consciousness so that there are no superpositions of different states of consciousness. A state reduction occurs at the level of the friend's consciousness [37]… Wigner's proposal is a move away from the Copenhagen idea that the quantum state represents knowledge available to a community of communicating observers, who have a common knowledge that is useful for making predictions about their combined future experiences. Wigner suggests that each conscious being is able to collapse one single objective quantum state, regardless of whether the information is actually physically shared. It is a move away from an essentially subjective pragmatic interpretation toward a more objective absolute one." From his later years Wigner's view matures in that he expresses a hope that a more general theory will be developed in the future, while being unsatisfied with the present incompleteness of his interpretation. In [38] it is cited: „In his later papers, Wigner expressed his "desire for a less solipsistic theory" but only in last papers did he consider the consequence of solipsism as a sufficient reason to repudiate his earlier views… At the end of a paper published in 1977, Wigner expresses the hope "that quantum mechanics will also turn out to be a limiting case, limiting in more than one regard, and that the philosophy which an even deeper theory of physics will support, will give a more concrete meaning to the word 'reality', will not embrace solipsism, much truth as this may contain, and will let us admit that the world really exists." [38]"

But what does this hope really mean in a more "realistic" description of the material world? Is it identical with Einstein's belief, quote – unquote: „Es gibt so etwas wie den 'realen Zustand' eines physikalischen Systems, was unabhängig von jeder Beobachtung oder Messung objektiv existiert und mit den Ausdrucksmitteln der Physik im Prinzip beschrieben werden kann." [40]? Primas & Esfeld [35] are convinced that the majority of contemporary scientists share this view with Einstein. Even if they are right, this is clearly not a good sign, since it means the comeback of 19th century materialism when classical physics fulfilled all requirements of "objective reality" as inferred above by Einstein.

The authors further polemize with Wigner, quoting some of their most interesting disagreements with him: „Wigner never discussed spontaneously broken symmetries, which are nowadays on all levels of physics, from elementary particle physics to molecular and solid state physics, of utmost importance… On the other hand, Wigner mostly analysed fictitious isolated systems with a finite number of degrees of freedom whose infinite environment he assumed to have a negligible effect. Yet, the interactions of an object system with its environment can have dramatic qualitative effects such as dressing, symmetry breakdown, and the emergence of qualitatively new properties. It is well known that the coupling of small molecular systems with the environment can lead to symmetry breakings which are of great biological significance such as molecular chirality."

Moreover: „Curiously for a scientist with an excellent background in chemistry, Wigner denies the objective reality of molecular chirality: "It should be admitted, first that the concept of the positions of atoms continues to be a useful concept but it has turned out to be an approximate one. It is useful under certain condition but it is not difficult to produce situations in which it is meaningless. A particular striking example is an optically active organic molecule in its normal state. This is neither right handed, nor left handed – in fact, the atoms have no clearly defined positions with respect to each other, not even approximately." [39]"

 – and: „Wigner showed much courage in relating the then unresolved questions of the measurement problem to the much deeper problem of consciousness. In view of this very unorthodox proposal it is astonishing that Wigner was very reactionary with respect of the dogmas of orthodox quantum mechanics. In contrast to von Neumann himself, he took the old von-Neumann codification of quantum mechanics as authoritative and not to be questioned. Much of the efforts to interpret the meaning of this codification and to prove no-go theorems, such as the insolubility of the measurement problem or the impossibility of a quantum theory of individual objects, are physically irrelevant since they are based on a codification of quantum mechanics that is valid only for strictly closed systems with finitely many degrees of freedom."

– and: „It is hard to believe that Wigner did not recognize the importance of the many possible physically inequivalent representations of the canonical commutation relations and the associated symmetry breakings for our understanding of molecular, mesoscopic, and macroscopic phenomena."

So where is the problem? Wigner knew of course what he was talking about. However, his final challenge for a more "realistic" theory does not meet the above mentioned counterarguments at all. Yet the incompleteness of the orthodox Copenhagen interpretation does not mean that it should be wrong, its incompleteness rather imparts a possible incompleteness of the concepts it is dealing with. It will be shown in this article that we must tackle the very incompleteness of our knowledge of (i) time irreversibility, (ii) the symmetry breaking, and (iii) the role of the environment and (iv) the quantum definition of individual objects. This means, that we cannot confront the incomplete orthodox interpretation by means of a reasoning based on incompletely defined concepts. In other words: these concepts must first be carefully re-examined before we can even start to construct any proposal regarding how to deal with an improved Copenhagen interpretation.

## 7. The paradox of quantum decoherence

In 1980 a novel decoherence program started to emerge, based on Everett's PhD thesis from 1957. Everett first summarised the Copenhagen interpretation in its final von Neumann - Wigner form [41]: „We take the conventional or "external observation" formulation of quantum mechanics to be essentially the following: A physical system is completely described by a state function $\psi$, which is an element of a Hilbert space, and which furthermore gives information only to the extent of specifying the probabilities of the results of various observations which can be made on the system by external observers. There are two fundamentally different ways in which the state function can change:

Process 1: The discontinuous change brought about by the observation of a quantity with eigenstates $\varphi_1$, $\varphi_2$, · · · , in which the state $\psi$ will be changed to the state $\varphi_j$ with probability $|(\psi, \varphi_j)|^2$.

Process 2: The continuous, deterministic change of state of an isolated system with time according to a wave equation $\partial\psi/\partial t = A\psi$, where $A$ is a linear operator. This formulation describes a wealth of experience. No experimental evidence is known which contradicts it."

Everett questioned the whole process of measurement - the problem already well-known at the time - namely the fact that the von Neumann view did only fit a certain type of measurement but is not able the explain all of them: „Von Neumann showed how to treat a special class of approximate measurements by the method of projection operators. However, a general treatment of all approximate measurements by the method of projections operators can be shown to be impossible… von Neumann's example is only a special case of a more general situation. Consider any measuring apparatus interacting with any object system. As a result of the interaction the state of the measuring apparatus is no longer capable of independent definition. It can be defined only relative to the state of the object system. In other words, there exists only a correlation between the two states of the two systems. It seems as if nothing can ever be settled by such a measurement." [41]

Furthermore Everett concentrated only on a set of measurement problems that differs from the von Neumann class, i.e. those excluding observers with a conscious mind and taking only into the consideration devices in the role of "observers". Quoting: „Not all conceivable situations fit the framework of this mathematical formulation. Consider for example an isolated system consisting of an observer or measuring apparatus, plus an object system. Can the change with time of the state of the total system be described by Process 2? If so, then it would appear that no discontinuous probabilistic process like Process 1 can take place. If not, we are forced to admit that systems which contain observers are not subject to the same kind of quantum-mechanical description as we admit for all other physical systems. The question cannot be ruled out as lying in the domain of psychology. Much of the discussion of "observers" in quantum mechanics has to do with photoelectric cells, photographic plates, and similar devices where a mechanistic attitude can hardly be contested. For the following one can limit himself to this class of problems, if he is unwilling to consider observers in the more familiar sense on the same mechanistic level of analysis." [41]

Up to this point the Everett's conclusion is certainly correct. But what about his proposal for a generalization of von Neumann's method based on projection operators? One should perhaps anticipate some generalizations where observers with a conscious mind together with the devices in the role of "observers" should be put side by side in some wider context. But in contrast he simply attempts to remove "Process l" from the Copenhagen postulates. In consequence it means to get rid of two axioms, i.e. the Born rule and the von Neumann - Wigner rule, downgrading the Born rule, not as an axiom but as an emergent rule from "Process 2". Everett finally arrives at his own eccentric and surreal interpretation of quantum mechanics, known now as the MWI (many-worlds interpretation): „We thus arrive at the following picture: Throughout all of a sequence of observation processes there is only one physical system representing the observer, yet there is no single unique state of the observer (which follows from the representations of interacting systems). Nevertheless, there is a representation in terms of a superposition, each element of which contains a definite observer state and a corresponding system state. Thus with each succeeding observation (or interaction), the observer state "branches" into a number of different states. Each branch represents a different outcome of the measurement and the corresponding eigenstate for the object-system state. All branches exist simultaneously in the superposition after any given sequence of observations." [41]

One can assume that Everett's understanding was inspired by Mach's principle of psychophysical paralysis [42] which otherwise also might have inspired Einstein in his theory of relativity. Everett assigns a similar role and position to the observer as ocurring in Einstein's theory of relativity. But there is one significant difference: Whereas in the theory of relativity the observer can be a "body without soul", quantum theory does not allow this in general. For instance the subject (the observer with conscious mind) and the object (the physical word) are not completely interchangeable, since there always must remain "something" from the object as well as from the subject. Goethe was not only a great poet but a great mystic as well. Once he said: „Alles, was im Subject ist, ist im Object und noch etwas mehr. Alles, was im Object ist, ist im Subject und noch etwas mehr." [43] And just this is the reason why the external reality hypothesis (ERH) should not be valid any more in quantum physics, which incorporates the roles of both - object and subject, in contrast with classical physics that concentrates fully only on the external material world. ERH was reformulated by Tegmark [44] as follows: „There exists an external physical reality completely independent of us humans." Hence Tegmark still believes in its validity and claims, and, if it is true, it is a sufficient condition for the justification of the MWI interpretation.

The MWI interpretation has recently inspired many physicists in their efforts to explain classicality as an emergent property of open quantum systems induced by their environments. This is best known as the quantum decoherence program. Zurek [45] introduced a new vernacular to describe the process of wavefunction collapse as well as elucidate the emergence of classical descriptions of reality appearing from quantum descriptions *viz.* einselection, the latter standing for environment - induced superselection. Although an excellent idea, it was inaptly set into the framework of the Everettian MWI. Such a concept will not work since it leads to a circular argument and therefore was rejected by a large part of the scientific community. Kastner [46] did performe a detailed analysis of this circularity quote, unquote: „In attempting to derive irreversible macroscopic thermodynamics from reversible microscopic dynamics, Boltzmann inadvertently smuggled in a premise that assumed the very irreversibility he was trying to prove: 'molecular chaos.' The program of 'Einselection' (environmentally induced superselection) within Everettian approaches faces a similar 'Loschmidt's Paradox': the universe, according to the Everettian picture, is a closed system obeying only unitary dynamics, and it therefore contains no distinguishable environmental subsystems with the necessary 'phase randomness' to effect einselection of a pointer observable. The theoretically unjustified assumption of distinguishable environmental subsystems is the hidden premise that makes the derivation of einselection circular. In effect, it presupposes the 'emergent' structures from the beginning. Thus the problem of basis ambiguity remains unsolved in Everettian interpretations."

It's surprising how many physicists, seeing the difficulties with the wave function collapse and the von Neumann - Wigner interpretation, yet jump happily into the above-mentioned decoherence program, not realizing the circularity of all these problems. For instance, before we can start to speak about some environmental subsystems, we must first attend to the third of Bohm's objections - wholeness versus fragmentation - discussed in section 4. Since quantum mechanics deals with isolated subsystems, hence we must decipher the puzzle formulated by Sutcliffe & Woolley, i.e. the isolated versus the individual, as considered in section 2. This is crucial to understanding the origin of 'emergent' structures, which if not will be falsely presupposed in the MWI interpretation, as Kastner correctly did notice. Only then can we begin a reasonable examination about the true meaning of Zurek's einselection. In passing one notes that Zurek later came forward with another promising idea, which he called quantum Darwinism [47] as a part of universal Darwinism. The thesis of universal Darwinism claims that Darwinian processes are not confined to biology and can be applied in other scientific branches. The Darwinian algorithm consists of three parts: 1) random variation in mutations or recombinations; 2) selection of the fittest variants; 3) heredity of these features in offsprings.

Darwin worried about randomness in the variation, since in his time classical physics determinism prevailed in philosophy and random processes descending from indeterministic principles were impermissible. This is not a problem today, when quantum physics has revealed such a principle. On the other hand, Darwin did not worry about the selective process, although he was convinced that it must be of a teleological origin, but in this case he was not in conflict with contemporary science. Exact sciences such as physics simply were and still are not able to incorporate teleology in an authentic, precise logical and mathematical manner, but at the same time it cannot be denied. In contrast to indeterminancy, which is opposite to determinancy, teleology is not opposite to causality (the opposite of "causal" is "acausal").

In a historical perspective Darwin was a teleologist. Lennox, for instance, shows in his paper [48] that Darwin uses the term 'Final Cause' consistently in his Species Notebook, the Origin of Species and also afterwards, and that evolution - as conceived by Darwin - is the result of mutations arising by chance and the selection which is teleological in nature. Here Lennox also quotes from the correspondence between Darwin and his close friend and colleague Asa Gray: „In an appreciation of Charles Darwin published in Nature in 1874, Asa Gray noted "Darwin's great service to Natural Science" lies in bringing back Teleology "so that, instead of Morphology versus Teleology, we shall have Morphology wedded to Teleology". Darwin quickly responded, "What you say about Teleology pleases me especially and I do not think anyone else has ever noticed the point."… Francis Darwin and T. H. Huxley reiterate this sentiment. The latter wrote that "…the most remarkable service to the philosophy of Biology rendered by Mr. Darwin is the reconciliation of Teleology and Morphology, and the explanation of the facts of both, which his view offers.""

Only a few biologists seem at present to be aware of the true meaning of teleology, at least  to the extent that was understood and expressed by Darwin. Today the majority of them appears quite confused. We can find many absurd propositions, such as "artificial selection is teleological and natural selection is not, since in the latter there is no intentional choice." Teleology is defined as downward causation or the final cause, and is quite independent of the fact whether there is some intentions behind or not. As examples one might mention two examples of statements from prominent biologists: a) the famous evolutionary biologist Ernst Mayr, although introducing a more explicit terminology, e.g. teleonomic processes as those owing its goal-directedness to the influence of an evolved program, expressed   [49]:

„adaptedness… is an *a posteriori* result rather than an a priori goal-seeking", b) Madrell [50]: „the proper but cumbersome way of describing change by evolutionary adaptation may be substituted by shorter overtly teleological statements for the sake of saving space, but this should not be taken to imply that evolution proceeds by anything other than from mutations arising by chance, with those that impart an advantage being retained by natural selection." One might wonder how much these attitudes of teleology of contemporary biologists really agrees with Darwin's original view.

The opinions and attitudes of most physicists regarding teleology are both negative and ignorant. In contrast Bohm (cf. the first objection mentioned in the section 4) warns us that disrespecting teleological principles prevent the construction of a full quantum theory. What is most surprising indeed, is that the teleology problem appeared for the first time already in classical physics. For instance Maxwell became immediately aware of it, when he studied the work of Boussinesq, Cournot, and St Venant dealing with the solution of hydrodynamic equations. Maxwell was fascinated by the occurrence of singular points leading to complete unpredictability of future states.

On 26 February 1879, he wrote a letter to Charles Darwin's half-cousin Francis Galton [51]: „Do you take any interest in Fixed Fate, Free Will &c. If so Boussinesq [of hydro dynamic reputation] 'Conciliation du véritable déterminisme mécanique avec l'existence de la vie et de la liberté morale' (Paris 1878) does the whole business by the theory of singular solutions of the differential equations of motion… There are certain cases in which a material system, when it comes to a phase in which the particular path which it is describing coincides with the envelope of all such paths may either continue in the particular path or take to the envelope (which in these cases is also a possible path) and which course it takes is not determined by the forces of the system (which are the same for both cases) but when the bifurcation of path occurs, the system, ipso facto, invokes some determining principle which is extra physical (but not extra natural) to determine which of the two paths it is to follow. When it is on the enveloping path it may at any instant, at its own sweet will, without exerting any force or spending any energy, go off along that one of the particular paths which happens to coincide with the actual condition of the system at that instant… But I think Boussinesq's method is a very powerful one against metaphysical arguments about cause and effect and much better than the insinuation that there is something loose about the laws of nature, not of sensible magnitude but enough to bring her round in time."

One must admire Maxwell's brilliant insights observing how he mentions teleological principles: principle which is extra physical but not extra natural; principle acting without exerting any force or spending any energy! Certainly, classical physics was unable to incorporate such a principle, but it begs the question: Is it possible to accomplish this task in quantum physics? Unfortunately most physicists are reluctant to tackle the problem, perhaps being victims of "causal fundamentalism", a wording introduced by Norton [52]: „Nature is governed by cause and effect; and the burden of individual sciences is to find the particular expressions of the general notion in the realm of their specialized subject matter."

Norton continues with a pessimistic feeling of vanity and emptiness that hovers over the current debate on the subject of teleology: „Aristotle described four notions of cause: the material, efficient, final and formal; with the efficient and final conforming most closely to the sorts of things we would now count as a cause. The final cause, the goal towards which a process moves, was clearly modelled on the analogy between animate processes and the process of interest. In the seventeenth century, with the rise of the mechanical philosophy, it was deemed that final causes simply did not have the fundamental status of efficient causes and that all science was to be reconstructed using efficient causes alone. Although talk of final causes lingers on, this is a blow from which final causes have never properly recovered."

Here lies a profound challenge for all of us. Until the teleological principle is adequately implemented in the mathematical formalism of quantum theory, the whole program of quantum decoherence, based on universal Darwinism, suggests nothing more than many years of squandering time. It imparts, e.g. that the von Neumann - Wigner interpretation can never be removed and that the causal Born rule cannot be replaced by einselection, the latter being inspired by teleological Darwinian adaptedness. Instead, the Born and einselection rules should coexist side by side, but first after the true teleological meaning and quantum origin of the einselection have been revealed.

## 8. The paradox of free will

What is meant by the notion of free will? Does free will really exist at all? Observe that in the era of classical physics it was generally accepted that all things in the universe do evolve in a strictly deterministic manner, and hence there was obviously no room for any justification of a free will. But what happened after the arrival of quantum mechanics, and the knowledge of the indeterministic laws of quantum physics? Is this agreement sufficient to give a reasonable explanation of free will? As will be seen below, the problem of a proper understanding of our free will still remains.

The information philosopher, Robert Doyle summarizes, in his 2011 book "The Scandal in Philosophy" [53], in great detail the whole history of opinions of famous philosophers and scientists on the topic of free will. He maintains that there prevails a sceptical view regarding possible satisfactory free will explanations, which he calls the "standard argument against free will" [53]: „The standard argument has two parts. 1) If determinism is the case, the will is not free. We call this the Determinism Objection. 2) If indeterminism and real chance exist, our will would not be in our control. We could not be responsible for our actions if they are random. We call this the Randomness Objection. Together, these objections can be combined as a single Responsibility Objection, namely that no Free Will model has yet provided us an intelligible account of the agent control needed for moral responsibility."

From the standpoint of a historical retrospective analysis, it seems that only the two-stage model could be a true candidate for the explanation of free will. This two-stage model consists of two processes: the process of random alternatives followed by the selection process of one choice. Doyle writes in his book that this idea appeared for the first time in 1884 and that it was formulated by William James. As a psychologist he felt no problem in denying determinism in the psychological domain of free will: „The stronghold of the determinist argument is the antipathy to the idea of chance... This notion of alternative possibility, this admission that any one of several things may come to pass is, after all, only a roundabout name for chance." [54]

It is interesting to note how physicists in general accepted the two-stage model. The first reaction came from Poincaré, at the time when quantum indeterminism had not yet arrived. Hence he does not refer to the laws of physics, concentrating mostly on the role of our unconscious mind as a source of random combinations and possibilities, but also on our conscious mind where the definite process of validation occurs [55]: „It is certain that the combinations which present themselves to the mind in a kind of sudden illumination after a somewhat prolonged period of unconscious work are generally useful and fruitful combinations... All the combinations are formed as a result of the automatic action of the subliminal ego, but those only which are interesting find their way into the field of consciousness... A few only are harmonious, and consequently at once useful and beautiful, and they will be capable of affecting the geometrician's special sensibility I have been speaking of; which, once aroused, will direct our attention upon them, and will thus give them the opportunity of becoming conscious... In the subliminal ego, on the contrary, there reigns what I would call liberty, if one could give this name to the mere absence of discipline and to disorder born of chance."

The physicist, Sir Roger Penrose shares the synoptical view of the two-stage model with Poincaré, particularly as concerns the role of our unconscious and conscious minds [56]: „In relation to this, the question of what constitutes genuine originality should be raised. It seems to me that there are two factors involved, namely a 'putting-up' and a 'shooting-down' process. I imagine that the putting-up could be largely unconscious and the shooting-down largely conscious. Without an effective putting-up process, one would have no new ideas at all. But, just by itself, this procedure would have little value. One needs an effective procedure for forming judgements, so that only those ideas with a reasonable chance of success will survive. In dreams, for example, unusual ideas may easily come to mind, but only very rarely do they survive the critical judgements of the wakeful consciousness. In my opinion, it is the conscious shooting-down (judgement) process that is central to the issue of originality, rather than the unconscious putting-up process; but I am aware that many others might hold to a contrary view."

A similar description of the two-stage model was also promoted by Compton [57]: „A set of known physical conditions is not adequate to specify precisely what a forthcoming event will be. These conditions, insofar as they can be known, define instead a range of possible events from among which some particular event will occur. When one exercises freedom, by his act of choice he is himself adding a factor not supplied by the [random] physical conditions and is thus himself determining what will occur." Compton was the first physicist, who discussed openly the relationship between free will and the quantum indeterminism and the randomness on the first stage of the two-stage model, i.e. what Penrose calls the 'putting-up' process. On the other hand, he was very pessimistic about the scientific explanation of the second stage - the act of choice, or Penrose's 'shooting-down' process.

It is notable that present-day quantum physics is still incomplete and unable to create a paradigmatic background for the justification of the second stage of free will [57]: „Let me then summarize how a physicist now views man's freedom. It is not from scientific observation that we know man is free. Science is incapable of telling whether a person's acts are free or not. Freedom is not something that one can touch or measure. We know it through our own innermost feelings. The first essential of freedom is the desire to attain something that one considers good. But desire lies outside the realm of science — at least outside of physics. You can't locate desire as somewhere in space. Similarly our recognition that within limits we can do what we try to do is not a matter of measurement or of external observation. It is a matter of immediate awareness. There is nothing in such awareness of freedom that is inconsistent with science. Freedom does, however, involve the additional determining factor of choice, about which science tells us nothing."

The citation above means that quantum theory, based on standard interpretations such as the orthodox Copenhagen view, forms only a necessary, but not a sufficient foundation for the explanation of the free will. This circumstance is clearly confirmed by Flanagan [58]: „Free actions, if there are any, are not deterministically caused nor are they caused by random processes of the sort countenanced by quantum physicists or complexity theorists. Free actions need to be caused by me, in a non-determined and non-random manner… I am open to there being genuine ontological indeterminacy at both the quantum level and the level of neural processing. But the attempt to gain free will from indeterminacy at the quantum level or at the level of global brain processes is a bad idea. The last thing anyone wants is for free will to be the result of random causal processes."

It is fetching to learn that Flanagan's representation of "non-determinism and non-randomness" as the main characteristics of free actions, appears under the name "adequate determinism" in Doyle's Cogito model [53]: „Free will is a combination of microscopic randomness and macroscopic adequate determinism, in a temporal sequence - first chance, then choice. Creative thoughts come to us freely. But our actions come from our adequately determined wills. Randomness is the "free" in free will. Adequate determinism explains the "will" in free will." One can understand adequate determinism as something indeterministic in principle, but unlike the laws of microscopic physics, it must not be random. Doyle thinks that a macroscopic ensemble, which statistically restrains randomness, fulfils this requirement. But our decisions are not statistical in character, since we often have to decide in a binary style such as "either - or". Moreover, our decisions have teleological traits, while the fortitude of macroscopic physics has nothing to do with it. One might think that Doyle is not aware of the incompleteness of quantum theory and therefore tries to explain free will at any cost, by only using the traditional understanding of quantum physics. The issue of adequate determinism must be solved exclusively on the quantum level. However, up til now, presently known interpretations of quantum theory are not capable of dealing with phenomena, which are simultaneously non-deterministic and non-random.

One notes that Compton's discussion of the closeness between quantum indeterminism and randomness in free will finally did convince Popper, who originally denied any influence of microphysical laws upon the psychological domain of free will. Besides, the similarity of the two-stage models, used in Darwin theory and in the concept of free will, caused a radical change of Popper's attitude towards natural selection, which he at first deemed to be a "tautology" making it unfalsifiable. In his 1977 Darwin Lecture, Darwin College Cambridge, Popper confessed his change of mind [59] by stating: „The selection of a kind of behavior out of a randomly offered repertoire may be an act of choice, even an act of free will. I am an indeterminist; and in discussing indeterminism I have often regretfully pointed out that quantum indeterminacy does not seem to help us; for the amplification of something like, say, radioactive disintegration processes would not lead to human action or even animal action, but only to random movements. This is now the leading two-stage model of free will. I have changed my mind on this issue. A choice process may be a selection process, and the selection may be from some repertoire of random events, without being random in its turn. This seems to me to offer a promising solution to one of our most vexing problems, and one by downward causation."

This formulation anticipates the concept of unification of a single two-stage model that holds both for Darwinism and free will. Eccles and Popper [60] formulate the view as: „New ideas have a striking similarity to genetic mutations. Now, let us look for a moment at genetic mutations. Mutations are, it seems, brought about by quantum theoretical indeterminacy (including radiation effects). Accordingly, they are also probabilistic and not in themselves originally selected or adequate, but on them there subsequently operates natural selection which eliminates inappropriate mutations. Now we could conceive of a similar process with respect to new ideas and to free-will decisions, and similar things. That is to say, a range of possibilities is brought about by a probabilistic and quantum mechanically characterised set of proposals, as it were − of possibilities brought forward by the brain. On these there then operates a kind of selective procedure which eliminates those proposals and those possibilities which are not acceptable to the mind."

Although one might, in such a way, achieve a unification of nature's laws of biology and psychology, there still remains a fundamental problem: There appears a void or lack of a hitherto unravelled fundamental natural law in science in general and quantum theory in particular, which is mandatory for a proper understanding and justification of the second part of two-stage model. This is the conundrum of the selection process in biological evolution and the choice process in the psychology of free will. Contemporary quantum theory allows only an explanation of the first part of two-stage model, which is not sufficient for a free will justification. In fact this is exactly in line with Bohr's opinion [61]: „To connect free will more directly with the limitation of causality in atomic physics, as it is often suggested, is, however, entirely foreign to the tendency underlying the remarks made here about biological problems." Bohr knew that quantum physics gives no answer to the selection process (the second part of two-stage model), and he therefore only called upon the recognition of relevant psychological experiences [61]: „To illustrate the argument, we may briefly refer to the old problem of free will. From what has already been said it is evident that the word volition is indispensable to an exhaustive description of psychical phenomena, but the problem is how far we can speak about freedom to act according to our possibilities. As long as unrestricted deterministic views are taken, the idea of such freedom is of course excluded. However, the general lesson of atomic physics, and in particular of the limited scope of mechanistic description of biological phenomena, suggests that the ability of organisms to adjust themselves to environment includes the power of selecting the most appropriate way to this purpose. Because it is impossible to judge such questions on a purely physical basis, it is most important to recognize that psychological experience may offer more pertinent information on the problems."

Bohr's statement above represents a synoptical vision, cf. the one of Eccles and Popper, unifying the principles behind biological and psychical phenomena. However, the problem of the second part of the two-stage model must indeed have a solution within a satisfactory extension of the present understanding of quantum theory, originally formulated for the description of inanimate nature, but such a solution does not yet exist on the level of quantum physics as restricted by the Copenhagen interpretation.

## 9. The paradox of the Meissner effect

The strange phenomenon of superconductivity was coined by Heike Kamerlingh Onnes in his two ground breaking papers [62], describing the complete disappearance of electrical resistance of mercury observed in 1911 by his assistant Gilles Holst at the Leiden research lab later known as the Kamerlingh-Onnes laboratory. Since then superconductors have been regarded as ideal conductors with zero resistance. The first theoretical articles on the subject of superconductivity, by Becker, Heller and Sauter [63], reflected this ideal conductivity, deriving the relation between the current and the electrical field, replacing Ohm's law in normal conductors. Starting with the expression for current

$$\mathbf{j} = en\mathbf{v} \tag{9.1}$$

where $e$ is the electron charge, $n$ the density of superconducting carriers and $\mathbf{v}$ the superfluid velocity, and further using Newton law for the carriers with mass $m$, they obtained the final expression, Becker's law

$$\frac{m}{ne^2}\frac{\partial \mathbf{j}}{\partial t} = \mathbf{E} \tag{9.2}$$

From (9.2) the expulsion of the magnetic field from superconductors can be predicted assuming that the external magnetic field was switched on after cooling below the critical temperature. However, if the order of these events was altered, the magnetic field was expected to remain inside the superconductor without any change. The subsequent experiment by Meissner and Ochsenfeld [64] brought a huge surprise, i.e. real superconductors, unlike ideal conductors, expelled the magnetic field from their interior even in the presence of a constant field with the succeeding transition into the superconducting state. The Meissner effect is hence in strong contradiction with Becker's law, since a constant field cannot produce any electromotive force bringing superconducting carriers in motion.

Two years later the so-called London equations were derived by Heinz and Fritz London [65] yielding phenomenological equations in compliance with the Meissner effect. First by using Faraday's law

$$\nabla \times \mathbf{E} = -\frac{1}{c}\frac{\partial \mathbf{B}}{\partial t} \tag{9.3}$$

they rewrote the Becker's equation (9.2) in its curl form

$$\frac{m}{ne^2}\nabla \times \frac{\partial \mathbf{j}}{\partial t} = -\frac{1}{c}\frac{\partial \mathbf{B}}{\partial t} \tag{9.4}$$

and then integrating it over time

$$\nabla \times \mathbf{j} = -\frac{ne^2}{mc}\mathbf{B} \tag{9.5}$$

putting the integration constant to zero as a specific ansatz for true superconductors, where the London equation (9.5) replaces Ohm's law in conductors. Applying Ampere's law

$$\nabla \times \mathbf{B} = \frac{4\pi}{c}\mathbf{j} \tag{9.6}$$

this equation can be written in the form

$$\lambda^2 \Delta \mathbf{B} = \mathbf{B} \tag{9.7}$$

where $\lambda$ stands for the London penetration depth

$$\lambda = \sqrt{\frac{mc^2}{4\pi ne^2}} \qquad (9.8)$$

The London equation (9.5) differs from the Becker's one (9.4) in one small detail: an ansatz in the form of zero integration constant, corresponding to the simple but important fact, that real superconductors, in contrast to ideal conductors, exhibits no memory effects.

An similar distinction is observed for the case of rotating objects, i.e. both ideal conductors and superconductors exhibit the same behavior, if they are accelerated below the critical temperature. Becker & al. [63] derived the relation between the angular velocity $\boldsymbol{\omega}$ and a generated magnetic field in the interior of ideal conductor. Substituting for $\mathbf{v}$ in Eq. (9.1) i.e.

$$\mathbf{v} = \boldsymbol{\omega} \times \mathbf{r} \qquad (9.9)$$

one obtains from the Becker's equation (9.4) after integration the final relation

$$\mathbf{B} = -\frac{2mc}{e}\boldsymbol{\omega} \qquad (9.10)$$

If an ideal conductor is first accelerated and subsequently cooled, instead of (9.10) after the integration, we get only the trivial solution with $\mathbf{B}$ equal to zero. Related to the Meissner effect, Fritz London [66] predicted an analogous effect also for rotating superconductors, where the Becker's relation (9.10) holds for superconductors regardless of its history, i.e. including the case when acceleration occurred first, followed by the cooling. London's derivation of Eq. (9.10) is based on a direct substitution in his Eq. (9.5), so that the generated magnetic field would always be the same, independent of the order of the acceleration and the cooling events. The generated magnetic moment is called the London moment. London's prediction for rotating superconductors was for the first time experimentally verified in 1964 by Hildebrandt [67].

It is intriguing to ask whether London's ansatz to Becker's equations is necessary? Alternatively, can we have an ansatz-free classical description of superconductors with an *ab-initio* derivation of the London equations based solely on the laws of classical mechanics and classical electrodynamics? This question has in fact divided the scientific community into two camps. For instance equation (9.5) was elegantly derived by de Gennes [68] by minimizing the total energy. He wrote the expressions for the kinetic energy and the energy of magnetic field:

$$E_k = \frac{1}{2}\int nm\mathbf{v}^2 dV = \frac{1}{2}\int \frac{m}{ne^2}\mathbf{j}^2 dV \qquad (9.11)$$

$$E_f = \frac{1}{8\pi} \int \mathbf{B}^2 dV \qquad (9.12)$$

Using Ampere's law (9.6) the total energy reads

$$E = E_k + E_f = \frac{1}{8\pi} \int \left[ \mathbf{B}^2 + \lambda^2 \left( \nabla \times \mathbf{B} \right)^2 \right] dV \qquad (9.13)$$

and by minimizing this expression with respect to $\mathbf{B}$ he got

$$\mathbf{B} + \lambda^2 \nabla \times \left( \nabla \times \mathbf{B} \right) = 0 \qquad (9.14)$$

which is exactly the London equation (9.7). This derivation was recapitulated by Essén and Fiolhais quoting [69]: „The conclusion of de Gennes is clearly stated in his 1965 book [68] (emphasis from the original): "The superconductor finds an equilibrium state where the sum of the kinetic and magnetic energies is minimum, and this state, for macroscopic samples, corresponds to the expulsion of magnetic flux." In spite of this, most textbooks continue to state that "flux expulsion has no classical explanation" as originally stated by Meissner and Ochsenfeld [64] and repeated in the influential monographs by Fritz London [66] and Nobel laureate Max von Laue [70].“ They continue with a list of authors preferring classical derivations of the Meissner effect [69]: „Many other authors have also reached the conclusion that the equilibrium state of a superconductor is in fact the state of minimum magnetic energy of an ideal conductor. Some of these are, in chronological order, Cullwick [71], Pfleiderer [72], Karlsson [73], Badía-Majós [74], Kudinov [75], and Mahajan [76]. For example, Cullwick writes "The well-known Meissner effect in pure superconductors is shown to be an expected rather than an unexpected phenomenon...". " And finally quoting [69]: „The reader may get the impression from our investigations above that we consider superconductivity to be a classical phenomenon. Nothing could be further from the truth... Since quantum physics must lead to classical physics in some macroscopic limit, it must be possible to derive our classical result from a quantum perspective. What we want to correct is the mis-statement that the Meissner effect proves that superconductors are "not just perfect conductors." According to basic physics and a large number of independent investigators, the specific phenomenon of flux expulsion follows naturally from classical physics and the zero resistance property of the superconductors — they are just perfect conductors.“

Summarizing this is the paradox of the Meissner effect. We have two camps of physicists - the first follows Meissner and London, and claims, that superconductivity in contrast to ideal conductivity has no classical explanation, and that quantum physics is necessary for its explanation. The second camp on the other hand identifies superconductors with perfect classical conductors. Who is right? A personal answer might appear shocking: no one! Perhaps a more correct resolution of this dilemma, caused by the two factions of physicists, might open the door for an understanding of the true nature of superconductivity, helping us to find accurate equations and precise descriptions and explanation.

As one can see, Becker's and London's equations are almost identical except one small but important difference: the latter ones have no history, indicating that causality is broken. Now, does some archetype of causality breaking occur in classical physics? Yes, it does and it refers back to Norton's dome, described in Ch. 3 "A-causality in Classical Physics", Ref. [52]: „The dome sits in a downward directed gravitational field, with acceleration due to gravity $g$. The dome has a radial coordinate $r$ inscribed on its surface and is rotationally symmetric about the origin $r = 0$, which is also the highest point of the dome. The shape of the dome is given by specifying $h$, how far the dome surface lies below this highest point, as a function of the radial coordinate in the surface, $r$. For simplicity of the mathematics, we shall set $h = (2/3g)r^{3/2}$. (Many other profiles, though not all, exhibit analogous a-causality.)" Norton's dome has two kinds of solutions. The first one is trivial and describes the situation where the mass simply remains at rest at the apex for all time:

$$r(t) = 0 \tag{9.15}$$

The second corresponds to another large class of unexpected solutions. For any radial direction the following equations hold:

$$r(t) = 0; \ \ t \leq T \tag{9.16a}$$

$$r(t) = (1/144)(t - T)^4; \ \ t \geq T \tag{9.16b}$$

where $T$ is an arbitrarily chosen, positive constant.

For this reason, the paradox associated with the Meissner effect appears to be of the same origin as the Norton dome paradox. From the point of view of classical physics ideal conductors and superconductors represent one common Norton-dome-like system with two solutions: one causal for ideal conductivity with the trivial solution, and one teleonomic for superconductivity, allowing the application of the principle of energy minimum reminiscent of de Gennes' derivation of London's equation. It is really a curiosity of classical physics, that both solutions - causal and telic - are still entirely consistent with the mathematics of Newton's laws of motion. Unfortunately, the teleological features of superconductivity have not yet been revealed, so a true understanding of this phenomenon has remained concealed.

The Nobel Laureates Ginzburg and Landau elaborated the description and the structure of the classical phenomenological description of superconductors by simulating the free energy $F$ in terms of a complex order parameter $\Psi$ [77]:

$$F_{GL} = \alpha |\Psi|^2 + \frac{\beta}{2} |\Psi|^4 + \frac{\hbar^2}{2m} \left| \left( \nabla - \frac{2ie}{\hbar c} \mathbf{A} \right) \Psi \right|^2 + \frac{1}{8\pi} |\mathbf{B}|^2 \tag{9.17}$$

which after minimizing with respect to variations in the order parameter $\Psi$ and the vector potential $\mathbf{A}$ leads to the celebrated Ginzburg-Landau equations. Eq. (9.17), where $\alpha$, $\beta$ are phenomenological parameters, contains both quadratic and biquadratic functions of the order parameter, forming the well-known Mexican hat, which is similar to the Norton dome, albeit being fully Lipschitz continuous. Actually, the solution of Eq. (9.17) based on the minimization process corresponds to the acausal Norton dome solution (9.16).

Returning to the controversy, related to the interpretation of the Meissner effect, it is true that both camps believe in the universal validity and consequences of Bohr's correspondence principle, i.e. that provided ideal conductors are (are not) identical with superconductors in classical physics, they are (are not) identical in quantum physics as well, and vice versa. Adding the fact, that superconductivity has a teleonomic component, Bohr's correspondence principle does not hold in this case, since quantum physics, at least in its Copenhagen version, is solely causal. This limitation of the Copenhagen interpretation was confronted by Bohm in his first critical point discussed in section 4. The consequences of this ignorance are significant. As already stated, quantum physics cannot deal with teleological phenomena implying that it, in view of what has been said above, cannot explain superconductivity. It can hence only describe superconductors and distinguish them from ideal conductors. Taking the point further, it cannot realize telicity in the superconducting phenomenon, indicating that the only possibility to describe superconductors within quantum physics would be to regard them as insulators. Of course, the official attitude of physicists today is different and the majority still believes that superconductivity is essentially explained within the framework of quantum physics. Nonetheless, such calculations were not fully logical since they conflated the classical teleological Norton-dome-like potentials with microscopic quantization procedures.

So we reach a fascinating conclusion: Classical physics does not distinguish superconductors from ideal conductors (it views them as one system with two possible solutions - one teleological, one causal), whereas quantum physics on the other hand does not distinguish them from insulators (since the Bohr principle of correspondence cannot be applied to teleological phenomena). We bring forward now three other confirmations that contemporary quantum physics, based on the Copenhagen interpretation, cannot really explain superconductivity.

Quantum theory of superconductivity must necessarily explain the Meissner effect. But it was never done: either in the first officially accepted microscopic BCS theory [78], or in any of its clones or replacements. The problem is that the electromotive forces described by Faraday's law of induction are equal to zero in stationary conditions of the Meissner effect, whereas the existing theories do not suggest any other electric forces needed to accelerate the electrons until the steady state supercurrent described by the London equation is achieved. Many physicists were inspired by this problem, and in the last decade we have observed several attempts to solve it, but such solutions go beyond the boundaries of contemporary classical or quantum physics. E.g. Kozynchenko [79] proposed a way out that requires an *ad-hoc* reformulation of the laws of classical electrodynamics for superconductors. On the other hand, Hirsch [82] did submit another solution, transcending *ad-hoc* laws of quantum physics. As he wrote many notable and significant articles devoted to this topic, we will quote and discuss some of them below.

In his paper "The origin of the Meissner effect in new and old superconductors" Hirsch [80] summarizes the position of all contemporary theories of superconductivity particularly regarding their inability to explain the Meissner effect: „In a somewhat circular argument, a 'conventional' superconductor is defined to be a superconductor described by BCS theory. In addition there are by now at least 10 different classes of materials that are generally believed to be 'unconventional', i.e. not described by BCS theory... And certainly there is no single 'unconventional mechanism' proposed to describe all unconventional superconductors: new mechanisms are being proposed that apply specifically to one family only, e.g. the cuprates, or the iron pnictides, or the heavy fermion superconductors. However all superconductors, whether conventional or not, exhibit the Meissner effect. I argue that BCS theory cannot explain the Meissner effect, so it cannot explain any superconductor. Furthermore, none of the unconventional mechanisms proposed to explain 'unconventional' superconductivity has addressed the question of how to explain the Meissner effect. I argue that none of these mechanisms describe any superconductor because they cannot explain the Meissner effect."

The problem of the transition into the superconducting state after cooling is not the only one. For instance, what goes on when, beyond the critical temperature, a superconductor becomes conductor again. What happens with the momentum of the supercurrent? Hirsch, in his paper "The disappearing momentum of the supercurrent in the superconductor to normal phase transformation" [81] writes: „A conundrum that didn't exist before was thus created by Meissner's discovery: if there are no collision processes that dissipate Joule heat in the superconductor-to-normal transition in the presence of a magnetic field, what happens to the mechanical momentum of the disappearing current? The kinetic energy of the current is 'stored' in the normal state electronic state, but its momentum is not. Of course the only possible answer is that the momentum of the current is transmitted to the body as a whole. But what is the physical mechanism by which this transfer of momentum happens with no energy transfer and no energy dissipation? Surprisingly this basic and fundamental question has never been asked (nor answered) in the extensive literature on superconductivity since 1933 (213,616 papers according to the Web of Science)."

In the author's opinion, Hirsch is absolutely right, as far as the two statements given above. Does it mean that the Meissner effect and superconductivity have no quantum physical solutions? Hirsch thinks that they have, and revives in his work "The Bohr superconductor" [82] some old ideas from the pre-BCS era. He shows that any superconductor, described classically, is equivalent to a system of charge carriers rotating in the whole bulk in orbits of radius $2\lambda$. He presents several ways of solution; the simplest and most transparent is the following: The total angular momentum of the Meissner current in a long cylinder of radius $R$ and height $d$ with applied magnetic field parallel to the cylinder axis can be written in the two equivalent forms

$$L = \left(2\pi R d \lambda n\right)\left(mvR\right) = \left(\pi R^2 dn\right)\left(mv\left(2\lambda\right)\right) \tag{9.18}$$

where the first expression describes the angular momentum of the supercurrent flowing within $\lambda$ of the surface, and the second one describes the angular momentum of all the charge carriers in the bulk in their orbits of radius $2\lambda$. The angular momentum of each electron in its circular orbit yields

$$l = mv\left(2\lambda\right) \tag{9.19}$$

From these results Hirsch infers: „It is natural to conclude that electrons reside in such orbits even in the absence of an applied magnetic field, as opposed to assuming that the $2\lambda$ orbits are somehow 'created' by the applied field... It should be noted that the hypothesis that superconducting electrons reside in large orbits was made by several researchers in the pre-BCS era [83–85]."

At this point a serious problem appears, i.e. when the classical concept of mesoscopic orbits is quantized (for details of the derivation of quantum equations see the original paper). This forces Hirsch to declare that such a solution exceeds the rules of quantum physics: „The Bohr atom starts from some simple assumptions and deduces that the angular momentum of the electron in Bohr orbits is quantized in integer units of $\hbar$... Similarly we point out here that from some simple assumptions it can be deduced that electrons in superconductors reside in mesoscopic orbits with orbital angular momentum $\hbar/2$... The fact that the orbital angular momentum in these orbits in the superconductor is found to be a half integer rather than an integer multiple of $\hbar$ is very remarkable. The correct interpretation of this finding could have profound implications. We have suggested that it indicates an intrinsic double-valuedness of the electron wavefunction, in contradiction with conventional quantum mechanics. Other less radical interpretations may be possible." Hirsch closes his paper, expressing his desire to find the wavefunction, that will obey the above mentioned concept of mesoscopic electron orbits: „What we are proposing is that the correct wavefunction of the superconductor, when it is found, will necessarily show physical properties consistent with the picture provided by quantized $2\lambda$ orbits with angular momentum $\hbar/2$ and the associated macroscopically inhomogeneous charge distribution. The BCS wavefunction does not."

No one up to now has been lucky to discover such a wavefunction, including Hirsch himself. According to Bohr's principle of correspondence, this wavefunction must exist being derived *ab initio* from first principles, i.e. only from the knowledge of the Schrödinger equation taking into account a certain number of nuclei and electrons. One should note that superconductors are crystals with the same discrete translational symmetry as other solids, yet from a mathematical point of view it is impossible to find some type of "rotational symmetry" leading to the solution of mesoscopic circle orbits. How is it possible to unravel such a puzzle? The answer is simple repeating the statement that Bohr's principle of correspondence for superconductors is broken and quantum physics is unable to explain them.

Continuing the focus will be set on the problem of rotating superconductors, especially on the relation between the magnetic field and the angular velocity (9.10). The obvious question befalls: what type of mass $m$ does appear in this equation? Is it a bare or an effective electron mass? Hirsch argues in his paper "The London moment: what a rotating superconductor reveals about superconductivity" [86] and quotes a huge number of experimental work dealing with measurements of both the low-$T_c$ as well as the high-$T_c$ superconductors. The results confirm without exception the validity of using the bare electron mass in the Eq. (9.10). Hirsch continues: „BCS theory does not describe any change in the character of the electronic states other than the pairing correlations, and the electric current in the normal state is carried by carriers with effective mass $m^*$ rather than bare mass $m$, so BCS theory is in disagreement with experiment. Furthermore, BCS theory attributes the pairing of the carriers to the electron-phonon interaction, which is completely inconsistent with the evidence that superfluid carriers become free of interactions with the ionic lattice."

Hirsch further proposes the concept of mesoscopic orbits as a possible solution for the mass problem in the London moment. As already mentioned, no one has found any microscopic wave function describing such mesoscopic orbits, since it is impossible to derive from the mathematical aspects of crystal symmetry. Does there exist some microscopic explanations in favor of the bare mass? Naturally one cannot avoid the universal concept of Bloch states for the description of carriers, and it does not matter if these carriers are paired or unpaired. If they represent original electrons or are replaced by polarons, bipolarons etc., we will always arrive at the effective mass which contradicts the experimental facts of the bare mass in the London moment. For the third time we conclude that Bohr's principle of correspondence is broken for superconductors and quantum physics is unable to explain them.

Finally we turn our attention to the experimental constituent in Eq. (9.1). While the macroscopic current is measurable, the density and velocity of carriers are not. In his last paper [87] Hirsch proposes a method for the velocity measurement: „Experimentally the speed of the supercurrent in superconductors has never been measured and has been argued to be non-measurable, however we point out that it is in principle measurable by a Compton scattering experiment. We predict that such experiments will show that superfluid carriers respond to an applied magnetic field according to their bare mass, in other words, that they respond as free electrons, undressed from the electron-ion interaction, rather than as Bloch electrons."

If superconductivity is going to have a microscopic explanation, Hirsch's request is absolutely legitimate. The velocity of carriers in superconductors must be measurable in the same manner as in conductors. Does this indicate that experimental physicists are lazy, incompetent or badly equipped? Certainly not! They must have deeper reasons why such measurements are infeasible. Hence, for the fourth time we conclude that for superconductors Bohr's principle of correspondence is broken and quantum physics is unable to explain them.

Since the discovery of high-$T_c$ superconductivity dozens, of various microscopic theories have been proposed. Every theory starts from some effective Hamiltonian, from which the macroscopic phase of the supercurrent should be derived.. To the present day there is no agreement among scientists, which one is natural, correct, proper, exact or authentic. Every theory attempts to see something in place of the supercurrent carriers. We argue that a true microscopic theory must not see anything, i.e. must describe superconductors as insulators, because the explanation of teleological traits in the phenomenon of superconductivity goes beyond quantum physics as being limited by the Copenhagen interpretation.

In this section we have discussed various attempts to explain and understand the Meissner effect, requiring always some *ad hoc* extension beyond either the classical or the quantum physical formulation. Moreover, if looking at any author, who tries to find some causal origin of the Meissner effect, one always see huge complicated drawings of possible mechanisms along with long causal chains. One can interminably ask what is behind the specification of the first link of the chain and what is its own cause. Such drawings remind us of medieval designs for the construction of a perpetuum mobile.

However, the problem of so-called teleological phenomena is quite different. Its characteristics are not *ad hoc* only for the Meissner effect and superconductivity. As we will demonstrate in the next section, it has played an enormous role in understanding the electroweak and the strong interactions, and, as we have mentioned in section 5, it also concerns Universal Darwinism. From here it has inspired quantum decoherence and, as outlined in section 6, contributed to the elucidation of our decisions in the problem of free will. Phrased differently, in the Meissner effect we have encountered a teleological principle, which exceeds the domains of superconductivity and solid state physics in general and, in an astonishing way, reaches three major provinces of the world; inanimate nature, animate nature, and psychology.

## 10. The paradox of Higgs bosons

As we have seen the teleological principle has not been acknowledged as a compelling argument for the understanding of the phenomenon of superconductivity, which led the scientific community to accepting causal quantum-physical explanations such as the BCS theory. This did impact the construction of elementary particle mechanisms, which, to a high degree, were inspired by the BCS theory of superconductivity. The most famous example is the celebrated Higgs mechanism, incorporated in the Standard model, with the aim to explain how particles in the electro-weak interaction gain their mass. Since teleological phenomena violate Bohr's principle of correspondence, it is not surprising, when Comay [88], after a careful analysis of Higgs equations, came to the conclusion that they violate the Bohr principle as well. Comay's analysis is based on well-known cornerstones such as the variational principle, special relativity, Maxwellian electrodynamics and the fundamental elements of quantum field theory. Additionally, Comay insists on the necessity to test every quantum-physical equation, according to a wider sense of Bohr's correspondence principle, stated by the present version of Wikipedia as: „The term is also used more generally, to represent the idea that a new theory should reproduce the results of older well-established theories (which become limiting cases) in those domains where the old theories work."

Three basic testings of the quantum-physical equations are mentioned, i.e. the equation for the action

$$S = \int L(\psi, \psi_{,\mu}) d^4x \qquad (10.1)$$

the Euler-Lagrange equation of the Lagrangian density $L$

$$\frac{\partial}{\partial x^\mu} \frac{\partial L}{\partial \frac{\partial \psi}{\partial x^\mu}} - \frac{\partial L}{\partial \psi} = 0 \qquad (10.2)$$

and the Hamiltonian density $H$, obtained from the Lagrangian density $L$ by means of the Legendre transformation

$$H = \sum_i \dot{\psi}_i \frac{\partial L}{\partial \dot{\psi}_i} - L \qquad (10.3)$$

Comay [88] made four main contentions against the quantum theoretical consistency of the Higgs mechanism. Using units, $\hbar = c = 1$, the Higgs boson Lagrangian density writes in the following compact form:

$$L_{Higgs} = \phi^+_{,\mu} \phi_{,\nu} g^{\mu\nu} + m^2 \phi^+ \phi + OT \qquad (10.4)$$

where $\phi$ is the scalar function of the Higgs boson, $m$ denotes the Higgs mass and $OT$ denotes other terms (cubic etc.).

The first Comay argument avers: The action $S$ in Eq. (10.1) is a dimensionless Lorentz scalar, while the Lagrangian density $L$ in (10.1) and (10.4) is a Lorentz scalar whose dimension is $[L^{-4}]$. Hence, the first term of (10.4) proves that the dimension of the Higgs function $\phi$ is $[L^{-1}]$. Since the density of a quantum particle is given in terms of its wave function

$$\rho = \left|\psi\right|^2 \tag{10.5}$$

the Schrödinger density satisfies the continuity equation, with the dimension of the wave function is $[L^{-3/2}]$. The difference in dimensions between the Higgs and the Schrödinger functions indicates that Higgs theory does not have a non-relativistic limit. This is an inconsistency in the broader sense of Bohr's correspondence principle.

The second Comay argument reads: The Higgs function $\phi$ is a scalar function whose dimension is $[L^{-1}]$. Hence, the product $\phi^{+}\phi$ is a scalar with dimension $[L^{-2}]$. On the other hand, the density is the zeroth component of a 4-vector whose dimension is $[L^{-3}]$. Hence, the product $\phi^{+}\phi$ does not represent the density of the Higgs boson. This is another inconsistency of the Higgs theory with the Bohr's correspondence principle.

The third Comay argument says: Examining the Euler-Lagrange equation of the Higgs boson, one finds, applying Eq. (10.2) to Higgs Lagrangian density (10.4), the following equation for $\phi$:

$$\Box\phi - m^2\phi + OT = 0 \tag{10.6}$$

The Euler-Lagrange equation of the particle's Lagrangian density must agree with the fundamental quantum mechanical equation, the Schrödinger equation, which is a first order differential equation with respect to the time variable. But the Euler-Lagrange equation of the Higgs boson is a second order differential equation with respect to the time variable. Hence, the Bohr's correspondence principle is violated again.

The fourth Comay argument yields: Applying (10.3) to (10.4), one obtains for the Higgs Hamiltonian density

$$H_{Higgs} = \dot{\phi}^{+}\dot{\phi} + (\nabla\phi^{+}).\nabla\phi - m^2\phi^{+}\phi + OT \tag{10.7}$$

The first term in (10.7) proves that the Higgs Hamiltonian must depend on a time-derivative of the Higgs function $\phi$. This is yet another inconsistency of Higgs theory in relation to the broader sense of Bohr's correspondence principle, since the Schrödinger Hamiltonian is independent of the time derivatives of the wave function.

Further arguments can be found in Comay's original paper, e.g. the analysis of the Higgs energy-momentum tensor that depends quadratically on its mass $m$ contradicting its classical correspondence of linear dependency on the mass. Comay himself views the main problem of Higgs theory by the fact that Higgs equation belongs to the Klein-Gordon family. He cites Dirac's opinion on this issue [89]: „I found this development quite unacceptable. It meant departing from the fundamental ideas of the non-relativistic quantum mechanics, ideas which demanded a wave equation linear in $\partial/\partial t$. It meant abandoning the whole beautiful mathematical scheme for the sake of introducing certain physical ideas." In fact Comay at last appeals: „The present work provides further arguments that support Dirac's opinion... The quite large numbers of contradictions of the Higgs boson theory which are described in this work make a basis for the expectation that this theory will be abandoned."

In fact physicists usually do concentrate on the problem of renormalizability of field theories, yet it seems that the significance of Comay's results is the proof of a total failure of Bohr's principle of correspondence in relation to the Higgs mechanism. It has remained unnoticed or fully ignored by the scientific community. Comay's analysis focuses on the final stage of the long-termed development of the Higgs mechanism in its resulting equations. It is clear that one have to find the crucial mistake in the whole process proving that quantum physics must be "on guard" as regards the principle of causality, and further not allowing the incorporation of the teleological "Mexican hat" potential into the quantum equation. This will prohibit the Higgs equation to follow from "true" quantum physics. We will begin with a brief recapitulation of the history from the main papers, the ideas of which were later used in the formulation of the Higgs mechanism, continuing with the simplest example of the Abelian version of the mechanism, which is now regarded as the model for the explanation of the phenomenon of superconductivity.

As we have mentioned in the previous section, the Mexican hat in a classical setting was for the first time adopted into a macroscopic formulation of superconductivity in 1950 by Ginzburg and Landau [77], leading to a spontaneous breaking of symmetry. The first accepted microscopic theory of superconductivity [78], the BCS model from 1957 did not mention the Mexican hat at all and was derived without any reference to any symmetry violations. The ideas leading to the phenomenological Ginzburg-Landau theory including the Mexican hat were for the first time translated into quantum field theory in 1961 by Nambu and Jona-Lasinio [90]. However, during the same year, serious problems surfaced when Goldstone formulated his theorem [91] stating: „The spontaneous breaking of a continuous symmetry can be associated with a massless and spinless particle, the so called Nambu-Goldstone boson". The Goldstone theorem, proved one year later by Goldstone, Salam and Weinberg [92], disqualified the Nambu–Jona-Lasinio model. Since it is difficult to find, from the general proof [92], the particles that could serve as 'Goldstone bosons', it was believed that the application of spontaneous symmetry violations in field theory, such as the model of Nambu and Jona-Lasinio, would be impossible.

In the same year, 1962, Schwinger [93] proposed the possibility of combining mass with gauge fields. It caused an immediate revival of the old gauge theories from 1954 of Yang and Mills [94], dealing with an extension of the concept of gauge theory considered for Abelian groups used in quantum electrodynamics, to non-Abelian groups, in an attempt to explain strong interactions. In order to preserve gauge invariance, the Yang–Mills field must only give rise to massless particles. One year later, Anderson being inspired by the Goldstone theorem, Schwinger's idea and the Yang-Mills theory, wrote an article [95] about the possible violation of Goldstone's theorem in superconductors: „Schwinger has pointed out that the Yang-Mills vector boson implied by associating a generalized gauge transformation with a conservation law... does not necessarily have zero mass... We show that the theory of plasma oscillations is a simple nonrelativistic example exhibiting all of the features of Schwinger's idea... The boson, which appears as a result of the Goldstone theorem and has zero unrenormalized mass, is converted into a finite-mass plasmon by the interaction with the appropriate gauge field, which is the electromagnetic field." In his article Anderson suggested that a similar process in particle physics could be responsible for giving mass to Yang-Mills gauge bosons, further adding: „It is likely, then, considering the superconducting analogue, that the way is now open for a degenerate-vacuum theory of the Nambu type without any difficulties involving either zero-mass Yang-Mills gauge bosons or zero-mass Goldstone bosons. These two types of bosons seem capable of "cancelling each other out" and leaving only finite mass bosons." It is symptomatic that Anderson's statement became the hint that soon led to the development of the Higgs mechanism.

It is straightforward to work out the derivation of the Abelian Higgs equations for superconductors, which the reader can find in most current textbooks dealing with the Higgs mechanism. The Langrangian density is formulated in a quantum version of the classical Ginzburg-Landau theory, with quadratic and quartic powers of the potential, imitating the Mexican hat.

$$L = -\frac{1}{4} F^{\mu\nu} F_{\mu\nu} + D_\mu^* \phi^* D^\mu \phi - \mu^2 \phi^* \phi - \lambda (\phi^* \phi)^2 \tag{10.8}$$

where $F$ and $D$ are defined by means of the electromagnetic potential $A$ and the charge $q$:

$$F^{\mu\nu} \equiv \partial^\mu A^\nu - \partial^\nu A^\mu; \quad D_\mu \equiv \partial_\mu + iqA_\mu \tag{10.9}$$

The complex field $\phi(x)$, simulating the Mexican hat, may then be expressed by means of two real fields $\eta(x)$ and $\xi(x)$ and a real number $v$:

$$\phi(x) = \frac{1}{\sqrt{2}} \big( \eta(x) + v \big) e^{i\xi(x)/v} \tag{10.10}$$

The Lagrangian density (10.8) then takes the form:

$$L = \left[ -\frac{1}{4} F^{\mu\nu} F_{\mu\nu} + \frac{q^2 v^2}{2} A_\mu A^\mu \right] + \left[ \frac{1}{2} \partial_\mu \eta \, \partial^\mu \eta + \mu^2 \eta^2 \right] + \frac{1}{2} \partial_\mu \xi \, \partial^\mu \xi + q v A^\mu \partial_\mu \xi + OT \tag{10.11}$$

where $OT$ stands for cubic and quartic terms. The terms between the second square brackets represent a massive scalar particle with mass $m_\eta = (-2)^{1/2}\mu$. The third term is a massless scalar particle. The occurrence of this particle was inferred from the shape of the Mexican hat potential with the degree of freedom connected to the angular displacement. A motion in this direction does not face any resistance since the energy in the adjacent state is the same. This field is therefore massless and was identified with the Goldstone boson.

After the application of the gauge transformation for the electromagnetic field

$$\phi \rightarrow \phi' = e^{i\theta(x)}\phi; \quad A \rightarrow A' = A_\mu - \frac{1}{q}\partial_\mu \theta \qquad (10.12)$$

with the specific substitution for $\theta(x)$

$$\theta = -\frac{\xi}{v} \qquad (10.13)$$

one arrives at the final Abelian Higgs equation for the Lagrangian density

$$L = \left[ -\frac{1}{4}F^{\mu\nu}F_{\mu\nu} + \frac{q^2 v^2}{2}A'_\mu A'^\mu \right] + \left[ \frac{1}{2}\partial_\mu \eta \partial^\mu \eta + \mu^2 \eta^2 \right] + OT \qquad (10.14)$$

The Higgs theory contains no massless particles, since the field $\xi(x)$ has entirely disappeared. The gauge symmetry, unlike any global symmetry, does not lead to massless Goldstone bosons, but instead ends up with a massive gauge field. The final Lagrangian contains only a massive gauge boson $A'_\mu(x)$ (first part) and a massive component of the Higgs field $\eta(x)$ (second part). The first part is viewed as the quantum representation of the classical London equation (9.7) and is usually interpreted as the formation of massive photons inside superconductors.

Starting the examination of the Higgs mechanism with Anderson's original paper [95], we will pay special attention to the last sentence in the following quote: „I should like to close with one final remark on the Goldstone theorem. This theorem was initially conjectured, one presumes, because of the solid-state analogues, via the work of Nambu and of Anderson. The theorem states, essentially, that if the Lagrangian possesses a continuous symmetry group under which the ground or vacuum state is not invariant, that state is, therefore, degenerate with other ground states. This implies a zero-mass boson. Thus, the solid crystal violates translational and rotational invariance, and possesses phonons;..."

One may ask whether Anderson, in view of what has been said here, really appreciated the content of the Goldstone theorem? Quasiparticles, representing collective oscillations in condensed matter were known long before the Goldstone theorem was formulated. When it was proved, it was sometimes easy to assign them to Goldstone bosons and corresponding broken symmetries, and sometimes not. For instance, magnons in ferromagnets are Goldstone bosons descending from broken rotational symmetry, or longitudinal phonons in liquids are Goldstone bosons of the broken Galilean symmetry. But what about solids? Anderson claims, that "the solid crystal violates translational and rotational invariance, and possesses phonons". However, not so, phonons are the consequence of a broken Galilean symmetry, and further, which Goldstone bosons correspond to the broken translational and rotational symmetry in crystals? Unfortunately one can find no answers to this question in Anderson's papers.

One might try to find an answer to the question: what are the Goldstone bosons in solids in Wikipedia. A recent reply reads: „In solids, the situation is more complicated; the Goldstone bosons are the longitudinal and transverse phonons and they happen to be the Goldstone bosons of spontaneously broken Galilean, translational, and rotational symmetry with no simple one-to-one correspondence between the Goldstone modes and the broken symmetries."

Although Wikipedia occasionally must be taken with a grain of salt, the sentence "No simple one-to-one correspondence between the Goldstone modes and broken symmetries" is surprising! Since the Goldstone theorem is a complement of Noether's first theorem, the claim is equivalent to saying that the conservation laws for linear and angular momentum may not be in a simple one-to-one correspondence with translational and rotational symmetry. Consequently a unique and unambiguous one-to-one correspondence between the Goldstone modes and spontaneously broken symmetries in solids must necessarily exist! The conclusion must be that phonons in solids do not represent the full set of Goldstone bosons, and some bosons are still missing. It is better if we in the future would add this issue to the list of unsolved problems of quantum physics rather than sweeping it under the carpet.

Principally we should concentrate our energies on hunting for the lost Goldstone bosons in solids. Supposing that a solid is composed of $N$ nuclei, the whole system has $3N$ degrees of freedom. The quantum mechanical treatment eliminates 6 degrees associated with the centre of mass, and the remaining $3N$ - 6 degrees after the Galilean symmetry breaking represent the vibrational modes, the phonons. At the first sight it seems that the violation of translational and rotational invariance has no counterpart in the form of the existence of the corresponding Goldstone bosons. One may ask whether this is really true. Does the quantum field share the centre of mass separation with quantum mechanics considering the same system? The answer is no. We will provide a proof in the next section and show: first, qualitatively the appearance of the 6 lost Goldstone bosons, and second, quantitatively their contributions to the quantum field equations together with the experimental evidence. It means that every solid composed of $N$ nuclei has precisely $3N$ spontaneously broken symmetries with the exact correspondence to the emerging $3N$ Goldstone bosons. This rule holds universally, regardless of the size of $N$, i.e. if it is small or great, if it represents finite molecule or infinite crystal, and irrespective of the solid character, i.e. if it is insulator, conductor, semiconductor, or superconductor.

Returning to Eq. (10.10) where the field $\xi(x)$ relates to the degrees of freedom connected with the angular displacement in the Mexican hat. The field corresponding to the Lagrangian density in Eq. (10.11) results in a massless scalar particle that was identified with the Goldstone boson, or is it really so? As just said above, every solid produces exactly $3N$ Goldstone bosons. Thus the massless scalar particle descending from the field $\xi(x)$ cannot really be a Goldstone boson! It rather looks like the following. Instead of loosing six authentic Goldstone bosons in the solid suddenly one imposter Goldstone boson pops up in the superconductor. Thus we end up with the illusive equation (10.11) without the chance to discuss the rest of the Higgs equations which became misleading as well. This should be the end of Anderson's textual description about the gauge bosons eating the Goldstone bosons, on the basis of which the anticipated Higgs mechanism was created.

We will close this section with a philosophical consideration of the Higgs mechanism. One of the best reflections I have found was written by Earman [96]: „Higgs further showed that the gauge could be chosen so that the Goldstone bosons are suppressed and that in this "unitary gauge" the new field had acquired a mass. As the semi-popular presentations put it, "Particles get their masses by eating the Higgs field." Readers of Scientific American can be satisfied with these just-so stories. But philosophers of science should not be. For a genuine property like mass cannot be gained by eating descriptive fluff, which is just what gauge is. Philosophers of science should be asking the Nozick question: What is the objective (i.e. gauge invariant) structure of the world corresponding to the gauge theory presented in the Higgs mechanism? ...consider the following three-tiered dilemma. First tier: Either the gauge invariant content of the Higgs mechanism is described by local quantum fields satisfying the standard assumptions of Poincaré invariance, local commutativity, spectrum condition, etc. or not. If not, then the implementation of the Higgs mechanism requires a major overhaul of conventional QFT. If so, go to the second tier. Second tier: Either the gauge invariant system admits a finite dimensional Lie group as an internal symmetry group or not. If so, Noether's first theorem applies again. But (since the other standard assumptions are in place) Goldstone's theorem also applies and, hence, Goldstone bosons have not been suppressed after all. If not, go to the third tier. Third tier: Either the gauge invariant system admits no non-trivial symmetries at all or else it admits only discrete symmetries. In either case Goldstone bosons are quashed. In the former case spontaneous symmetry breaking is not an issue since there is no symmetry to break. In the latter case it is possible that the discrete symmetry is spontaneously broken. But the usual argument for symmetry breaking using the conserved Noether current does not apply. And while it is possible that some completely different sort of construction will demonstrate the spontaneous breakdown of the hypothesized discrete symmetry there are no extant demonstrations that have more than a hand waving force."

If we compare our analysis of the Goldstone theorem with the Earman's multi-tier reflection, it is obvious that the whole chain ends at the second tier, i.e. the Goldstone bosons have not been suppressed after all. The rest is nothing but metaphysics. In fact Higgs named his Nobel lecture "Evading the Goldstone theorem" [97], but this was no evasion rather a total misreading of this theorem. Anderson's stimulating paper [95] was indeed an incorrect marking, misleading Higgs and his co-authors on a lost course.

## 11. The paradox of the centre of mass of quantum systems

In this section we will show an independent proof of the Goldstone theorem for condensed matter, molecules and solids. Since the original Goldstone-Salam-Weinberg proof [92] is admittedly the most general, it is not, for the purpose of this study, sufficiently transparent regarding the fundamental question of the one-to-one correspondence between broken symmetries and associated massless and spinless bosons. The problem of translational and rotational symmetry violations in solids is a nice example.

Solid state physics has adopted some adequate patterns from quantum electrodynamics. The quantum field electron-photon Feynman diagrammatic technique has found many successful applications in solid state electron-phonon treatments. Many years ago, when I finished my studies in solid state physics, I did start to work in a quantum chemistry department. In this environment quantum field methods are not so often exercised in comparison to its use in solid state physics, and moreover mostly on the fermionic electron-hole level. At that time I was familiar with the works of Czech emigrants Paldus and Čížek, who were inspired by Goldstone's revision [99] of Brueckner's many-body theory [98] with the solution expressed in a manner that avoids the problem of unlinked clusters, introducing Goldstone's diagrammatic electron-hole mechanism into quantum chemistry [100].

The goal of my PhD thesis [101] was to introduce, in a diagrammatic treatment, the full electron-phonon interactions into quantum chemistry, i.e. to elaborate the mathematical framework for second quantization of the full electron-vibrational Hamiltonian. Goldstone's theorem was at first beyond my interest, since it was mostly discussed in the domain of elementary particle physics, but hardly not mentioned at the time in solid state physics and quantum chemistry. As a result, I thought that electron-phonon mechanisms really constituted a complete description of solids, i.e. exactly how we learned in school as it was described in all textbooks.

Let us start with the full Hamiltonian of the system of nuclei and electrons in second quantization electron-hole formalism, as it is conventionally used in quantum chemistry.

$$H = T_N(\dot{\mathbf{R}}) + E_{NN}(\mathbf{R}) + \sum_{PQ} h_{PQ}(\mathbf{R}) a_P^+ a_Q + \frac{1}{2} \sum_{PQRS} v_{PQRS}^0 a_P^+ a_Q^+ a_S a_R \tag{11.1}$$

Here $T_N$ stands for kinetic energy of nuclei, $E_{NN}$ for nuclear potential energy, $h_{PQ}$ for one-electron matrix elements and $v_{PQRS}^0$ for two-electron matrix elements. The terms $E_{NN}$ and $h_{PQ}$ can be expanded in the Taylor series

$$E_{NN}(\mathbf{R}) = \sum_{n=0}^{\infty} E_{NN}^{(n)}(\mathbf{R}) \tag{11.2}$$

$$h_{PQ}(\mathbf{R}) = h_{PQ}^0 + \sum_{n=1}^{\infty} u_{PQ}^{(n)}(\mathbf{R}) = h_{PQ}^0 + \sum_{n=1}^{\infty} \langle P | \sum_i \frac{-Z_i e^2}{|\mathbf{r} - \mathbf{R}_i|} | Q \rangle^{(n)} \tag{11.3}$$

where $E_{NN}^0$ and $h_{PQ}^0$ are nuclear potential and one-electron terms for fixed (equilibrium) nuclear coordinates. In full analogy with the effective electron-phonon solid state Hamiltonian

$$H = \sum_{\mathbf{k},\sigma} \varepsilon_{\mathbf{k}} a_{\mathbf{k},\sigma}^+ a_{\mathbf{k},\sigma} + \sum_{\mathbf{q}} \hbar \omega_{\mathbf{q}} \left( b_{\mathbf{q}}^+ b_{\mathbf{q}} + \frac{1}{2} \right) + \sum_{\mathbf{k},\mathbf{q},\sigma} u^{\mathbf{q}} \left( b_{\mathbf{q}} + b_{-\mathbf{q}}^+ \right) a_{\mathbf{k}+\mathbf{q},\sigma}^+ a_{\mathbf{k},\sigma} \tag{11.4}$$

we can rewrite the Hamiltonian (11.1) in a complete second quantization form for electronic and as well as vibronic modes:

$$H = T_N(\tilde{B}) + E_{NN}(B) + \sum_{PQ} h_{PQ}(B) a_P^+ a_Q + \frac{1}{2} \sum_{PQRS} v_{PQRS}^0 a_P^+ a_Q^+ a_S a_R \qquad (11.5)$$

where $B$ and $\tilde{B}$ represent the coordinate and momentum oscillator operators. If we require that these and all following equations to be applicable to both quantum chemistry and solid state physics, we need to introduce a cross-platform notation for $B$ and $\tilde{B}$, viz $B_r = b_r + b_{\check{r}}^+$ and $\tilde{B}_r = b_r - b_{\check{r}}^+$. In quantum chemistry $r = \check{r}$ holds, while in solid state physics one assumes that for any vibrational mode $r$ there exists corresponding mode $\check{r}$ that fulfills the identity $\omega_r = \omega_{\check{r}}$. Comparing this notation with the usual solid state notation the simple transition $P \rightarrow \mathbf{k}, \sigma, r \rightarrow \mathbf{q}, \check{r} \rightarrow \mathbf{-q}$ is supposed.

The potential energy of the motion of the nuclei is defined through the quadratic part of the internuclear potential plus some additive term representing the self-consistent influence of the electron-nuclear potential

$$E_{pot} = E_{NN}^{(2)}(B) + V_N^{(2)}(B) \qquad (11.6)$$

In the adiabatic limit the kinetic energy $E_{kin}$ is identical to the kinetic energy of the nuclei $T_N$. If we want to include the nonadiabatic case in our treatment, it is necessary to incorporate a new additive kinetic term originating from the kinetic energy of electrons. The resulting kinetic energy of the system has the form

$$E_{kin} = T_N(\tilde{B}) + W_N^{(2)}(\tilde{B}) \qquad (11.7)$$

so that the vibrational part of the total Hamiltonian can be expressed as

$$H_V = E_{kin}(\tilde{B}) + E_{pot}(B) \qquad (11.8)$$

where the kinetic and potential energies are given by

$$E_{pot} = \frac{1}{4} \sum_{r \in V} \hbar \omega_r B_r^+ B_r \qquad (11.9)$$

$$E_{kin} = \frac{1}{4} \sum_{r \in V} \hbar \omega_r \tilde{B}_r^+ \tilde{B}_r \qquad (11.10)$$

Finally we get the well-known vibrational Hamiltonian

$$H_V = \frac{1}{4} \sum_{r \in V} \hbar \omega_r \left( B_r^+ B_r + \tilde{B}_r^+ \tilde{B}_r \right) = \sum_{r \in V} \hbar \omega_r \left( b_r^+ b_r + \frac{1}{2} \right) \qquad (11.11)$$

Then the total Hamiltonian of the system reads:

$$H = E_{NN}(B) - E_{NN}^{(2)}(B) - V_N^{(2)}(B) - W_N^{(2)}(\tilde{B}) + \sum_{PQ} h_{PQ}(B) a_P^+ a_Q$$

$$+ \frac{1}{2} \sum_{PQRS} v_{PQRS}^0 a_P^+ a_Q^+ a_S a_R + \frac{1}{4} \sum_{r \in V} \hbar \omega_r \left( B_r^+ B_r + \tilde{B}_r^+ \tilde{B}_r \right) \tag{11.12}$$

However there is one obstacle here. The Hamiltonian (11.12) cannot be directly used for quantum chemical calculations like the Hamiltonian (11.4) in solid state physics. Both are in the form of a "crude" representation, to be precise, the representation with fixed nuclear positions. Although this makes no problems in solids, where only valence and conducting bands are to be taken into account, in molecular calculations all electrons must be incorporated, including also the inner shell ones. Unfortunately direct application of the Hamiltonian (11.12) leads to a divergent series expansion.

Nevertheless there is a straightforward way to solve this difficulty, namely one must find a suitable renormalization where all the divergent terms vanish. Instead of the original electrons and bosons (vibrational modes or phonons) we introduce a new set of fermionic and bosonic operators, with the transformed Hamiltonian optimized on the Born-Oppenheimer (B-O) level for adiabatic molecules.

The most general quasiparticle transformation, which "simulates" the B-O approximation, can be introduced in the following form:

$$\bar{a}_P = \sum_Q c_{PQ}(B) a_Q \qquad \bar{b}_r = b_r + \sum_{PQ} d_{rPQ}(B) a_P^+ a_Q \tag{11.13}$$

with the unitary conditions

$$\sum_R c_{PR}(B) c_{QR}^+(B) = \delta_{PQ} \qquad d_{rPQ} = \sum_R c_{RP}^+(B)[b_r, c_{RQ}(B)] \tag{11.14}$$

For nonadiabatic systems we add to the previous $Q$ (coordinate)-dependent transformation yet another $P$ (momentum)-dependent one:

$$\bar{a}_P = \sum_Q \tilde{c}_{PQ}(\tilde{B}) a_Q \qquad \bar{b}_r = b_r + \sum_{PQ} \tilde{d}_{rPQ}(\tilde{B}) a_P^+ a_Q \tag{11.15}$$

with the unitary conditions

$$\sum_R \tilde{c}_{PR}(\tilde{B}) \tilde{c}_{QR}^+(\tilde{B}) = \delta_{PQ} \qquad \tilde{d}_{rPQ} = \sum_R \tilde{c}_{RP}^+(\tilde{B})[b_r, \tilde{c}_{RQ}(\tilde{B})] \tag{11.16}$$

The successive application of both transformations, adiabatic and nonadiabatic ones is equivalent to the unitary transformation of the whole Hamiltonian

$$\bar{H} = e^{-S_2(P)} e^{-S_1(Q)} H e^{S_1(Q)} e^{S_2(P)} \tag{11.17}$$

Note that I have preferred quasiparticle transformations to unitary transformation of the Hamiltonian for the sake of greater transparency. Incidentally, Fröhlich used a similar unitary transformation in his derivation of the effective two-electron interaction [105] that was later incorporated into the BCS theory:

$$\bar{H} = e^{-S(Q,P)} H e^{S(Q,P)} \qquad (11.18)$$

For details of the derivation of the field equations, as based on the quasiparticle transformations (11.13-16), see the book [102]. For a long time I considered this theory to be essentially correct. To sum up, there were four well-known formulae, one in quantum mechanics, and three in quantum field theory, which could have been rederived in exact agreement:

1) the Pople's equations for the ab-initio calculation of vibrational frequencies [103]
2) the Fröhlich's expression for the correction of the ground state energy [104,105]
3) the Fröhlich's effective two-electron interaction [105]
4) the Lee-Low-Pines polarons and their self-energy [106]

During the numerical tests in 1998 the abovementioned theory totally failed in the calculations of the first correction beyond the B-O approximation, i.e. the adiabatic correction known as the Born-Huang ansatz [7]:

$$\Delta E_{0(ad)} = \left\langle \psi_0\left(\mathbf{R}\right) \middle| T_N \middle| \psi_0\left(\mathbf{R}\right) \right\rangle_{\mathbf{R}_0} \qquad (11.19)$$

This was a great disappointment for me. In order to directly compare the field formula for the adiabatic correction with the mechanical Born-Huang ansatz, I did rewrite the latter into its field form. In the adiabatic case the N-electron function $\psi_0\left(\mathbf{R}\right)$ can be expanded as a single Slater determinant though the one-electron functions $\varphi_I\left(\mathbf{R}\right)$:

$$\psi_0\left(\mathbf{R}\right) = \frac{1}{\sqrt{N!}} \left\| \prod_I^N \varphi_I\left(\mathbf{R}\right) \right\| \qquad (11.20)$$

The function $\varphi_P\left(\mathbf{R}\right)$ can be expanded in terms of the coefficients $c_{PQ}\left(\mathbf{R}\right)$, dependent on the nuclear coordinates, and the orthonormal set of one-electron wavefunctions defined in the equilibrium position $\mathbf{R}_0$:

$$\varphi_P\left(\mathbf{R}\right) = \sum_Q c_{PQ}\left(\mathbf{R}\right) \varphi_Q\left(\mathbf{R}_0\right) \qquad (11.21)$$

In the following work we will use the following notation for the spin-orbitals: I, J, K, L – occupied; A, B, C, D – virtual (unoccupied); P, Q, R, S – the arbitrary ones. One obtains then a very simple expression for the Born-Huang ansatz (11.19)

$$\Delta E_{0(ad)} = \sum_{Ali\alpha} \frac{\hbar^2}{2M_i} \left| c_{AI}^{i\alpha} \right|^2 \tag{11.22}$$

where $M_i$ stands for the nuclear mass and $\alpha$ for the Cartesian coordinates.

It is now useful to proceed from the Cartesian to the normal coordinate system using the following transformation

$$c_{PQ}^r = \sum_{i\alpha} c_{PQ}^{i\alpha} \alpha_{i\alpha}^r \tag{11.23}$$

One wants to substitute for $c_{PQ}^{i\alpha}$ in (11.22), so we need to know the inverse matrix $\beta_{i\alpha}^r$. From

$$\alpha^+ \beta = I \tag{11.24}$$

Eq. (11.22) can be expressed in normal coordinates as

$$\Delta E_{0(ad)} = \sum_{Alrsi\alpha} \frac{\hbar^2}{2M_i} c_{AI}^r c_{AI}^{s\,*} \beta_{i\alpha}^{r\,*} \beta_{ia}^s \tag{11.25}$$

The matrices $\alpha$ and $\beta$ have dimension $3N$ x $3N$. The first one diagonalizes the potential energy

$$\alpha^+ E_{pot} \alpha = \left\{ \frac{1}{2} \hbar \omega, 0, 0 \right\} \tag{11.26}$$

resulting in $3N - 5$ or $3N - 6$ vibrational frequencies (the first holds for diatomic molecules) and the 5 or 6 accounting for zero energy. On the other hand, the second matrix diagonalizes the kinetic energy

$$\beta^+ E_{kin} \beta = \left\{ \frac{1}{2} \hbar \omega, \rho, \tau \right\} = \sum_{i\alpha} \frac{\hbar^2}{4M_i} \beta_{i\alpha}^{\,*} \beta_{i\alpha} \tag{11.27}$$

with all $3N$ nonzero values: aside the $3N - 5$ or $3N - 6$ vibrational modes we have 2 or 3 nonzero values for rotational modes and 3 nonzero values for translational ones. So the final field expression for the Born-Huang ansatz reads:

$$\Delta E_{0(ad)} = 2 \sum_{AI} \left( \sum_{r \in V} \frac{1}{2} \hbar \omega_r + \sum_{r \in R} \rho_r + \sum_{r \in T} \tau_r \right) \left| c_{AI}^r \right|^2 \tag{11.28}$$

It is now evident that all electron-vibrational and electron-phonon field theories fail beyond the B-O approximation, because they do not pass the Born-Huang test. In Eq. (11.28) they are only able to justify the first term that consists of vibrational modes, but they are unable to arrive at the second and third terms dealing with rotational and translational quanta. The electron-phonon mechanism has its roots in the B-O approximation, where only the diagonalization procedure (11.26) of the potential energy via the α matrices is relevant.

Therefore, in this specific case the quantum field can only share with quantum mechanics the centre-of-mass (COM) separation of the external and the internal degrees of freedom.

Even though I immediately discovered this serious problem of the quantum field formulation after the 1998 numerical tests, I have hesitated for many years to publish the revised version of employing quantum field theories for condensed matter systems such as molecules and solids. Naturally I did not want to make a fool of myself by claiming that there are some new particles in physics, and that there is a new type of covariance in the quantum field, binding together the internal and external degrees of freedom in a way similar to the way Lorentz covariance binds together space and time. Finally I did present this topic in Cambridge in 2010, and this lecture was published in 2012 [107].

Without any knowledge of the Goldstone theorem, we can see that the existence of $3N - 5$ or $3N - 6$ massless spinless bosons - vibrations/phonons – results from Eq. (11.26). The rest of the massless spinless bosons which I have called rotons and translons, follows from Eq. (11.27). However, the complete lack of interest from the scientific community in this topic has indeed been astonishing, but the reason may be quite banal. Whenever I speak with solid state physicists, they all admit that they have always thought that the adiabatic approximation is synonymous with the B-O approximation and that they see no difference between them, but it is in actual fact the almost negligible difference between them that generates the 5 or 6 lost Goldstone bosons! As a consequence the original concept of the electron-vibrational field theory [102] must be abandoned and replaced by a new field theory with two major upgrades:

The first upgrade: Eqs. (11.9-10) for the potential and the kinetic vibrational energies, must be rewritten using a complete set of $3N$ Goldstone bosons - phonons, rotons and translons - in agreement with the Born-Huang field transcription (11.28). We shall call these bosons hyperphonons or hypervibrations.

$$E_{pot} = \frac{1}{4} \sum_{r \in V'} \hbar \omega_r B_r^+ B_r \qquad (11.29)$$

$$E_{kin} = \frac{1}{2} \left( \frac{1}{2} \sum_{r \in V'} \hbar \omega_r + \sum_{r \in R} \rho_r + \sum_{r \in T} \tau_r \right) \tilde{B}_r^+ \tilde{B}_r \qquad (11.30)$$

Defining the hyper-vibrational double-vector:

$$\boldsymbol{\omega} = \begin{pmatrix} \omega_r \\ \tilde{\omega}_r \end{pmatrix} = \begin{pmatrix} \omega_r & 0 & 0 \\ \omega_r & \dfrac{2}{\hbar} \rho_r & \dfrac{2}{\hbar} \tau_r \end{pmatrix} \qquad (11.31)$$

we get covariant expressions for the boson part of the hyper-vibrational Hamiltonian with respect to all $3N$ Goldstone modes:

$$H_B = \frac{1}{4} \sum_r \left( \hbar \omega_r B_r^+ B_r + \hbar \tilde{\omega}_r \tilde{B}_r^+ \tilde{B}_r \right) \tag{11.32}$$

Instead of the original total Hamiltonian (11.12) we obtain

$$H = E_{NN}(B) - E_{NN}^{(2)}(B) - V_N^{(2)}(B) - W_N^{(2)}(\tilde{B}) + \sum_{PQ} h_{PQ}(B) a_P^+ a_Q$$

$$+ \frac{1}{2} \sum_{PQRS} v_{PQRS}^0 a_P^+ a_Q^+ a_S a_R + \frac{1}{4} \sum_r \left( \hbar \omega_r B_r^+ B_r + \hbar \tilde{\omega}_r \tilde{B}_r^+ \tilde{B}_r \right) \tag{11.33}$$

The second upgrade: We will apply the same quasiparticle transformations (11.13-16) as in the original electron-vibrational concept, but with entirely different interpretation. In full analogy with Lorentz covariance, binding together space and time coordinates, we introduce a new covariance, binding together the internal and external degrees of freedom, i.e. Eqs. (11.13-16) will comprise all $3N$ degrees of freedom without the mechanical COM separation. We will denote the quality of this unitary transformation as the field COM covariance.

Here we will only sketch the derivation, since it is, however, very time-consuming. Details of the derivation have been given in previous work [107]. The final formula for the change of the ground state energy has a surprisingly simple analytical form:

$$\Delta E_0 = \sum_{AIr} \left( \hbar \tilde{\omega}_r \mid c_{AI}^r \mid^2 - \hbar \omega_r \mid \tilde{c}_{AI}^r \mid^2 \right) \tag{11.34}$$

where the summation refers to virtual spin-orbitals $A$, occupied spin-orbitals $I$, and all hyper-vibrational modes $r$, $r \in \{V,R,T\}$. The coefficients $c$ resp. $\tilde{c}$ are related to the adiabatic and the non-adiabatic transformation, respectively, and determined by the set of equations

$$u_{PQ}^r + (\varepsilon_P^0 - \varepsilon_Q^0) c_{PQ}^r + \sum_{AI} [(v_{PIQA}^0 - v_{PIAQ}^0) c_{AI}^r - (v_{PAQI}^0 - v_{PAIQ}^0) c_{IA}^r] - \hbar \omega_r \tilde{c}_{PQ}^r = \varepsilon_P^r \delta_{PQ} \tag{11.35}$$

$$(\varepsilon_P^0 - \varepsilon_Q^0) \tilde{c}_{PQ}^r + \sum_{AI} [(v_{PIQA}^0 - v_{PIAQ}^0) \tilde{c}_{AI}^r - (v_{PAQI}^0 - v_{PAIQ}^0) \tilde{c}_{IA}^r] - \hbar \tilde{\omega}_r c_{PQ}^r = \tilde{\varepsilon}_P^r \delta_{PQ} \tag{11.36}$$

where $u$ are the coefficients of the electron-hyperphonon interaction, $\varepsilon^0$ are one-electron energies, and $v^0$ two-electron potential energies. Finally the set of equations in the second order of the Taylor expansion results in the ab-initio self-consistent equations for hyper-vibrational frequencies $\omega$ and $\tilde{\omega}$, namely for the unknown potential and kinetic matrix elements in the Eqs. (11.6-7) and in the total Hamiltonian (11.33):

$$V_N^{rs} = \sum_I u_{II}^{rs} + \sum_{AI} [(u_{IA}^r + \hbar \omega_r \tilde{c}_{IA}^r) c_{IA}^s + (u_{IA}^s + \hbar \omega_s \tilde{c}_{IA}^s) c_{IA}^r] \tag{11.37}$$

$$W_N^{rs} = 2 \hbar \tilde{\omega}_r \sum_{AI} c_{AI}^r \tilde{c}_{IA}^s \tag{11.38}$$

The fermionic one-particle correction $\Delta H'_F$ is more complex and therefore we select only those terms which are decisive for excitation mechanism

$$\Delta H'_F = \sum_{PQr} \left[ \hbar \tilde{\omega}_r \left( \sum_A c^r_{PA} c^{r*}_{QA} - \sum_I c^r_{PI} c^{r*}_{QI} \right) - \hbar \omega_r \left( \sum_A \tilde{c}^r_{PA} \tilde{c}^{r*}_{QA} - \sum_I \tilde{c}^r_{PI} \tilde{c}^{r*}_{QI} \right) \right] N[a^+_P a_Q]$$
$$+ \sum_{PRr} \left[ \left( \varepsilon^0_P - \varepsilon^0_R \right) \left( | c^r_{PR} |^2 + | \tilde{c}^r_{PR} |^2 \right) - 2\hbar \tilde{\omega}_r \, \mathrm{Re} \left( \tilde{c}^r_{PR} c^{r*}_{PR} \right) \right] N[a^+_P a_P] \tag{11.39}$$

The first part (11.39) is of a pure one-fermion origin and has a non-diagonal form. The second part is a vacuum value of type $\langle 0 | B_r B_s | 0 \rangle$ and/or $\langle 0 | \tilde{B}_r \tilde{B}_s | 0 \rangle$ of the mixed fermion-boson terms, where the bosonic part is of the quadratic form of coordinate and/or momentum operators.

We can now demonstrate the simplest applications of these general equations where the field COM covariant theory reduces to the classical mechanical COM separation limit, i.e. $\omega = \tilde{\omega}$ and only the Goldstone bosons descending from the Galilean broken symmetries - vibrations/phonons - are taken into consideration. We will present the limits of the four examples leading up to Pople's equations, Fröhlich's ground state energy correction and the effective two-electron Hamiltonian, and the Lee-Low-Pines polarons. These four cases were particularly mentioned as they immediately follow from the pure electron-vibrational theory [102]. We start with the four cases and continue with three more related to the Born-Huang ansatz.

1) Pople's equations for the *ab initio* calculation of vibrational frequencies [103]:
From the Eqs. (11.35) and (11.37) in the adiabatic limit, where the coefficients $\tilde{c}$ equal zero we get

$$u^r_{PQ} + (\varepsilon^0_P - \varepsilon^0_Q) c^r_{PQ} + \sum_{AI} [(v^0_{PIQA} - v^0_{PIAQ}) c^r_{AI} - (v^0_{PAQI} - v^0_{PAIQ}) c^r_{IA}] = \varepsilon^r_P \delta_{PQ}; \quad c^r_{PP} = 0 \tag{11.40}$$

$$V^{rs}_N = \sum_I u^{rs}_{II} + \sum_{AI} \left( u^r_{IA} c^s_{AI} + u^s_{IA} c^r_{AI} \right) \tag{11.41}$$

One observes that Ref. [103] contains these equations, not in the MO (molecular orbital) basis, but for programming purposes in the LCAO (linear combination of atomic orbitals) basis of "moving" atomic orbitals that follow the adiabatic "motion" of the nuclei. Both approaches are fully equivalent; the MO basis notation is simpler and shorter, and therefore more suitable for a theoretical treatment, whereas the LCAO basis is more practical in numerical calculations. One can look at the equations (11.35-38) as a generalization of Pople's CPHF equations [103] for the case of a general field COM covariant theory, which includes also the case of the break-down of the B-O approximation.

2) Fröhlich's expression for the correction to the ground state energy [104,105]:

Neglecting two-electron terms, the formula for the change of the ground state energy (11.34) takes the following explicit simple form:

$$\Delta E_0 = \sum_{Al, r \in V} \left( \hbar \omega_r \mid c_{Al}^r \mid^2 - \hbar \omega_r \mid \tilde{c}_{Al}^r \mid^2 \right) = \sum_{Al, r \in V} \mid u_{Al}^r \mid^2 \frac{\hbar \omega_r}{(\varepsilon_A^0 - \varepsilon_I^0)^2 - (\hbar \omega_r)^2} \tag{11.42}$$

After a changeover from a quantum chemical to a solid state physics notation, one obtains exactly the same result as originally derived by Fröhlich [104] from perturbation theory and then rederived by him again on the basis of the unitary transformation (11.18) [105].

$$\Delta E_0 = 2 \sum_{\mathbf{k}, \mathbf{k}'; \mathbf{k} \neq \mathbf{k}'} \mid u^{\mathbf{k}' - \mathbf{k}} \mid^2 f_{\mathbf{k}} (1 - f_{\mathbf{k}'}) \frac{\hbar \omega_{\mathbf{k}' - \mathbf{k}}}{(\varepsilon_{\mathbf{k}'}^0 - \varepsilon_{\mathbf{k}}^0)^2 - (\hbar \omega_{\mathbf{k}' - \mathbf{k}})^2} \tag{11.43}$$

In his first paper [104] Fröhlich tried to interpret the new state as a superconducting one, since the optimization of the occupation factors $f_{\mathbf{k}}$ yields a decrease of the total energy. But as soon the experimental verification of the existing gap appeared Fröhlich, in response in his second paper [105] admittedly rederived this result once more, but this time laying stress on the effective two-electron terms resulting from electron-phonon interactions as a possible source of the gap formation. He did publish it as a challenge that somehow, by means of a true many-body treatment, going beyond the Hartree-Fock approximation, the expected gap would be achieved.

3) Fröhlich's effective two-electron interaction (Fröhlich Hamiltonian) [105]:

The electron-hyper-vibrational Hamitonian in the electron-vibrational limit leads to the following result for the effective two-electron interaction [107]:

$$\Delta H_F'' = \sum_{\substack{PQRSr \\ P \neq R, Q \neq S}} u_{PR}^r u_{SQ}^{r*} \frac{\hbar \omega_r [(\varepsilon_P^0 - \varepsilon_R^0)(\varepsilon_S^0 - \varepsilon_Q^0) - (\hbar \omega_r)^2]}{[(\varepsilon_P^0 - \varepsilon_R^0)^2 - (\hbar \omega_r)^2][(\varepsilon_S^0 - \varepsilon_Q^0)^2 - (\hbar \omega_r)^2]} N[a_P^+ a_Q^+ a_S a_R] \tag{11.44}$$

In solid state notation this sum reads ($r \rightarrow \mathbf{q}$; $P \rightarrow \mathbf{k+q}, \sigma$; $Q \rightarrow \mathbf{k'}, \sigma'$; $R \rightarrow \mathbf{k}, \sigma$; $S \rightarrow \mathbf{k'+q}, \sigma'$):

$$\Delta H_F'' = \sum_{\substack{\mathbf{k}, \mathbf{k'}, \mathbf{q}, \sigma, \sigma' \\ \mathbf{q} \neq 0}} \frac{\mid u^{\mathbf{q}} \mid^2 \hbar \omega_{\mathbf{q}} [(\varepsilon_{\mathbf{k+q}}^0 - \varepsilon_{\mathbf{k}}^0)(\varepsilon_{\mathbf{k'+q}}^0 - \varepsilon_{\mathbf{k'}}^0) - (\hbar \omega_{\mathbf{q}})^2]}{[(\varepsilon_{\mathbf{k+q}}^0 - \varepsilon_{\mathbf{k}}^0)^2 - (\hbar \omega_{\mathbf{q}})^2][(\varepsilon_{\mathbf{k'+q}}^0 - \varepsilon_{\mathbf{k'}}^0)^2 - (\hbar \omega_{\mathbf{q}})^2]} N[a_{\mathbf{k+q}, \sigma}^+ a_{\mathbf{k'}, \sigma'}^+ a_{\mathbf{k'+q}, \sigma'} a_{\mathbf{k}, \sigma}] \tag{11.45}$$

In comparison to the expression obtained by Fröhlich, known as the Fröhlich Hamiltonian

$$\Delta H_{F(Fr)}'' = \sum_{\substack{\mathbf{k}, \mathbf{k'}, \mathbf{q}, \sigma, \sigma' \\ \mathbf{q} \neq 0}} \mid u^{\mathbf{q}} \mid^2 \frac{\hbar \omega_{\mathbf{q}}}{(\varepsilon_{\mathbf{k+q}}^0 - \varepsilon_{\mathbf{k}}^0)^2 - (\hbar \omega_{\mathbf{q}})^2} a_{\mathbf{k+q}, \sigma}^+ a_{\mathbf{k'}, \sigma'}^+ a_{\mathbf{k'+q}, \sigma'} a_{\mathbf{k}, \sigma} \tag{11.46}$$

one can see that both expressions are equivalent, but not exactly identical. As the coordinate and momentum operators do not commute, the Hamiltonians (11.17) and (11.18) lead to a slightly different expression for the Fröhlich effective two-electron interaction. This type of ambiguity of the Fröhlich Hamiltonian was extensively discussed by Lenz and Wegner [109] by means of continuous unitary transformations.

4) Lee-Low-Pines polarons and their self-energy [106]:

Neglecting two-electron terms in the equation (11.39), one obtains a simple analytical expression for the fermionic one-particle excitation energies

$$\Delta\varepsilon_P = \sum_{r\in V}\left(\sum_{A\neq P}\frac{|u_{PA}^r|^2}{\varepsilon_P^0 - \varepsilon_A^0 - \hbar\omega_r} + \sum_{I\neq P}\frac{|u_{PI}^r|^2}{\varepsilon_P^0 - \varepsilon_I^0 + \hbar\omega_r}\right)$$

$$= \sum_{r\in V}\left(\sum_{R\neq P}|u_{PR}^r|^2\frac{1}{\varepsilon_P^0 - \varepsilon_R^0 - \hbar\omega_r} - 2\sum_{I\neq P}|u_{PI}^r|^2\frac{\hbar\omega_r}{(\varepsilon_A^0 - \varepsilon_I^0)^2 - (\hbar\omega_r)^2}\right) \qquad (11.47)$$

and in the solid state notation ($r\rightarrow\mathbf{q}$; $P\rightarrow\mathbf{k},\sigma$; $R\rightarrow\mathbf{k}\text{-}\mathbf{q},\sigma$; $I\rightarrow\mathbf{k}\text{-}\mathbf{q},\sigma$ with the occupation factor $f_{\mathbf{k}\text{-}\mathbf{q}}$)

$$\Delta\varepsilon_{\mathbf{k}} = \sum_{\mathbf{q}\neq0}|u^{\mathbf{q}}|^2\frac{1}{\varepsilon_{\mathbf{k}}^0 - \varepsilon_{\mathbf{k}\text{-}\mathbf{q}}^0 - \hbar\omega_{\mathbf{q}}} - 2\sum_{\mathbf{q}\neq0}|u^{\mathbf{q}}|^2 f_{\mathbf{k}\text{-}\mathbf{q}}\frac{\hbar\omega_{\mathbf{q}}}{(\varepsilon_{\mathbf{k}}^0 - \varepsilon_{\mathbf{k}\text{-}\mathbf{q}}^0)^2 - (\hbar\omega_{\mathbf{q}})^2} \qquad (11.48)$$

The electron energies $\varepsilon_{\mathbf{k}}^0$ with the corrections (11.48) represent the quasiparticles/polarons that were originally derived via the Lee-Low-Pines transformation [106]. The first part of (11.48) refers to individual polarons, whereas the second part represents correlations originating from the effective field of other polarons.

The previous examples were based on a simplistic model of the electron-vibrational/electron-phonon Hamiltonian, taking only Galilean broken symmetries from the Goldstone theorem into account. Investigating the consequences of including all Goldstone bosons, descending from all three types of symmetry breakings in condensed matter, our attention will be focused on three most interesting and important examples: the Born-Huang ansatz, and ground state energy and excitation spectra of the B-O degenerate systems [107,108].

5) The Born-Huang ansatz [7]:

In the adiabatic limit, which means that all non-adiabatic coefficients $\tilde{c}$ will be equal to zero, the change of the ground state energy (11.34) yields the adiabatic correction

$$\Delta E_{0(ad)} = \sum_{AIr}\hbar\tilde{\omega}_r\,|c_{AI}^r|^2 = 2\sum_{AI}\left(\sum_{r\in V}\frac{1}{2}\hbar\omega_r + \sum_{r\in R}\rho_r + \sum_{r\in T}\tau_r\right)\!\Big|c_{AI}^r\Big|^2 \qquad (11.49)$$

that is exactly identical with the Born-Handy ansatz (11.28). We see that only the field COM covariant theory can pass the test of the Born-Huang ansatz. Noting that the roton terms in (11.49) has nothing to do with the energies of the rotational degrees of molecular freedom as well as the translon terms with the de Broglie wave of the translational freedom of the whole system, the contribution of roton and translon quanta occurs even if the molecule or crystal is completely in rest and does not rotate or move.

6) The ground state energy of the B-O degenerate systems:

Let us consider the complete non-adiabatic and field COM covariant case, where we only omit two-electron terms in order to obtain transparent analytical expressions:

$$\Delta E_0 = \sum_{AIr} |u_{AI}^r|^2 \frac{\hbar \tilde{\omega}_r}{(\varepsilon_A^0 - \varepsilon_I^0)^2 - (\hbar \omega_r)^2} \tag{11.50}$$

which in the form of the sum of vibrational, rotational and translational parts finally reads

$$\Delta E_0 = \sum_{AI, r \in V} |u_{AI}^r|^2 \frac{\hbar \omega_r}{(\varepsilon_A^0 - \varepsilon_I^0)^2 - (\hbar \omega_r)^2}$$
$$+ 2 \sum_{AI, r \in R} |u_{AI}^r|^2 \frac{\rho_r}{(\varepsilon_A^0 - \varepsilon_I^0)^2} + 2 \sum_{AI, r \in T} |u_{AI}^r|^2 \frac{\tau_r}{(\varepsilon_A^0 - \varepsilon_I^0)^2} \tag{11.51}$$

After the rewriting Eq. (11.51) in solid state notation one obtains

$$\Delta E_0 = 2 \sum_{\mathbf{k}, \mathbf{k'}} |u^{\mathbf{k'}\text{-}\mathbf{k}}|^2 \frac{\hbar \omega_{o, \mathbf{k'}\text{-}\mathbf{k}}}{(\varepsilon_{c, \mathbf{k'}}^0 - \varepsilon_{v, \mathbf{k}}^0)^2 - (\hbar \omega_{o, \mathbf{k'}\text{-}\mathbf{k}})^2}$$
$$+ 4 \sum_{\mathbf{k}, r \in R} |u^r|^2 \frac{\rho_r}{(\varepsilon_{c, \mathbf{k}}^0 - \varepsilon_{v, \mathbf{k}}^0)^2} + 4 \sum_{\mathbf{k}, r \in T} |u^r|^2 \frac{\tau_r}{(\varepsilon_{c, \mathbf{k}}^0 - \varepsilon_{v, \mathbf{k}}^0)^2} \tag{11.52}$$

where $o$ denotes the optical branches and $c$, $v$ the conducting and the valence bands respectively. This is a fascinating result. It demonstrates how the true quantum field, respecting fully the Goldstone theorem, copes with the B-O degenerate systems such as J-T molecules and superconductors: the superposition principle for degeneracy removal is simply bypassed, since rotons and translons are actually responsible for symmetry breaking. In a superconductor, such symmetry breakings produce several geometrically different symmetry broken states, and split the original half-occupied conducting band of a conductor into two bands - one fully occupied valence band and one empty conducting band.

7) The excitation spectra of B-O degenerate systems:

Looking at the degeneracy removal and a gap formation in the B-O degenerate systems, we will only take the diagonal form in Eq. (11.39) into account. Further we neglect the second part of this equation, which does not depend on the electron distribution defined by the occupation/virtual states as does the first part. After omitting of two-electron terms we get:

$$\Delta \varepsilon_P = \sum_r \hbar \tilde{\omega}_r \left( \sum_{A \neq P} \frac{|u_{PA}^r|^2}{(\varepsilon_P^0 - \varepsilon_A^0)^2 - (\hbar \omega_r)^2} - \sum_{I \neq P} \frac{|u_{PI}^r|^2}{(\varepsilon_P^0 - \varepsilon_I^0)^2 - (\hbar \omega_r)^2} \right) \tag{11.53}$$

The diagonal form of the J-T one-particle excitation expression (11.53) is fully justified in solid state physics where translational symmetry is supposed. Since we have two bands, in solid state notation the one-particle Hamiltonian (11.53) reads

$$\Delta\varepsilon_{v,\mathbf{k}} = \sum_{\mathbf{q}\neq 0}|u^{\mathbf{q}}|^2\left(\frac{\hbar\omega_{o,\mathbf{q}}}{(\varepsilon^0_{v,\mathbf{k}}-\varepsilon^0_{c,\mathbf{k}-\mathbf{q}})^2-(\hbar\omega_{o,\mathbf{q}})^2}-\frac{\hbar\omega_{a,\mathbf{q}}}{(\varepsilon^0_{v,\mathbf{k}}-\varepsilon^0_{v,\mathbf{k}-\mathbf{q}})^2-(\hbar\omega_{a,\mathbf{q}})^2}\right)$$
$$+2\sum_{r\in R}|u^r|^2\frac{\rho_r}{(\varepsilon^0_{v,\mathbf{k}}-\varepsilon^0_{c,\mathbf{k}})^2}+2\sum_{r\in T}|u^r|^2\frac{\tau_r}{(\varepsilon^0_{v,\mathbf{k}}-\varepsilon^0_{c,\mathbf{k}})^2} \qquad (11.54)$$

$$\Delta\varepsilon_{c,\mathbf{k}} = -\sum_{\mathbf{q}\neq 0}|u^{\mathbf{q}}|^2\left(\frac{\hbar\omega_{o,\mathbf{q}}}{(\varepsilon^0_{c,\mathbf{k}}-\varepsilon^0_{v,\mathbf{k}-\mathbf{q}})^2-(\hbar\omega_{o,\mathbf{q}})^2}-\frac{\hbar\omega_{a,\mathbf{q}}}{(\varepsilon^0_{c,\mathbf{k}}-\varepsilon^0_{c,\mathbf{k}-\mathbf{q}})^2-(\hbar\omega_{a,\mathbf{q}})^2}\right)$$
$$-2\sum_{r\in R}|u^r|^2\frac{\rho_r}{(\varepsilon^0_{c,\mathbf{k}}-\varepsilon^0_{v,\mathbf{k}})^2}-2\sum_{r\in T}|u^r|^2\frac{\tau_r}{(\varepsilon^0_{c,\mathbf{k}}-\varepsilon^0_{v,\mathbf{k}})^2} \qquad (11.55)$$

leading to two sets of one-particle corrections, one set for the valence band electronic corrections and the latter set for the conducting ones.

Taking notice of the inner-band frequencies $\omega_{a,\mathbf{q}}$ that are not involved in the ground state energy equation, but are present in one-particle correction terms, these terms are the same as those in the Fröhlich's Hamiltonian (11.46), i.e. the denominators of them can achieve both positive and negative values. On the other hand the terms with inter-band optical frequencies $\omega_{o,\mathbf{q}}$ are optimized by means of equation (11.52), and therefore the negative denominators will be prevailing. This will result in negative values of $\Delta\varepsilon_{v,\mathbf{k}}$ and positive values of $\Delta\varepsilon_{c,\mathbf{k}}$. Of course, from the general form of the equations (11.54-55) one cannot uniquely predict the existence of a gap. Not all conductors might become necessary superconductors at absolute zero. It depends on many factors but the most important factor is the bandwidth. It is clear from (11.54-55) that the narrow bands (high $T_C$ superconductors) result in greater gaps than broad bands (low $T_C$ superconductors).

We have presented here an independent proof of the Goldstone theorem with a one-to-one correspondence between broken symmetries and associated massless and spinless bosons in condensed matter, such as molecules and solids, and we have found the lost bosons, i.e. the rotons and the translons. We have shown that a field theory based only on phonons is insufficient for the description of B-O degenerate states such as J-T systems and superconductors where the general field COM covariant theory, incorporating all Goldstone bosons - phonons, rotons and translons is unavoidable. The Born-Huang ansatz plays the same role for the field COM covariance as the Maxwell equations do for Lorentz covariance.

One small question remains: Is the Born-Huang ansatz experimentally credible to the same extent as the Maxwell equations? If so we can authorize the formulation of a new type of field covariance. It was the merit of Handy [110] and many others, that thousands of Born-Huang ansatz calculations and their comparison with experimental data were performed, since for a long time a reasonable doubt prevailed regarding the equivalency of these results, and those of the exact quantum mechanical COM separation and the Born-Huang ansatz as the first correction to the B-O approximation. Kutzelnigg therefore renamed the Born-Huang ansatz as the Born-Handy ansatz [8]: „Handy and co-workers have never claimed to have invented the ansatz referred to here as the "Born-Handy ansatz", but they certainly convinced a large audience that this ansatz is of enormous practical value, even if it has not been completely obvious why it leads to correct results. Handy and co-workers realized that the difficulties with the traditional approach come from the separation of the COM motion (and the need to define internal coordinates after this separation has been made). They therefore decided to renounce the separation."

Once the Born-Huang (Born-Handy) ansatz was experimentally confirmed, we have subsequently experimental evidence of the lost Goldstone bosons - rotons and translons - as well. Their role in quantum systems is quite curious. Their contribution to the corrections of the ground state energy is already significant in small molecules like hydrogen molecule with only one vibrational mode and the five lost Goldstone boson modes, and is even several times greater than the contribution of the pure vibrations. The bigger the molecule, the lesser effect have these rotons and translons. In most of solids, with a huge amount of phonons, such as conductors, semiconductors or insulators, the effect of the six rotons/translons becomes negligible and the electron-phonon field theory is sufficient. However, in B-O degenerate systems suddenly a surprise appears. We have arrived at the same formula for the ground states of finite as well as infinite systems, both molecules and crystals, in the case of a B-O electronic degeneracy. However, Goldstone bosons arising from the violation of rotational and translational symmetries, rotons and translons, produce singularities in the original symmetrical positions, and the system is forced to avoid them, removing the degeneracy in an effectively one-particle manner at new asymmetric positions. This principle unifies the formation of the ground states of J-T molecules and superconductors, showing the way how quantum field formulations deals with virtual degeneracies originating from approximative B-O solutions. They are profoundly different from real degeneracies, where the principle of superposition takes place after the removal by some external perturbation (Stark or Zeeman effect). Quantum field theories simply do not share the centre of mass defined in quantum mechanics but solves the centre of mass problem "on its own".

If the lost Goldstone bosons - rotons and translons - are responsible for the mechanism of the formation of quantum states with spontaneously broken symmetries, it means, that the spontaneous symmetry breaking (SSB) was only phenomenologically described by classical physics, but it was completely misunderstood on the quantum level. We will continue with the analysis of the SSB in the next section.

## 12. The paradox of spontaneous symmetry breaking

Goldstone's theorem implies a quantum field reformulation of the J-T effect which is also valid for superconductors [107]: "Molecular and crystallin entities in a geometry of electronically degenerate ground states are unstable at this geometry except for the case when all matrix elements of the electron-rotational and electron-translational interactions equal zero." When I published this definition [107], I did not, at first, realize that it was a direct consequence of the Goldstone theorem, and secondly I did look upon it as an alternative to the official quantum mechanical definition. Later I saw that only the field J-T definition should be correct, since it leads to a truly broken symmetry for the ground state.

The official mechanical definition starts with the clamped-nuclei concept of the B-O approximation (see section 2), whereas the exact mechanical solution, based on the Monkhorst-Cafiero-Adamowitz shell method for nuclei, leads to no symmetry breaking since the nuclei can be delocalized. On the other hand, the BCS theory of superconductivity is a quantum field theory where a Bloch-type field is used that does not respect the Goldstone theorem, and therefore the BCS theory cannot be correct. Yet the Bloch field was successfully applied to insulators, conductors, and semiconductors, but from the perspective of materials with B-O degenerate ground states and broken symmetries, such a field will be crippled and not usable. In such a case it permits linear superpositions of degenerate states, as it was originally described in the BCS paper [78]: „The normal phase is described by the Bloch individual-particle model. The ground state of a superconductor, formed from a linear combination of normal state configurations in which electrons are virtually excited in pairs of opposite spin and momentum, is lower in energy than the normal state by amount that is proportional to the average $(\hbar\omega)^2$, which is consistent with the isotope effect. A mutually orthogonal set of excited states, in one-to-one correspondence with those of the normal phase, obtains by specifying the occupation of certain Bloch states and by using the rest to form a linear combination of virtual pair configurations."

It is interesting to observe that, the BCS paper [78] does not mention any symmetry breakings. Four decades later, Weinberg wondered how Bardeen, Cooper, and Schrieffer could not see it [111]: „A superconductor is simply a material in which electromagnetic gauge invariance is spontaneously broken. This is not the way that most experts have historically thought about superconductivity. Early phenomenological theories were known to violate electromagnetic gauge invariance, but this was regarded as more annoying than enlightening. Broken symmetry is never mentioned in the seminal paper by Bardeen, Cooper, and Schrieffer [78] that first gave us a microscopic theory of superconductivity. Anderson [95] subsequently stressed the important role of broken symmetry in superconductors, but even today most textbooks explain superconductivity in terms of detailed dynamical models, with broken symmetry rarely mentioned."

Looking closely at the structure of the BCS theory, there is no asymmetry input/output, and hence there was absolutely no reason for Bardeen, Cooper, and Schrieffer to discuss any violations of symmetry. But we have already seen above that using the correct field, fully respecting the Goldstone theorem, the microscopic theory for superconductors will yield true asymmetric ground states, which are geometrical in accordance with the J-T effect. It means, that superconductors also share the common spontaneous symmetry breaking (SSB) with the J-T molecules. The quantum chemist and expert on the J-T effect, Isaac Bersuker came exactly to the same conclusion, writing in his book [112]: „Moreover, since the JTE has been shown to be the only source of spontaneous distortion of high-symmetry configurations, we come to the conclusion that the JTE is a unique mechanism of all the symmetry breakings in condensed matter." Bersuker further restated the prominence and role of the J-T effect in connection with the discovery of high-Tc superconductivity [112]: „The next significant resurgence of interest in the Jahn–Teller effect is related to the late 1980s and is still continuing. It was triggered by one of the most important Nobel Prize discoveries in physics of our times inspired by the Jahn–Teller effect: the high-temperature superconductivity. As explained by the authors of this discovery, ''the guiding idea in developing this concept was influenced by the Jahn–Teller polaron model'' (J.G. Bednorz and K.A. Müller, in Nobel Lectures: Physics, Ed. G. Ekspong, World Scientific, Singapore, 1993, p. 424)."

Since the mechanism of SSB lies behind both the J-T effect and superconductivity, it entails that both phenomena must be described by similar mathematical equations, cf. the presentation given in the previous section. It is therefore somewhat surprising that Bersuker ignored this fact accepting the Cooper pair based explanation for superconductors. In his book [112] he suggests further efforts to find a bridge between J-T distortions and Cooper pairing citing: „However, the electron–phonon coupling is not the only factor that influences the superconductivity, and it does not determine the SC transition temperature directly. The path from local JT distortions to formation of polarons via lattice dynamics, to formation of Cooper pairs, their density, stability, and mobility, temperature dependence, to perfect diamagnetism, to structural phase transition, and finally to SC, is very long and thorny, and therefore a comprehensive JT theory of this important phenomenon has not yet been accomplished."

In this connection it should be worth recalling Newton's second antimetaphysical rule [113]: „Hypoth. II. Ideoque effectuum naturalium ejusdem generis eædem sunt causæ. Uti respirationis in Homine & in Bestia; descensus lapidum in Europa & in America; Lucis in Igne culinari & in Sole; reflexionis lucis in Terra & in Planetis." It means that to the same natural effect the same causes must be assigned. Hence J-T distortions are the common causes of both phenomena, the J-T effect and superconductivity, which are observed on systems with B-O degenerate ground states and broken symmetries. Then it is not necessary to prompt the "very long and thorny path" from J-T distortions to the formation of Cooper pairs. In order to improve the understanding of the identity of these two phenomena, we will concentrate on their common cause, i.e. the J-T distortions, and especially on their underlying intrinsic concepts, the B-O approximation and the SSB, which both need to be carefully revised.

Starting first with the SSB and the problem of broken symmeties in the early period of classical physics, it led to a theological controversy about the question of its origin, between Newton and Leibniz. Hermann Weyl in his book [114], describes the conflict in detail: „The net result is that in all physics nothing has shown up indicating an intrinsic difference of left and right. Just as all points and all directions in space are equivalent, so are left and right. Position, direction, left and right are relative concepts. In language tinged with theology this issue of relativity was discussed at great length in a famous controversy between Leibniz and Clarke, the latter a clergyman acting as the spokesman for Newton [115]. Newton with his belief in absolute space and time considers motion a proof of the creation of the world out of God's arbitrary will, for otherwise it would be inexplicable why matter moves in this rather than in any other direction. Leibniz is loath to burden God with such decisions as lack "sufficient reason." Says he, "Under the assumption that space be something in itself it is impossible to give a reason why God should have put the bodies (without tampering with their mutual distances and relative positions) just at this particular place and not somewhere else; for instance, why He should not have arranged everything in the opposite order by turning East and West about. If, on the other hand, space is nothing more than the spatial order and relation of things then the two states supposed above, the actual one and its transposition, are in no way different from each other... and therefore it is a quite inadmissible question to ask why one state was preferred to the other.""

Weyl attempts to reconcile these opposite views from a new perspective of the theory of relativity after three centuries: „If nature were all lawfulness then every phenomenon would share the full symmetry of the universal laws of nature as formulated by the theory of relativity. The mere fact that this is not so proves that contingency is an essential feature of the world. Clarke in his controversy with Leibniz admitted the latter's principle of sufficient reason but added that the sufficient reason often lies in the mere will of God. I think, here Leibniz the rationalist is definitely wrong and Clarke on the right track. But it would have been more sincere to deny the principle of sufficient reason altogether instead of making God responsible for all that is unreason in the world. On the other hand Leibniz was right against Newton and Clarke with his insight into the principle of relativity. The truth as we see it today is this: The laws of nature do not determine uniquely the one world that actually exists, not even if one concedes that two worlds arising from each other by an automorphic transformation, i.e., by a transformation which preserves the universal laws of nature, are to be considered the same world."

Does Weyl's statement hold in universally physics? In classical physics including Einstein's relativity, we need not have to care about Leibniz' principle of sufficient reason, since this principle is of teleological origin. We have also seen that in the classical solution of superconductivity this principle is hidden behind a multiplicity of solutions, the first one leading to ideal conductivity and the second one to superconductivity, in full agreement with Weyl's interpretation. However, this construal is not valid in quantum physics, since quantum physics is causal and hence cannot observe teleological phenomena. After a rigorous use of Goldstone's theorem, we have obtained a superconducting ground state, which is identical to the ground states of insulators with J-T distorted nuclear positions. This is of course not enough for a quantum explanation of superconductivity. The quantum nature of Leibniz' principle must be first discovered, and only then might the quantum essence of superconductivity and all other teleological phenomena be revealed and seriously discussed.

This evokes the question: Does there exist a general formulation of the relationship between broken symmetries and the respective physical phenomena, valid both in classical and quantum physics? Pierre Curie proposed already in 1894 a formulation [116], fully based on causality, predating an analysis, which according to Bohr's principle of correspondence would apply to an extension from classical to quantum physics. The English translation of Curie's formulation with further extensive analysis can be found in the book of Brading and Castellani [117]. We quote here only the essential part: „Curie was led to reflect on the question of the relationship between physical properties and symmetry properties of a physical system by his studies on the thermal, electric and magnetic properties of crystals, these properties being directly related to the structure, and hence the symmetry, of the crystals studied... His conclusions, systematically presented in his 1894 work "Sur la symétrie dans les phénomènes physiques", can be synthesized as follows:

(a) A phenomenon can exist in a medium possessing its characteristic symmetry or that of one of its subgroups. What is needed for its occurrence (i.e. for something rather than nothing to happen) is not the presence, but rather the absence, of certain symmetries: "Asymmetry is what creates a phenomenon".

(b) The symmetry elements of the causes must be found in their effects, but the converse is not true; that is, the effects can be more symmetric than the causes.

Conclusion (a) clearly indicates that Curie recognized the important function played by the concept of symmetry breaking in physics (he was indeed one of the first to recognize it). Conclusion (b) is what is usually called "Curie's principle" in the literature, although one should notice that (a) and (b) are not independent of each other."

One might observe that Curie uses solely the pair of concepts: symmetry - asymmetry, and does not explicitly speak about SSB. Therefore Curie's formulation reveals only half of the truth about SSB, i.e. it deals only with causal phenomena arising from the multiplicity of asymmetric solutions of physical equations. It says nothing about the teleological phenomenon descending from the transitions between these asymmetric states. As an example the causal effect, related to ferromagnetism, is well explained in quantum physics. On the other hand, the associated teleological Einstein-de Haas effect has no quantum physical explanation, even if it was well described in the framework of classical physics in 1915 [118], before the concept of the electron spin was known, leaving only Ampère's hypothesis that magnetism is caused by microscopic circular motions of electric charges as a sufficient background. Superconductivity is an another example: Quantum physics can only solve the causal condensation of a conductor to the superconducting phase with multiple asymmetric ground states, but the explanation of the teleological transition between these ground states lies beyond the limits of quantum physics.

The SSB is primarily associated with two phenomena, i.e. one causal and one teleological. Curie's formulation deals with the first one. The phenomena in question corresponds to the following symmetry argument that was formulated by van Fraassen [119]: „There are two forms of argument which reach their conclusion 'on the basis of considerations of symmetry'. One, the symmetry argument proper, relies on a meta-principle: that structurally similar problems must receive correspondingly similar solutions. A solution must 'respect the symmetries' of the problem. The second form, rather less important, assumes a symmetry in its subject, or assumes that an asymmetry can only come from a preceding asymmetry. Both exert a strong and immediate appeal, that may hide substantial tacit assumptions."

The first part of van Fraassen's argument is quite clear and there is full agreement in the scientific community regarding its interpretation. Even if we get several asymmetric states as solutions of symmetric equations, i.e. if the original symmetry is broken, the asymmetric states together form a set which respects the original symmetry of the problem. However, the second form is not clear at all, and there is no unified view how to interpret it. One group of physicists attempts to explain it within the causal laws of physics. Usually imperfections and random fluctuations are taken into account. Stewart and Golubitsky in their book [120] describe the role of imperfections as follows: „We've said that mathematically the laws that apply to symmetric systems can sometimes predict not just a single effect, but a whole set of symmetrically related effects. However, Mother Nature has to choose which of those effects she wants to implement. How does she choose? The answer seems to be: imperfections. Nature is never perfectly symmetric. Nature's circles always have tiny dents and bumps. There are always tiny fluctuations, such as the thermal vibration of molecules. These tiny imperfections load Nature's dice in favour of one or other of the set of possible effects that the mathematics of perfect symmetry considers to be equally possible."

There is no doubt that imperfections play an important role in macroscopic objects. The Norton dome is a nice example of this. Nontheless, Norton emphasises the difference between the real and ideal dome [121]: „However, we do not even have a falsified prediction. The dome is not intended to represent a real physical system. The dome is purely an idealization within Newtonian theory. On our best understanding of the world, there can be no such system. For an essential part of the setup is to locate the mass exactly at the apex of the dome and exactly at rest. Quantum mechanics assures us that cannot be done. What the dome illustrates is indeterminism within Newtonian theory in an idealized system that we do not expect to be realized in the world."

Such idealizations are of course not convincing enough for many physicists to take the Norton dome seriously, particularly when there are no applications to the real world. Norton reckons all objections against the dome and the indeterminism in classical physics [121]: „What's wrong with the dome? Many believe that the dome somehow lies outside what is proper in Newtonian theory. Three distinct bases for this judgment are described below, along with my reasons for finding them unconvincing. 1. Does the dome employ an incomplete formulation of Newtonian physics? 2. Is the dome "unphysical"? a) Unphysical as gauge (over-description). b) Unphysical as false. c) Unphysical as pathological. d) Unphysical through under-description. 3. Does the dome use inadmissible idealizations?"

Although Norton disproves very carefully the above mentioned objections, a curious situation arises: On one hand Norton is correct in his arguments against his opponents, but on the other his opponents are right about the determinism of classical physics. Malament [122] has expressed this quandary in a very elegant way: „But I am not sure that the full complexity and interest of the breakdown is adequately captured by saying, either, "Newtonian particle mechanics is an indeterministic theory" (full stop) or "Norton's example is not a well-defined Newtonian system" (full stop). Indeed, I am not convinced we have clearly posed alternatives here – because we do not have a sufficiently clear idea in the first place what should count as a "Newtonian system" (or count as falling within the "domain of application" of Newtonian theory). My inclination is to avoid labels here and direct attention, instead, to a rich set issues that the example raises."

In order to gain a deeper insight into this problem one should look for analogies such as the Norton dome. A historical review of comparable examples was written by van Strien. In her article "The Norton Dome and the Nineteenth Century Foundations of Determinism" [123] she says: „Lipschitz-indeterministic systems played an important role in the work of Boussinesq, who, in 1878, discussed several such systems, one of which was similar to the Norton dome [124]." This paper contains also an observation of an unexpected importance: „Though determinism as a general metaphysical principle was seldom doubted, it was not necessarily founded on a physical theorem about the uniqueness of solutions to certain differential equations."

It's quite astonishing to notice the difference in the way 19th century scientists were thinking! Maxwell, in his letter to Darwin's half-cousin Galton (quoted in the section 7), being fascinated by the works of Boussinesq & al., does not speak about the problem of indeterminism in classical physics at all. Yet he mentions "some determining principle which is extra physical (but not extra natural)" instead. It indicates that Maxwell was well aware of a missing law of nature, when he continues: „Boussinesq's method is a very powerful one against metaphysical arguments about cause and effect and much better than the insinuation that there is something loose about the laws of nature, not of sensible magnitude but enough to bring her round in time."

It appears that Norton already used the correct word "acausal" in his first paper dealing with the dome [52]. Using the word "acausal" as a proper opposite to the word "causal" is to be preferred, because it leaves the door open for teleological phenomena. Any attempt to analyse indeterminism on the classical level is meaningless since classical physics is unable to deal with it, perhaps at best can somehow "tolerate" it as in the case of Norton's dome. We may not even need to discuss the problem of Lipschitz continuity: the Mexican hat in the Ginzburg-Landau equations has no such discontinuity at the apex and, after all, we observe the same type of double solutions for the Norton dome and ideal conductors (first solution) and superconductors (second solution) the latter leading to the Meissner effect. Classical physics is able to describe the Norton dome as well as the Meissner effect, but is unable to determine what evoked the phenomenon.

If we take into account all the known examples of SSB in classical physics, i.e. the Mexican hat, Norton's dome, or the systems discussed by Boussinesq, we conclude in relation to the SSB archetype, forming the "falling down from the apex" in the classical physics, that the trigger of falling down is unknown. According to Maxwell's contemporary, mathematician and philosopher Peirce, no physics based on a corpuscular philosophy can solve the problem of the SSB trigger [125]: „Now I maintain that the original segregation of levo-molecules, or molecules with a left-handed twist, from dextro-molecules, or molecules with a right-handed twist, is absolutely incapable of mechanical explanation... The three laws of motion draw no dynamical distinction between right-handed and left-handed screws, and a mechanical explanation is an explanation founded on the three laws of motion. There then is a physical phenomenon absolutely inexplicable by mechanical action. This single instance suffices to overthrow the Corpuscular Philosophy."

Although these lines were written in the era of classical physics, Peirce's argument is applicable to quantum physics as well: It is built on the atomic ideas of Democritus, with the same corpuscular philosophy as in classical physics. It means that quantum physics in its Copenhagen interpretation is incapable of solving the problem of the SSB trigger. As we have realized this is a teleological issue, related to Leibniz' principle of sufficient reason, the second form of van Fraassen's argument, and Peirce's corpuscular philosophy argument, which can be fully ignored in classical physics. Unfortunately physicists usually ignore this difficulty also in quantum physics, and this attitude is a mistake, since it is motivated by the incorrect belief, that quantum physics employs the same SSB archetype as the classical one.

On the quantum level, physical systems can be described in a dual way: either mechanically or as a field system, in accordance with the Weinberg's statement [126]: „If it turned out that some physical system could not be described by a quantum field theory, it would be a sensation; if it turned out that the system did not obey the rules of quantum mechanics and relativity, it would be a cataclysm." Note that we demonstrated a similar picture in section 11 particularly for condensed matter cases (molecules and crystals): Side by side the precise Monkhorst's quantum mechanical approach to the solution of the Schrödinger equation we present the true quantum field solution fully respecting the Goldstone theorem.

We are now in the position to ask whether this dual description of physical systems would lead to the same result. The answer is certainly not, because different results obtain for symmetry broken states. Quantum mechanics admits the superposition of multiple degenerate states, while Weinberg, on the other hand, has proven their orthogonality for quantum fields within the infinite volume limit. The Weinberg's argument is often quoted by various authors, e.g. in the article of Brading and Castellani [127]: „In quantum physics SSB actually does not occur in the case of finite systems: tunnelling takes place between the various degenerate states, and the true lowest energy state or "ground state" turns out to be a unique linear superposition of the degenerate states. In fact, SSB is applicable only to infinite systems — many-body systems (such as ferromagnets, superfluids and superconductors) and fields — the alternative degenerate ground states being all orthogonal to each other in the infinite volume limit and therefore separated by a "superselection rule" (see for example Weinberg, 1996, pp. 164–165)."

Though this citation reflects Weinberg's opinion, one can also find other interpretations, emphasizing the difference between the mechanical and the field descriptions, rather than the condition of the volume infinity of fields, as e.g. in the paper of van Dam [128]: „For the SSB in the rod and in the ferromagnet two comments were made that apply equally to this case: each of the ground states has an equal chance to be the ground state of the physical system, and the ground states are related to each other by the U(1) symmetry of the Lagrangian. Relating this situation to quantum mechanics might provoke suspicion: through tunnelling the real ground state could surely be a superposition of the individual ground states, which would not be degenerate. But although such a situation could occur in ordinary quantum mechanics, it does not apply to quantum field theory. The fields live in an infinite volume and thus have infinitely many degrees of freedom. Tunnelling cannot happen in this case. All vacuum states are orthogonal. This is extensively discussed by, for example, Weinberg [111, pp. 163-167]."

The results presented in the previous section, testify in favour of van Dam's interpretation: there is no SSB in quantum mechanics, it can solely be described in quantum field theory. Moreover, one might not need to rely on Weinberg's proof, which is valid only for infinite-volume fields, since the Goldstone boson, associated with broken translational and rotational symmetries, earlier in this work called translons and rotons [107], is responsible for singularities at symmetric points. Hence the manifestation of SSB, in order to avoid these singularities, does appear for any system, regardless of the system being finite or infinite.

As a matter of fact, if we have two legitimate physical descriptions, the quantum mechanical and the quantum field one, these two formulations must necessarily yield the same energies for the system under examination. Therefore the classical SSB archetype, here denoted as "falling down from the apex", is not workable for quantum systems. Quantum physics can only calculate the symmetry broken states, while the SSB transition must somehow follow from the transitions between the mechanical and the field states of the system proceeding without any energy gain or loss, since symmetrical mechanical points and symmetry broken field points represent two descriptions with the same energy. A new law of nature is indispensable for this purpose, since this type of transitions has no support within the Copenhagen interpretation. As a consequence, from this new law, the origin of asymmetry must emerge, and evidently, according to the second form of the van Fraassen's argument, this asymmetry cannot arise *ex nihilo*.

So where should one start searching for this new law of nature? One way would be to focus on the simplest cases of SSB. We will, however, not start with complex nonadiabatic phenomena such as superconductivity or the J-T effect, nor with the adiabatic, but infinite systems like ferromagnets. The simplest example of SSB for our purpose would be small adiabatic systems such as the formation of isomers. The exact quantum mechanical solution, according to the Monkhorst-Cafiero-Adamowitz approach [6,9], cannot lead to isomerism. The latter appears only after the introduction of the B-O approximation, or the clamped-nuclei concept. It seems like this approximation somehow mimics the quantum field properties where SSB can only appear. This is probably why Sutcliffe and Woolley, see section 2, have put questions related to the transition from the isolated to the individual systems on the list of unsolved problems in quantum theory. In fact the solution of this problem, as presented here, is the key to providing a description of the SSB archetype on the quantum level.

As stated here the application of the classical SSB, i.e. the Mexican hat, to quantum equations, such as the J-T effect, superconductivity or the Higgs mechanism, is a disparity. This results in a misinterpretation of related phenomena, and it reminds of the reincarnation of the notorious medieval puzzle, namely how many angels can dance on the apex of a needle, i.e. a comparable mismatch by combining a material piece of a needle with the spiritual world of angels. Consequently quantum equations, grounded on the classic SSB, lead to a paradoxical situation. Hence the scientific community appears divided in regard to the question whether SSB is presently accounted for or not. According to Bersuker, the J-T effect is a unique mechanism of all existing SSBs in condensed matter, while, according to Weinberg, quantum mechanics cannot lead to any SSB. Bardeen, Cooper and Schrieffer did not find any SSB in their theory of superconductivity, yet Weinberg wondered how they could miss it. In standard text books, the Higgs mechanism is described as a SSB mechanism, but gradually a strong opposition against this view has emerged, usually quoting the Elitzur's theorem [129], maintaining that local gauge symmetries cannot be spontaneously broken. The tip of the iceberg is the statement by Anderson, that ferromagnetism is not a case of SSB. His opinion was condemned by Peierls and Kaplan, see their discussions in Ref. [130].

Finally, Unzicker in his book "The Higgs Fake: How Particle Physicists Fooled the Nobel Committee" [131] quotes the prophetic pronouncement of Rudolf Clausius: „Once the error is based like a foundation stone in the ground, everything is built thereupon, nevermore it returns to light." In the following section we will return to the beginning of the many-body treatment in quantum physics, revealing an overlooked secret regarding the B-O approximation. Here one may anticipate the starting point of the whole inconsistent chain that culminates in the uncertainty of many physicists, which phenomenon is of the SSB type and which one is not.

## 13. The paradox of one Bohr complementarity

We begin by looking more carefully at the qualitative nature of the B-O approximation. Even if it has been successfully proven in most molecular many-body calculations as well as its accuracy verified times and again, it is nevertheless a very non-standard approximation. In its original derivation, via the $m/M$ (electron mass/nuclear mass) series expansion, it is unique in the sense that no competing series exists, cf. the Brillouin-Wigner or the Rayleigh-Schrödinger perturbation theories. In contrast to the latter, the B-O approximation is not only the basis of almost all molecular quantum chemical calculations, but it provides also a fundamental concept for molecular structure [132]. Since one knows that physical quantities, such as mass, charge, energy, momentum etc., need exact definitions, how is it then possible that molecular and crystalline structures are defined by an approximation. The latter must either be erroneous or alternatively not principally an approximation at all, hiding a precise and fundamental meaning. One reason for accepting the second choice, would be the argument by Sutcliffe and Woolley, see the section 2, regarding the simplest case of broken symmetries. For instance the Monkhorst-Cafiero-Adamowitz approach, dealing with isolated molecules, does not recognize any isomerism. They appear only on the B-O level, that deals exclusively with individual molecules. The conclusion is clear, no approximation will lead to new phenomena such as symmetry violations or even to making an ontological shift from isolated to individual order of the systems under investigation. Quoting again the challenge by Sutcliffe & Woolley, see section 2: „The interesting question is how to get from the quantum theory of an Isolated Molecule to a quantum theory of an individual molecule by rational mathematics."

Pioneering quantum mechanics, describing a system of nuclei and electrons, cannot answer the abovementioned dilemma. We need a field theoretical description of the fermions as renormalized electrons and the nuclei replaced by Goldstone bosons as vibrational, rotational and translational modes. As was shown in section 11, the field Hamiltonian (11.33) satisfies this request finally resulting in the clamped-nuclei concept, the first step of the B-O approximation, yielding all the equations known from this approximation, such as those of Pople for the ab-initio calculation of vibrational frequencies (11.40,11.41) and the Born-Huang ansatz (11.49). However, the clamped-nuclei concept is a *contradictio in adjecto*. For instance either one keeps the nuclei in mind, which can never be "clamped" since the nuclear positions do not commute with the total Hamiltonian, or one becomes fixed on the adjective "clamped", emphasizing instead of nuclei rather to speak about some traces or footprints, such as e.g. traces of electrons on the screen or traces of photons on a photographic plate. In a previous paper [108] I did introduce the notion of property-object dualism as follows: In quantum mechanics, nuclei and electrons represent objects to be described on an equal footing, and the vibrational modes are their common property. On the other hand, in quantum field theories the objects are represented by electrons and the Goldstone bosons, and "nuclear" positions or "clamped nuclei" are properties of the pertinent field equations.

As far as adiabatic systems are concerned, we have shown a proof of the equivalency between the field equations and the mechanical equations based on the B-O concept. What happens for nonadiabatic systems? While the practice of the B-O model is limited to adiabatic systems, there are no such limitations for the field equations. They are valid on the whole scale from adiabatic to nonadiabatic. For example we proved in section 11 how the true field equations, based on Goldstone's theorem, do lead to symmetry broken ground states with the degeneracy removed on the one-particle level in both J-T systems and superconductors. In contrast the B-O approach to nonadiabatic systems leads to metaphysical solutions that are based on linear combinations of degenerate states. For the case of an infinite number of degenerate states the B-O concept gives rise to Mexican hats and the Berry-phase problem in so-called J-T molecules, see Bersuker [112]: „…one of the simplest JT $E \otimes e$ problems with linear vibronic coupling yields an APES (Adiabatic Potential Energy Surface) in the form of a ''Mexican hat''… For a long time, up to the last decade, not very much attention was paid to these conical intersections; rather they were viewed largely as a ''theoretical curiosity.'' This perception changed recently in view of the latest achievements in the treatment of such systems. One of these achievements is a generalization directly related to the JTE and now known as the topological (geometric) phase, or the Berry-phase problem… An important feature of the Berry-phase implications in JT problems is that the peculiar phase factor that changes the sign of the electronic wavefunctions and makes the ground vibronic state degenerate occurs only when one or an odd number of conical intersections are encircled, while it retains the same sign if an even number (including zero) are encircled. This was shown by direct calculation of the phase in the $E \otimes e$ problem. In fact, the phase is $\phi_0 = n\pi$, where $n$ is the number of conical intersections encircled; for $n = 0, 2, 4, \ldots$ the sign of the wavefunction does not change."

The Berry-phase was mentioned above. The exact definition can be found in Berry's original paper [133]: „A quantal system in an eigenstate, slowly transported round a circuit $C$ by varying parameters $\mathbf{R}$ in its Hamiltonian $H(\mathbf{R})$, will acquire a geometrical phase factor $\exp[i\gamma(C)]$ in addition to the familiar dynamical phase factor… If $C$ lies near a degeneracy of $H$, $\gamma(C)$ takes a simple form which includes as a special case the sign change of eigenfunctions of real symmetric matrices round a degeneracy." However, there is a problem, since the Berry-phase, based on the definition above, cannot be applied to J-T molecules at all. For instance the B-O approximation can either be derived as a result of the $m/M$ expansion or in an equivalent form using the hierarchical quantization process, i.e. first quantize the electronic motion at fixed classical nuclear positions and then *a posteriori* quantize the nuclear motion. Applying the Berry-phase to J-T molecules is just a product of this type of hierarchical quantization. A more exact simultaneous quantization method, i.e. Monkhorst's approach, see section 2, or the quantum field approach using the full set of Goldstone bosons, presented here section 11, can never lead to a meaningful Berry-phase in J-T systems.

In section 12 we mentioned the often quoted Weinberg attitude to the occurrence of SSB: Only infinite many-body systems and fields can be spontaneously broken, whereas finite systems due to the possible tunnelling between degenerate states cannot. But from the perspective of our analysis this attitude does not reflect the whole truth. Physicists often reject considerations that has even if only a little sniff of chemistry and do not realize that the simplest examples of SSB, such as isomerism arising on the platform of quantum chemistry in finite systems and that a revealing of the true nature of the B-O approximation is the correct response to a correct understanding of SSB. In fact, fields with infinite degrees of freedom are not necessary for SSB occurance at all.

We have here argued that a field theoretical formulation are valid for molecules with finite degrees of freedom as well. Moreover, this theory is able to describe isomerism, i.e. SSB. On the other hand, many-body quantum mechanical applications, regardless of being finite or infinite, can never lead to any SSB. Based on Monkhorst's quantum mechanical approach, fully ignoring the B-O concept, one concludes that there is no isomerism, no ferromagnetism, no J-T effect, and no superconductivity. Ferromagnets, in calculations similar to those of Cafiero and Adamowitz, would look like giant molecules with full spherical symmetry and therefore without any SSB. J-T molecules would look like any other molecules with a non-degenerate electronic spectrum and superconductors looking like insulators. Only after the introduction of the B-O approximation, here a "dirty trick", the isomerism appears, and we can construct the simplest Heisenberg model of ferromagnetism. But this simplification makes only sense for adiabatic systems. However, in cases such as J-T molecules and superconductors, it yields ontologically incorrect results and here field methods based on the full set of Goldstone bosons must be applied.

In agreement with Weinberg's statement that every quantum system can be described by two profoundly different descriptions, either as a mechanical system or as a field system, and that only the second one recognizes SSB, one may ask if these two descriptions are still equivalently valid or whether only one of them is fundamentally correct. There is also the temptation to believe that the field approach supersedes the mechanical one, i.e. that the quantum field description should be seen as a generalization of quantum mechanics in the same way as general relativity becomes a generalization of special relativity. But listen to what Einstein said about this topic [18]: „Newton's theory deserves the name of a classical theory. It has nevertheless been abandoned since Maxwell and Hertz have shown that the idea of forces at a distance has to be relinquished and that one cannot manage without the idea of continuous "fields." The opinion that continuous fields are to be viewed as the only acceptable basic concepts, which must also be assumed to underlie the theory of the material particles, soon won out. Now this conception became, so to speak, "classical"; but a proper, and in principle complete, theory has not grown out of it. Maxwell's theory of the electric field remained a torso, because it was unable to set up laws for the behaviour of electric density, without which there can, of course, be no such thing as an electro-magnetic field. Analogously the general theory of relativity furnished then a field theory of gravitation, but no theory of the field-creating masses."

Einstein's quote above indicates that we must accept both the mechanical- and the field description as equally valid. However, the mechanical description pictures every system as isolated with no SSB, while the field formulation allows SSB at the same time being responsible for the individual character of the system. The question becomes whether there is any mathematical transformation between them, as inquired by Sutcliffe & Woolley. This reminds on the first stage of the development of quantum physics, when it turned out that elementary entities had both mechanical properties (particles) and field properties (waves). Even if Bohr had introduced the concept of complementarity into physics, some physicists developed a negative attitude towards it and attempted to find direct transformations between particles and waves without the need for any complementarity, such as Einstein or de Broglie. In analogy with just how a century ago the different aspects of the mechanical and the field descriptions of elementary entities led to Bohr's concept of complementarity, we are now in a similar situation with respect to the descriptions of the many-body system. Hence a second type of Bohr complementarity for the many-body level is requested.

This suggestion does not sound weird, since a second type of complementarity was already requested a long time ago by one of the co-founder of quantum mechanics Pascual Jordan. In his now almost forgotten paper [134] he wrote: „We assume here an idealised photographic plate: Each photon hitting it will be absorbed, and a single photon will with certainty activate a certain silver grain... If we assume the exposed silver grain to be indeed in a state of well-defined decision as to its developability, then we must conclude that it is not merely voluntary resignation on our part if we do not describe the silver grain in terms of wave functions of its single atoms. Doing so would entangle us in contradictions, as we have already seen above. Therefore the physical situation itself must contain guarantees that such contradictions cannot take place - and only a second type of complementarity can give this guarantee. There must exist in the silver grain a certain situation by which its description in terms of atomic wave functions is made impossible - only in this manner can the grain function as it does... It is then apparent that the situation - though it is clear to a certain extent - does not allow a complete and final analysis; there remain open certain questions. For one cannot avoid the difficulties merely by describing the silver grain (or an analogous part of any observational instrument) as a "mixture" of the form of the statistical matrix; this would not help us much, for it cannot describe an increase of entropy any better than the Schrödinger equation of a "pure case." It seems to me that entirely new conceptions are necessary."

This infers an extra quality of complementarity imparting a new type of quantum transitions on the many-body level, which cannot be deduced from any known rules of quantum physics. It entails a new axiom, and this axiom was requested by Jordan in the same paper [134]: „This leads us to acknowledge that it is both possible and necessary to formulate a physical axiom not formulated hitherto. Above we held it to be part of the definition of macrophysics, to show no complications in the manner of complementarity, but to allow a complete "objectivation" of phenomena in space and time. But usually one defines macrophysics only by stating that it deals with great numbers of microphysical individuals - and this is another and a different definition. We need therefore a special axiom to express the empirical fact that these two definitions define the same thing - that really each large accumulation of microphysical individuals always shows a well-defined state in space and time - that a stone never, unlike an electron, has indeterminate coordinates. One often vaguely believes this to be guaranteed already by Heisenberg's $\Delta p \cdot \Delta q > h$; but in fact this relation only provides a possibility and not a necessity for the validity of our axiom. Let us assume that, in our experiment involving the photon, the photographic plate be removed, but that we have an arrangement whereby a macro-physical stone will fall according to the decision of the photon. Then, if we strictly assume v. Neumann's view, the stone comes to possess a wave function which makes it undecided whether it does fall or does not, and an observer has the opportunity to compel the stone to a decision by the mental process of forgetting that interference between the two wave functions of the falling stone would be possible. Schrödinger's famous cat is another illustration of this point. I think we can summarize the situation by saying that indeed a new feature - to be formulated by a new axiom - lies in the fact that such things do not happen; all formulations of quantum mechanics hitherto given do not suffice to exclude them. We are unable to make a clock with a hand which does not always point to a definite figure on the dial. This is a well-known fact, but a fact of which present theory gives no sufficient account."

The suggested axiom above, is in our case relevant to describe quantum jumps between the isolated and individual characteristics of the many-body system, and provides a response to the request of Sutcliffe and Woolley. There is, however, one main obstacle regarding the definition of an isolated system. Every finite isolated system is subjectively defined and since transitions between isolated and individual systems are in principle immeasurable, one must necessarily exclude the subjectivity of the observer. As a consequence the isolated system must be identical with the whole Universe, with the individual systems then identical with its fragmentations. We have arrived at the problem of fragmentation and wholeness, cf. how it was framed by Bohm in the 3rd statement quoted in section 4. The second complementarity, embracing the whole Universe, is the megascopic mirror of the first Bohr microscopic complementarity. We have just touched ancient knowledge, i.e. presenting the Universe as a mosaic where the smallest one resembles the Greatest One.

The new axiom presented here is necessary for the justification of quantum jumps of the Universe. From its state of total wholeness to its total fragmented states and vice versa, it certainly goes beyond the scope of the Copenhagen interpretation. Yet it neither contradicts nor denies this interpretation, but rather appears to be its natural extension. Then again, acceptance of this axiom implies the end of Everett's MWI (discussed in section 7). The MWI was promoted as a deterministic theory for the physical Universe and attempted to explain why a world appears to be indeterministic for human observers. The basic stipulation of MWI was to represent the whole Universe by a deterministic wave function. Although still popular, many physicists now have doubts about the completeness of such a description. Vaidman comments on this problem, i.e. whether the wave function is or not sufficient, in the following way [135]: „As mentioned above, the gap between the mathematical formalism of the MWI, namely the wave function of the Universe, and our experience is larger than in other interpretations. This is the reason why many thought that the ontology of the wave function is not enough. Bell 1987 (p.201) felt that either the wave function is not everything, or it is not right. He was looking for a theory with local "beables" [136]."

Surely, the ontology of the wave function is not enough. It does not mean that it is not right, since no known experiment is in conflict with it. In fact the wave function is an incomplete description of the Universe. Most probably the protagonists of the MWI never gave a thought to the ontological problem of the B-O approximation, and therefore do not realize that their wave function is unable to describe real objects like molecules or solids, but only the subjectively defined isolated objects of different ontological meanings, where, according to the language of Cafiero and Adamowitz, instead of individual molecules one has only molecular atoms or atomic molecules, and instead of crystals only crystalline atoms or atomic crystals.

However, the incomplete character of the wave function of the Universe should not set us off to look for some local "beables". As we have seen, there is a complementary field description of the Universe, leading ultimately to real objects, the symmetries of which can be broken. The second complementarity has an original significance for a proper understanding of all SSB phenomena. These processes have a sign of complexity and emergence, and therefore one must ask which fundamental principle lies behind. Rowlands has adequately formulated this request [137]: „This is symmetry, which we see all about us in the laws of physics, the fundamental interactions and the fundamental particles. Finding symmetries can help us to decomplexify our explanations, and, if we can understand where symmetry comes from, lead to more profound understanding. We should also note that some symmetries are broken; that is, what is fundamentally symmetric appears, under certain conditions, to display some asymmetry. The reason for this cannot be arbitrary, and if we can discover it, along with the reason for symmetry, this will be a big step in our understanding of the foundations. One thing it can't be is fundamental because nature never acts in such an arbitrary way. It has to be, in some way, a sign of complexity or emergence."

With the up till now known laws of quantum physics one can simply only arrive at a mechanical many-body description with no symmetries violated and a field description exhibiting possible broken symmetries. But to suggest an explanation for the broken symmetry, understood as a transition from unbroken to broken states, on needs to accept the new axiom. We continue quoting Rowlands giving his opinion [137]: „Again, if we decide that simplicity is to be preferred over complexity, we should be looking for something that is staggeringly simple, yet somehow capable of generating complexity. If we think our basic idea is a complicated construction, say a 10-dimensional space-time, then we have no way of knowing how this breaks up into the simpler component parts that must exist because we have no fundamental mechanism for doing this. We should certainly take notice of what the string theorists say about the symmetries required by nature, and we should expect to find them, even those expressed in 10 dimensions, but we should expect to find them by working out how such a complex idea emerges from simpler ones, in which the structures of the components reveal them as diverse in origin, the 'brokenness' of the larger symmetry coming from its inherent complexity, not by some arbitrarily-imposed concept of 'symmetry-breaking'. Broken symmetries are a sure signature of complexity, not of simplicity."

In the above citation Rowlands has touched upon the crucial problem how symmetry breaking is comprehended today as some arbitrarily imposed conception. Yet if the second complementarity is original, then every SSB is an emergent process, i.e. the succession of events, appearing either on the microscopic or the macroscopic level, as a consequence of the perpetual sequence of megascopic quantum jumps between states that represent the wholeness and the fragmentation of the Universe.

Unfortunately megascopic quantum jumps are not measurable. This fact is reflected in emergent SSB phenomena, such as superconductivity and the J-T effect being the best-known examples. Knowing, that the density and velocity of superconducting carriers are non-measurable in principle, see section 9, Bersuker describes a similar situation in J-T systems [112]: „Among other things Van Vleck [138] wrote that "it is a great merit of the J-T effect that it disappears when not needed." This declaration reflects the situation when there was very poor understanding of what observable effects should be expected as a consequence of the J-T theorem. The point is that the simplified formulation of the consequences of the J-T theorem as "spontaneous distortion" is incomplete and therefore inaccurate, and may lead to misunderstanding. In fact, there are several (or an infinite number of) equivalent directions of distortion, and the system may resonate between them (the dynamic J-T effect)... It does not necessarily lead to observable nuclear configuration distortion, and this explains why such distortions often cannot be observed directly... Even in 1960 Low in his book [139] stated that "it is a property of the J-T effect that whenever one tries to find it, it eludes measurements.""

Bersuker's countering argument against Van Vleck and Low can only be acceded from the point of view of the microscopic origin of the J-T effect. But as we have realized, the J-T effect is a megascopic phenomenon. Any other attempt to explain this effect microscopically leads inevitably to paradoxes. From a microscopic perspective quantum tunnelling between differently distorted J-T states restores the original symmetry provided one considers a sufficiently long period of time. And this opens directly the problem discussed at the end of the previous section, i.e. is there any SSB or not. Moreover considering the simplest microscopic example of quantum tunnelling, the particle in a box, divided into two parts by a barrier. If the particle is initially located in the first part, after an extended time period it will be found with the same probability in both parts. But in the case of the J-T effect, there is no observer who selects one of the possible distorted J-T states. Therefore the microscopic concept of quantum tunnelling cannot be applied to the J-T effect, and Van Vleck and Low were absolutely right, when they challenged its direct measurability. Incidentally microscopic processes are causal, while megascopic ones are teleological. Teleology interpreted as the final cause can be then understood as downward causality, descending from the Greatest One, the whole Universe, in contrast to the upward causality, ascending from the smallest entities, the elementary particles. In this way teleology finally gains a primary significance, as it was requested by Bohm, quoted in his 1st statement in section 4. Causality in physics has its well-established empirical basis, but the principal non-measurability of megascopic phenomena might perhaps be the main reason why teleology is not till now explicitly incorporated in the exact sciences.

Approaching the next questions, i.e. what identifies the probabilities of megascopic quantum jumps, and what are the megascopic mirrors of the microscopic projection and the Born rule, Bohm in his 5th cited statement, see above, speaks about projections and injections. Indeed the latter phrase agrees perfectly with a requested megascopic mirror of the former. One can observe in various international forums that some scientists do feel that there is a connection between the mechanism of SSB and Zurek's einselection, but that no one knows exactly how the latter might explicitly be incorporated. The quantum decoherence program is focused on replacing the Born rule by einselection and entirely removing the von Neumann - Wigner rule, see section 7. This program, however, was inspired by Universal Darwinism, and Darwin was himself a teleologist, with a telic notion of adaptation. Therefore an environmentally induced superselection rule must be of teleological origin and co-exists with the Born law being its megascopic mirror.

Unlike the Born rule einselection is not defined by some simple formula. If the environmental perturbation is negligible, the probability of transition into each of the $n$ distorted J-T states equals $1/n$. Alternatively, if we have an isomer with two possible configurations, e.g. a right-handed and left-handed molecule, and the environmental influence prefers the right-handed one, then the transition probability for the latter equals one and for the other equalling zero. In superconductors transition probabilities between distorted ground states are given by the magnitude of the external magnetic field.

Note that Bohm's 2nd statement, implicates a limited world view, based on microscopic quantum physics, dealing with elementary particles as 'basic building blocks' out of which everything is made, but it also promotes an image of the old materialistic idea of Democritus. In contrast, megascopic quantum phenomena can be seen as reflections of Plato's Forms. As recognized, the world of Plato and his a-spatial and a-temporal Forms are transcendental to our own world of substances, i.e. space and time respectively. It is superordinate to matter.

Since megascopic quantum jumps are not directly measurable, one may wonder whether the claim has any meaning at all. For instance do they proceed at some given time as a microscopic quantum jump or is there any relationship to time at all? An answer might emerge by looking carefully what the attribute "a-temporal" in world of Plato exactly means: A Form does not exist within any time period, rather it provides the formal basis for time. One should perhaps ask whether quantum physics uses a similar time concept as classical/relativistic physics. Here is the view of Lee Smolin on the nature of time [141]: „ More and more, I have the feeling that quantum theory and general relativity are both deeply wrong about the nature of time. It is not enough to combine them. There is a deeper problem, perhaps going back to the beginning of physics." In other words it means going back to Plato and Aristotle. In Plato's Forms we will finally find the true relationship between megascopic quantum jumps and time, i.e. these jumps do not proceed in time, rather time is created by them. At the megascopic level this gives rise to the concept of quantum of time, which otherwise is unreachable at the microscopic level. Microscopic quantum theory yields quanta of energy, momentum, angular momentum etc., but never a quantum of time or of space. Different objects have different energies, momenta etc., but the Universe is one, the time is one, and the space is one. In this sense megascopic quantum jumps are the true clock of the Universe.

Focusing now on the relationship time - events. In Newtonian physics time is foundational, and events that can happen at any moment of time are subordinate. Despite the fact that microscopic quantum physics shares the concept of time with classical physics, events do not play the subordinate role there, but are original as well as time and independent on it. On one hand we have time evolution of the wave function, representing only Aristotelian potentiality, and on the other we have reality created during the measurement process by timeless events - microscopic quantum jumps. And just this separation of the events from time is the source of almost all philosophical problems of contemporary quantum physics. Bell wrote a very poignant parody relevant to this topic [136]: „It would seem that the theory is exclusively concerned with "results of measurement" and has nothing to say about anything else. When the "system" in question is the whole world where does one find the "measurer"? Inside, rather than outside, presumably. What exactly qualifies some subsystems to play this role? Was the world wave function waiting to jump for thousands of millions of years until a single-celled living creature appeared? Or did it have to wait a little longer for some more highly qualified measurer - with a Ph.D.? If the theory is to apply to anything but idealized laboratory operations, are we not obliged to admit that more or less "measurement-like" processes are going on more or less all the time more or less everywhere? Is there ever then a moment when there is no jumping and the Schrödinger equation applies?"

Bell was certainly correct in his view regarding "measurement-like" processes above. Our "megascopic quantum jumps" do correspond to Bell's request, and since the origin of time is rooted in megascopic events, we are able to finally reunite time and events on the quantum level. However, unlike events on the classical level those in the quantum case will be foundational and time subordinate. This idea, in fact, goes back to ancient times. Aristotle expressed this opinion in his Physics, Book I, Part 14, where he defined time as "the number of movement in respect of before and after". Thus it cannot exist without a succession, and it does not exist on its own rather it is relative to the motions of things [142]. Moreover, Aristotle's concept of time is more suitable for quantum theory than Newtonian classical mechanics, when he says that time, in order to exist, requires the presence of a soul capable of "numbering" the movement. This statement complies with the von Neumann - Wigner rule.

One might wonder why the Newtonian and Aristotelian concepts of time are so different. Pondering this issue, one must not forget that the Greek language has two expressions for time: χρόνος, chronos, and καιρός, kairos, while the Latin language only one expression *tempus*. Our way of thinking is to a large extent influenced by our spoken language. The Newtonian time concept is basically pragmatic, but it is unable to explain the time arrow problem (see section 3). If we accept the premise that all processes in the Universe consist only of microscopic and megascopic events and therefore that irreversibility is foundational, then the arrow of time naturally develops from Aristotle's definition of time as "the number of movement in respect of before and after". Emergent time reversibility can only appear in subsystems with no SSB (mostly adiabatic systems), where a one-to-one correspondence between the mechanical and the field states holds. These subsystems can then be described by reversible evolution as formulated by the Schrödinger equation. This contradicts Santilli's claim, see section 3, since irreversibility in the Universe has its origin in irreversible microscopic and megascopic events.

Before ending this section we will focus on the second Bohr complementarity asking: why did only Jordan called for it and not Bohr himself? Actually, Bohr did put forward another type of complementarity between observational conditions of animate and inanimate nature [61]: „In this promising development we have to do with a very important and, according to its character, hardly limited extension of the application of purely physical and chemical ideas to biological problems, and since quantum mechanics appears as a rational generalization of classical physics, the whole approach may be termed mechanistic. The question, however, is in what sense such progress has removed the foundation for the application of so-called finalistic arguments in biology. Here we must realize that the description and comprehension of the closed quantum phenomena exhibit no feature indicating that an organization of atoms is able to adapt itself to the surroundings in the way we witness in the maintenance and evolution of living organisms. Furthermore, it must be stressed that an account, exhaustive in the sense of quantum physics, of all the continually exchanged atoms in the organism not only is infeasible but would obviously require observational conditions incompatible with the display of life."

This can undoubtedly be understood as the third complementarity, which, although Bohr mentions here teleological (finalistic) arguments, has nothing to do with the second one. Life is a mystery, and no hitherto available knowledge, as e.g. the discovery of DNA, and no finalistic arguments are sufficient to explain it. However, Bohr becomes here aware of the teleological origin of Darwinian adaptation, in stark contrast to his contemporary quantum scientists who did not see it at all. He, on the other hand, was convinced that teleological adaptation only appeared in living organisms and he never considered the possibility of its occurrence in inanimate matter, such as in all SSB phenomena. Yet by realizing the distinction between causal and teleological phenomena, Bohr was very close to formulating the second kind of complementarity [61]: „In biological research, references to features of wholeness and purposeful reactions of organisms are used together with the increasingly detailed information on structure and regulatory processes that has resulted in such great progress not least in medicine. We have here to do with a practical approach to a field where the means of expression used for the description of its various aspects refer to mutually exclusive conditions of observation. In this connection, it must be realized that the attitudes termed mechanistic and finalistic are not contradictory points of view, but rather exhibit a complementary relationship which is connected with our position as observers of nature."

Bohr states correctly, that attitudes termed mechanistic (causal) and finalistic (teleological) are not contradictory, but fails, in my opinion, when saying that they are in a complementary relationship. Actually there is no complementarity between them because they do co-exist side by side. Summarizing, the first complementarity represents a causal microscopic relationship between mechanical and field descriptions of elementary entities (the particle-wave dualism), the second complementarity epitomizes the teleological megascopic relationship between the mechanical and the field descriptions of the whole Universe, where these two descriptions deal with the centre of gravity in different ways. This conclusion is a direct consequence of the Goldstone theorem, proved by Goldstone, Salam, and Weinberg, incidentally in the same year when Bohr died. Unfortunately, as already pointed out, a year later the Goldstone theorem was misconstrued by Anderson, see section 10, and as a result there appeared an inconsistent Higgs mechanism, which would never reveal a second complementarity.

## 14. The second quantum floor

The transition from classical to microscopic quantum physics is nicely distinguished if one compares the Rutherford and the Bohr atomic models. In the former an electron needs to accelerate its movement in order to achieve a higher energy level around the nucleus. In the Bohr model there is no need for any movement or acceleration; all electrons occupy certain quantum states and transitions to another level occur by way of quantum jumps.

In addition to the microscopic quantum domain, mentioned above and which we name the first quantum floor, one also has the whole scale of megascopic phenomena, operating on a second quantum floor. Surprisingly we are here in a similar situation compared to a century ago. The microscopic quantum description all of a sudden has become insufficient. Note that we have repeatedly discussed this issue in previous sections, e.g. for transitions between isomers and distorted J-T states in molecules, and in superconductors, and found that microscopic tunnelling, represented by the direct movement of elementary particles, is entirely outside the scope of the given problem. We need an additional concept of megascopic quantum jumps appearing either on the microscopic or the macroscopic level, which is in a similar relation to microscopic tunnelling as the concept of microscopic quantum jumps, in the Bohr model, relates to the classical electronic movement in the Rutherford model.

Concentrating briefly on superconductivity and its megascopic origin, we recapitulate the three main reasons why superconductivity does not have any microscopic explanation, see also sections 9 and 11:

1) Every microscopic theory must result in Born's rule for the probabilities related to the density and the velocity of the superconducting carriers, and these quantities have to be experimentally measurable. But as we know, they are in principle non-measurable.

2) No microscopic theory allows us to avoid the universal concept of Bloch states for the description of superconducting carriers. It means that the carrier mass must take the effective mass of the electrons into account. This, however, is in direct contradiction with measurements of the London moment, where only bare electronic masses are reported.

3) According to the second form of van Fraassen's argument, asymmetry cannot arise ex nihilo. Although we know that the original asymmetry around the superconductor can be present in the form of an external magnetic field, no microscopic theory is able to explain the Meissner effect mechanism, when the superconductor is cooled below the critical temperature and the constant magnetic field cannot bring about any acceleration of the superconducting carriers. It means that a microscopic theory is unable to implement the original asymmetry, and this fact contradicts the mentioned second form of the van Fraassen argument.

Nevertheless, superconductors can be described macroscopically on the quantum level. The concept of macroscopically-occupied quantum states was proposed by London [66]. The macroscopic wave function obtains as

$$\Psi(\mathbf{r}) = |\Psi_0(\mathbf{r})| \exp\left(i\theta(\mathbf{r})\right) = \sqrt{n_s} \exp\left(i\theta(\mathbf{r})\right) \qquad (14.1)$$

with the phase $\theta$ and the amplitude $\Psi_0$ characterizing the density, $n_s$, of the superconducting carriers. From this equation, using the well-known relation between the kinetic and the canonical momentum, one gets the expression for the velocity of superconducting carriers

$$\mathbf{v}_s = \frac{\hbar}{m_s} \nabla . \theta - \frac{e_s}{m_s c} \mathbf{A} \qquad (14.2)$$

and after taking the curl of the velocity

$$\nabla \times \mathbf{v}_s = \frac{-e_s}{m_s c} \mathbf{B} \qquad (14.3)$$

and substituting from (9.1) one finally obtains the phenomenological London equation (9.5). It is interesting to note that from London's macroscopic wave function (14.1) one can derive Josephson's equations [143]. Hence in this formulation one does not need any microscopic concepts of superconductivity, as was already shown by Feynman [144]. Considering the solution of an easier case of a SIS (Superconductor-Insulator-Superconductor) junction, both "superconductor bodies" start to 'feel' each other at a certain distance, e.g. of the order of nanometres, because of the estimated coherence length consistent with their wave functions. Feynman proposed two coupled time dependent equations

$$i\hbar \frac{\partial \Psi_1}{\partial t} = E_1 \Psi_1 + K \Psi_2; \qquad i\hbar \frac{\partial \Psi_2}{\partial t} = E_2 \Psi_2 + K \Psi_1 \qquad (14.4)$$

where $K$ is a phenomenological parameter which describes the properties of the insulating barrier. From the Eqs. (14.1) and (14.4) one simply obtains the first Josephson equation

$$I(t) = I_c \sin\left(\theta_2(t) - \theta_1(t)\right) \tag{14.5}$$

and the second Josephson equation

$$U(t) = \frac{\hbar}{2e} \frac{\partial\left(\theta_2(\mathbf{r}) - \theta_1(\mathbf{r})\right)}{\partial t} \tag{14.6}$$

where $U(t)$ and $I(t)$ are the voltage across and the current through the Josephson junction, $\theta_2 - \theta_1$ is the phase difference across the junction, and $I_c$ is a phenomenological constant, representing the "critical current" of the junction.

One should observe that the ground and the excitation states of the superconductors have a microscopic explanation, as well as all solids etc. Previously we have derived the J-T distorted ground state of superconductors, represented by the Eq. (11.52), and their excited states, Eqs. (11.54,55). However, superconductivity itself, as caused by the transitions between distorted J-T states, has no microscopic explanation. At this point microcosm and megacosm does shake hands.

Whereas Eqs. (11.52,54,55) are of the Bloch type, the transitions between the distorted J-T states, the Fourier mirror of the Bloch functions - the Wannier functions - is more transparent. From the Bloch functions

$$\psi_{\mathbf{k}}(\mathbf{r}) = \exp(i\mathbf{k}\mathbf{r}) u_{\mathbf{k}}(\mathbf{r}) \tag{14.7}$$

and after the application of the Fourier transformation we get the Wannier functions

$$\chi_{\mathbf{R}}(\mathbf{r}) = \frac{1}{\sqrt{N}} \sum_{\mathbf{k}} \psi_{\mathbf{k}}(\mathbf{r}) \exp(-i\mathbf{k}\mathbf{R}) \tag{14.8}$$

where $N$ is the number of primitive cells in the crystal and $\mathbf{R}$ is any lattice vector. There is one Wannier function for each Bravais lattice vector. The sum on $\mathbf{k}$ includes all the values of $\mathbf{k}$ in the Brillouin zone. The Wannier functions fulfil the relation

$$\chi_{\mathbf{R}}(\mathbf{r}) = \chi_{\mathbf{R}+\mathbf{R}'}(\mathbf{r} + \mathbf{R}') \tag{14.9}$$

Using Wannier functions instead of the Bloch ones, the electronic wave function can be then rewritten as

$$\Psi(\mathbf{r}) = \frac{1}{\sqrt{N!}} \left\| \prod_I^N \chi_I(\mathbf{r}) \right\| = \frac{1}{\sqrt{N!}} \left\| \prod_{i\sigma}^N \chi_{i\sigma}(\mathbf{r}) \right\| \tag{14.10}$$

In this notation the spin orbitals $I$ (or orbitals $i$ and spins $\sigma$) are related to the lattice vectors $\mathbf{R}$.

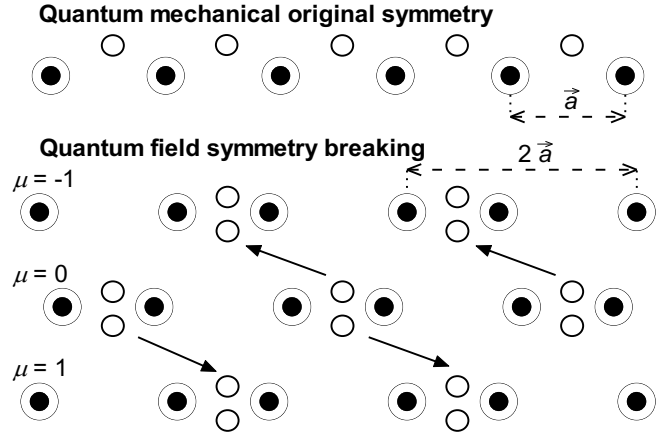

**FIG 1**: Megascopic quantum jumps between the quantum mechanical and quantum field states of the Universe, leading to the megascopic tunnelling of superconducting carriers.

● nuclei with electronic core
○ valence electrons in symmetric quantum mechanical and asymmetric quantum field positions
$\mu$ einselection (environment-induced superselection) factor
$\vec{a}$ original lattice constant before the symmetry breaking

Fig. 1 shows two microscopic solutions for the superconductor ground state: the mechanical one based on the Monkhorst-Cafiero-Adamowitz approach with no broken symmetry and the same lattice constant **a** as in the conducting state above the critical temperature; and the field one with J-T broken symmetry and the lattice constant 2**a**. As said in previous sections, one needs a new rule allowing direct transitions from a mechanical state reflecting the whole Universe, into the field states that descends from the fragmentation of the Universe – and vice versa. During these megascopic quantum jumps the electron pairs, valence orbitals occupied by two electrons with opposite spins (no Cooper pairs) have the chance either to be relocated in the right or left directions, or to return into their original position, depending on the external magnetic field that determines the einselection factor $\mu$. In the simplest one-dimensional case there are only three possible discrete values of $\mu$, namely -1, 0 and 1.

In passing we note that only the idea of megascopic quantum jumps is able to overcome the three abovementioned main problems regarding any attempt to explain superconductivity microscopically. First, immeasurable megascopic quantum jumps imply non-measurability of the density and the velocity of superconducting carriers. Second, the "teleportation" of superconducting carriers, a new kind of non-local quantum phenomenon, guarantees the bare electronic mass in the London moment measurements. Third, only the holistic incorporation of the environment by way of megascopic einselection is capable of responding to the asymmetry in the form of an external magnetic field. Henceforth there is also no contradiction to the second form of van Fraassen's argument, and the Meissner effect can thus be explained.

Returning to the field solutions of Fig. 1, depicting distorted nuclear positions with the adequate electronic redistribution one realizes that the electrons form some kind of "chemical bonds" occupied by two electrons with opposite spins. As it was explained in the previous section, megascopic quantum jumps do not occur in time, but time is created by them. Denoting the quantum time scale as $\tau$, the time $t = n\tau$ will be understood as the interval of $n$ megascopic events, i.e. $n$ transitions from the field state to the complementary mechanical state and back. After two megascopic events, considering only a one-dimensional chain, the valence orbitals occupied by the two electrons will be relocated to new positions, which are in the following relation to the original positions

$$\left(\chi_{i\uparrow}(\mathbf{r})\chi_{i\downarrow}(\mathbf{r})\right)_{2\tau} = \left(\chi_{i\uparrow}(\mathbf{r}-2\mu\mathbf{a})\chi_{i\downarrow}(\mathbf{r}-2\mu\mathbf{a})\right)_0 \exp\left(i\theta(\mathbf{r})\right); \qquad \mu \in \{-1,0,1\} \qquad (14.11)$$

Here the phase $\theta$ appears at the macroscopic level, reflecting the "teleportation" of superconducting carriers, i.e. the valence orbitals occupied by two electrons. In our megascopic interpretation, unlike the BCS theory which associates this phase with the motion of carriers in Bloch's $\mathbf{k}$ space, the phase is associated with the macroscopic $\mathbf{l}$ space, orthogonal to the Bloch $\mathbf{k}$ space:

$$\theta(\mathbf{r}) = \mathbf{l}.\mathbf{r}; \qquad \mathbf{l} = \frac{m_s \mu \mathbf{a}}{\hbar \tau}; \qquad m_s = 2m_e \qquad (14.12)$$

This equation gives us a relation between the macroscopic momentum, and the carrier mass and the velocity. The carrier mass corresponds to the double valued electronic mass, as well as the carrier charge to the double valued electronic charge. The superconducting carriers appears without the mentioning of Cooper pairs; they are identical with the smallest relocated entities - double occupied valence orbitals. Using Eq. (14.9) one gets

$$\chi_{i\sigma}\left(\mathbf{r}-2\mu\mathbf{a}\right) = \chi_{j\sigma}(\mathbf{r}); \qquad \mathbf{R}_j = \mathbf{R}_i + 2\mu\mathbf{a} \qquad (14.13)$$

and after two megascopic events Eq. (14.10) transforms into the new expression:

$$\Psi(\mathbf{r})_{2\tau} = \frac{1}{\sqrt{N!}}\left\|\prod_{j\sigma}^{N}\chi_{j\sigma}(\mathbf{r})\right\|\exp\left(i\theta(\mathbf{r})\right) = \Psi(\mathbf{r})_0 \exp\left(i\theta(\mathbf{r})\right) \qquad (14.14)$$

Finally generalizing Eq. (14.13) for an arbitrary time $t = 2n\tau$ and to the three-dimensional case, where instead of discrete values of the einselection factor we have the real interval $\mu \in \langle -1,1 \rangle$, one obtains the relations

$$\chi_{i\sigma}\left(\mathbf{r}-2n\mu\mathbf{a}\right) = \chi_{j\sigma}(\mathbf{r}); \qquad \mathbf{R}_j = \mathbf{R}_i + 2n\mu\mathbf{a}; \qquad t = 2n\tau; \qquad \mu \in \langle -1,1 \rangle \qquad (14.15)$$

Note that Eq. (14.14) can be generalized for arbitrary times, meaning, that in this way we have arrived at London's macroscopic wave function (14.1).

There arises an interesting question, whether the quantum of time from the Eq. (14.12) can be measured. Since we do not know the value of the einselection factor $\mu$, an extremely large magnetic field should be necessary in order to achieve the maximal value of $\mu$, i.e. -1 or 1. Unfortunately there is a limit for the magnitude of applied external magnetic fields, because the induced field destroys the superconducting state above the critical value of the current. This critical value is certainly much smaller than the one given by the maximal value of the einselection factor $\mu$.

We should perhaps salute Peirce's prophetic statement, see section 12, regarding classical mechanics and symmetry breakings, namely that the three laws of motion draw no dynamical distinction between right-handed and left-handed screws, and that there are physical phenomena, standing entirely beyond all corpuscular philosophy, yet absolutely inexplicable by mechanical action. Though his claim dates back to the pre-quantum era, it is surprisingly still valid for microscopic quantum physics, based on Bohr's correspondence principle. Therefore contemporary quantum physics has never succeeded to explain the Meissner effect, and if not the Meissner effect, it also concerns superconductivity. Any attempt to find their microscopic solutions inevitably implicates the breakdown of Bohr's correspondence principle, see section 9. Thus the BCS theory and all theories, based on the microscopic dynamics of the superconducting carriers, whatever these carries may be, whether unpaired or Cooper paired electrons, polarons, bipolarons etc., are still inadequate. Only a holistic megascopic quantum theory is able to overthrow the traditional corpuscular philosophy, and to explain the Meissner effect and superconductivity.

Most physicists agrees, that superfluidity should have an analogous explanation as compared to superconductivity. However if superconductivity has no microscopic explanation, superfluidity does not, too. Superfluidity shares with superconductivity the macroscopic wave function (14.1). In a similar way, cf. Eq. (14.2), follows from (14.1) in the case of superconductors, that we can write

$$\theta(\mathbf{r}) = \frac{M_s \mathbf{v} \cdot \mathbf{r}}{\hbar} \qquad (14.16)$$

where the magnetic field $\mathbf{A}$ is not present as in Eq. (14.2) and the superconducting carrier mass $m_s$ is replaced with the superfluid particle mass $M_s$. This leads to the quantization of the superfluid hydrodynamic circulation $\kappa$:

$$\kappa = \int \mathbf{v} \cdot d\mathbf{r} = \frac{2\pi n\hbar}{M_s} \qquad (14.17)$$

By analogy physicists attempt to see the common feature of both phenomena employing some type of Bose-Einstein condensation, involving atoms or pairs of atoms or pairs of electrons. In this way the interpretations of superfluidity in helium-4 and helium-3 atoms differ: The first one is treated as a direct Bose-Einstein condensation of bosonic particles, whereas the latter one deals with the formation of bosons only by pairing of two fermionic atoms similar to the BCS mechanism of electronic Cooper pairing. This sounds a bit odd - why should the different isotopic number change the whole mechanism of superfluidity?

We may actually impose a stronger requirement for superconductivity and superfluidity: Let they have, not only a common explanation, but let they also be two manifestations of the same phenomenon. Superfluids consist of two components - normal and superfluid. Let us suppose that the latter can form microscopic superconducting grains. Consider a model of a caterpillar truck. The superconducting carriers - "teleported" valence orbitals are temporarily "glued" on the surface of tube or capillary, exactly like the caterpillar on the ground. The nuclei then act like the truck. It is all about the relative relocation of the nuclei with the electronic core and the superconducting carriers: In superconductors the nuclei with the electronic core are macroscopically in rest and the superconducting carriers are relocated. In superfluids the opposite is true.

Let us now discuss other megascopic phenomena. Ferromagnetism can be understood as a macroscopic analogy of microscopic isomerism. Both phenomena have a microscopic quantum explanation. But the tunnelling between different isomers and between ferromagnets with different spin orientation is of megascopic origin and has no microscopic clarification. The latter case is known as the Einstein-de Haas effect [118] or Richardson effect [145]. Richardson was the first to conclude, from the principle of the conservation of angular momentum that the change in the magnetic moment of a free body causes this body to rotate. At that time no electron spin was known, so Richardson derived a prequantum relation between the magnetization $\mathbf{M}_o$ and total orbital angular momentum $\mathbf{J}_o$ of the motion

$$\mathbf{M}_o = \frac{e}{2m}\mathbf{J}_o \qquad (14.18)$$

where $e$ and $m$ stand for electron charge and mass. But as soon as the electron spin was discovered in 1925, it was clear that its contribution to the Einstein-de Haas effect in ferromagnets significantly predominates over the contribution of the angular momentum. So instead of Eq. (14.18) one has

$$\mathbf{M}_s = \frac{e}{m}\mathbf{J}_s \qquad (14.19)$$

with a gyromagnetic factor equal to 2. For deeper insight into the exact value of the gyromagnetic factor, see e.g. Maruani's article [146]. The similarity between Eqs. (14.18) and (14.19) demonstrates that the spin momentum should be of the same nature as the angular momentum of rotating bodies as conceived in classical mechanics. Quoting Maruani regarding the explanation he says in his work [147]: „The Dirac equation, which was derived by combining the relativistic invariance condition with the quantum probability principle, showed its fecundity by explaining the half-integer spin of fermions and by predicting antiparticles. In previous papers, we conjectured that the spinning motion of the electron was that of a massless charge moving at light velocity, this internal motion being responsible for the electron rest mass involved in external motions and interactions."

Proceeding further to non-equilibrium phenomena with the focus on chemical reactions. Quantum chemists usually and successfully formulate the problem and solve them with the help of microscopic quantum physical equations often without further thinking. They calculate potential energy surfaces (PES) of interacting molecules finding energy landscapes including energy minima and maxima, construing their results to interpret the investigated chemical reactions. This concept, however, does not answer the question, what a chemical reactions really amounts to. The critical point comes when PES's cross. This metaphysical concept is a fundamental consequence of the application of the B-O approximation, exactly as the misunderstood J-T effect. True quantum mechanical solutions, treating electrons and nuclei on the same footing, never lead to any symmetry breaking in J-T systems, as well as to any chemical reactions, since they do not recognize the individuality of molecules and their constituents. Only the true field solution based on the Goldstone theorem implies symmetry broken J-T states and recognizes the individual reacting molecules. If we have a reaction of the type A + B $\longleftrightarrow$ C (in both directions), then the left side contains $n_A + n_B$ - 12 phonons, 6 rotons and 6 translons, and the right side $n_C$ phonons, 3 rotons and 3 translons. The number of Goldstone bosons during the reaction is preserved, and the chemical reaction happens only if 3 rotons and 3 translons are annihilated and 6 phonons are created, or vice versa. This is a megascopic irreversible process - a quantum jump from the fragmented Universe, containing individual molecules A and B, into the whole Universe, where the molecular individuality is lost, and then again in the fragmented Universe with the molecule C of the same energy as that of A + B in the same configuration at which the reaction proceeds, or vice versa.

We know well the irreversible nature of chemical reactions, and Santilli was a half century ago absolutely right with his "no reduction theorem", see section 2. It means, that contemporary quantum physics with its reversible equations can never justify this irreversibility, and therefore is unable to explain a single chemical reaction. It is able to explain only the physical pre- and postprocesses occurring during reactions, but not the chemical reactions themselves. Santilli still believes in the microscopic explanation of chemical reactions, and has proposed an irreversible mathematical framework for physical equations. We have shown that chemical reactions belong to the group of megascopic phenomena, where irreversibility descends from the megascopic quantum jumps, and thus no irreversible equations are necessary.

If physical entropy is assigned to the multitude of equivalent isomeric molecular ground states there appears another paradox of residual entropy arising from chemistry. One of the first examples of residual entropy was already pointed out by Pauling to describe water-ice [148]. In water, each oxygen atom is bonded to two hydrogen atoms. For a large number of water molecules in this configuration, the hydrogen atoms have a large number of possible configurations that meet the 2-in 2-out rule, since each oxygen atom must have two 'near' (or 'in') hydrogen atoms, and two far (or 'out') hydrogen atoms. This freedom exists down to absolute zero. The existence of these multiple configurations amounts to randomness, or in other words, entropy. Thus systems that can take multiple configurations at or near absolute zero are said to have residual entropy. Of course, residual entropy is in strong contradiction with the third law of thermodynamics that states: the entropy of a perfect crystal of any pure substance approaches zero as the temperature approaches absolute zero. Here one might suggest a megascopic solution to this paradox. Realizing, that possible tunnelling between different isomeric configurations of ice is of megascopic origin, i.e. microscopically inaccessible, then it is convenient to introduce a separate "chemical" entropy for megascopic processes, so that the physical third law of thermodynamics remains still valid.

A second example of the illusive violation of the third law of thermodynamics is the liquid–glass transition with actually a negative residual entropy. There are several differences between liquid-crystal and liquid–glass transitions. In the first case the arrangement of atoms and molecules differs from that of a liquid, but in the second case this arrangement is indistinguishable from it. In the first case there is a first-order phase transition involving discontinuities in thermodynamic and dynamic properties such as volume, energy (latent heat), and viscosity. This is a transition between states of thermodynamic equilibrium at a single, well-defined temperature and is time reversible. In the second case there is no first-order phase transition, no latent heat and no well-defined temperature: the slower the cooling, the lower the transition temperature. This is a non-equilibrium and irreversible process forming a quenched disorder state, kinetically locked. Its entropy, density and structure depend on the thermal history. By extrapolating the heat capacity of the supercooled liquid below its glass transition temperature, it is possible to calculate the Kauzmann temperature [149] at which the difference in entropy between the liquid and solid phase becomes zero. Cooling below this temperature leads to the Kauzmann paradox: A supercooled liquid displays a lower entropy than the crystal phase. This paradox was not resolved till now, although there are several proposals how to deal with it. Kauzmann himself postulated that all supercooled liquids must crystallize before the Kauzmann temperature is reached. But the scientific community has no unique opinion.

The most striking enigma of the liquid–glass transition is this: If we use classical physics for its calculation, i.e. classical thermodynamics and thermodynamics of irreversible processes, we end up with the Kauzmann paradox and the negative residual entropy at absolute zero. But on the other hand if we perform the computation of energy landscapes based on the concept of broken ergodicity in the configurational phase space with an entropy loss, a glass does not appear to have any residual entropy at absolute zero. Anderson was inspired by this controversy so utterly that he wrote [150]: „The deepest and most interesting unsolved problem in solid state theory is probably the theory of the nature of glass and the glass transition."

Different results from macroscopic and microscopic calculations imply that the physical microscopic description in this case does not reveal all facts about the macroscopic world. Simply the macroscopic and microscopic entropies are not the same. The irreversible and the non-equilibrium processes behind the glass transition are chemical reactions, giving rise to the creation of anisotropic covalent bonds in polymorphic forms of glass. Analogous to the case of Pauling's ice, we need here a separation of the physical microscopic entropy from the megascopic chemical one, in order not to violate the third law of thermodynamics.

Another paradox arises from the query: do chemical reactions exist on the macroscopic level? A nice example of the type $C \rightarrow A + B$ is brittle fracture. Unlike ductile fracture this effect has no microscopic quantum explanation and is described phenomenologically only by classical physics. Once a crack starts to propagate in a brittle solid its velocity may quickly reach unbelievable several thousands of m/s [151], whereas in ductile fracture there is only a slow propagation of the crack. Further differences are [152]: Ductile material fractures after plastic deformation and its surface obtained at the fracture is dull or fibrous in appearance. A ductile crack will usually not propagate unless an increased stress is applied and generally cease propagating when loading is removed. On the other hand, brittle fractures are characterised as having little or no plastic deformation prior to failure and the fracture surface of a brittle failure is usually reasonably smooth. The cracks that propagate in a brittle material will continue to grow once initiated. Although brittle fracture has no microscopic explanation, there are nevertheless two categories of cracks from the microscopic perspective [152], (i) a crack that passes through the grains within the material undergoing transgranular fracture, (ii) a crack that propagates along the grain boundaries termed an intergranular fracture.

The most interesting fracture feature process is a ductile to brittle transition. At low temperatures some materials that would be ductile at room temperature become brittle. It means that at higher temperatures the fracture is of microscopic origin, but under the critical temperature it is of megascopic origin. Compare for instance the relation as microscopic normal conductivity under the critical temperature becomes megascopic superconductivity. This similarity is certainly not accidental, since materials that usually fracture in a brittle manner are glasses, ceramics, and some polymers and metals. This also explains why good superconductors are usually formed with brittle ceramics on the verge of rupture.

**Table 1**: Megascopic phenomena

| megascopic phenomena | equilibrium processes | | non-equilibrium processes |
|---|---|---|---|
| | adiabatic systems | non-adiabatic systems | |
| microscopic level | isomeric transitions | Jahn-Teller effect | chemical reactions |
| macroscopic level | Einstein-de Haas effect | superconductivity and superfluidity | brittle fracture |

In Table 1 we have summarized the mentioned teleological megascopic phenomena, appearing on both the microscopic and the macroscopic level. This can be considered the fulfilment of Norton's request to overcome the restricted domain of contemporary physics that concentrates solely on causal processes [52]: „In the physical sciences, an important reason for choosing a restricted domain is to fence off processes that are acausal. A second reason to which I gave less attention, is that different domains manifest different sorts of causes... I expect the case of final causes to be prevalent in chemistry and non-physical sciences - that the restriction to different domains will divide different types of causation, as opposed to fencing off acausal processes."

This is also a fulfilment of Jordan's request, see section 13, to introduce the second complementarity in order to explain classicality. Megascopic chemical processes are responsible for the quantum decoherence of the wave function, and the classical appearance of the silver grain on the photographic plate follows accordingly. From the same reason Schrödinger's cat can never be in some state of superposition. Either the irreversible megascopic event, representing the chemical reaction between the poison and the cat happens or it does not, and nothing in between.

To conclude, this might also be an answer to von Weizsäcker's problem, cited in section 6, where he correctly recognized irreversibility as a necessary condition for classicality, but left open the question of sufficiency. We know now, that this condition is not fulfilled in the framework of the microscopic Copenhagen interpretation, but within the conception of megascopic irreversible processes it is fully sufficient.

## 15. Matter-mind dualism

We start with a short recapitulation of the megascopic mirrors of the microscopic quantum concepts and associated rules as they were introduced in section 13.

1. The first complementarity versus the second: while microscopic quantum theory is based on Bohr's complementarity between the mechanical (particle) and field (wave) picture of elementary entities, megascopic quantum theory is founded on two megascopic complementary accounts of the whole Universe, namely the mechanical description representing its total wholeness, and the field formulation representing its total fragmentation. The implementation of a second complementarity was for first time requested by Jordan.

2. Causality versus teleology: microscopic quantum physics inherits causality (or upward causality) from classical physics due to Bohr's correspondence principle. Its megascopic mirror is teleology (downward causation or the final cause), mentioned already for first time by Maxwell, see section 7, as it proceeds from a principle that is extra physical, but not extra natural, and is acting without exerting any force or spending any energy.

3. Projections versus injections: transitions between different states in microscopic quantum physics are expressed as vector projections in Hilbert space. The megascopic mirror of these projections is represented by injections, the latter responsible for the transitions from the fragmented Universe into its wholeness and vice versa. It was proposed for the first time by Bohm.

4. The Born rule versus the einselection rule: The probability of transitions based on microscopic projections is calculated by means of Born's law. Its megascopic mirror is the einselection rule, which takes into account the influence of all environments. The concept of einselection was originally coined by Zurek in his quantum decoherence program with the aim to replace the Born rule. As we have shown in the section 7, it cannot be replaced, and in section 13 we also revealed the true meaning of einselection as the megascopic counterpart of the microscopic Born rule. It means that the latter remains to be tied to the microscopic projections, while the einselection principle determines the probabilities of the megascopic injection transitions.

The abovementioned four concepts and rules form, in the microscopic world, the basic axioms of the Copenhagen interpretation of quantum physics. But one axiom is still missing, i.e. the von Neumann - Wigner rule. At this point we have reached a very important but most commanding point: What is its megascopic mirror? Before we try to respond to this challenge, one must first deliberate more deeply on the problem of the matter-mind dualism that is necessary for the understanding of the von Neumann - Wigner rule. The matter-mind dualism had a most significant historical development after the Cartesian split. Heisenberg paid special attention to this problem in the Chapter 5 entitled "Die Entwicklung der philosophischen Ideen seit Descartes im Vergleich zu der neuen Lage in der Quantentheorie" in his book "Physik und Philosophie". Quoting directly the original version in German [153]: „Der Ausgangspunkt der cartesianischen Philosophie ist völlig verschieden von dem der antiken griechischen Philosophen. Hier wird nicht mit einem Grundprinzip oder einem Grundstoff begonnen, sondern mit dem Versuch, ein grundlegendes, unbestreitbares Wissen zu erwerben. Descartes erkennt, daß unser Wissen über unser eigenes Denken sicherer ist als unser Wissen über die äußere Welt. Aber schon seine Ausgangsposition mit dem Dreieck: Gott, Welt und Ich vereinfacht die Grundlage für das weitere Philosophieren in einer gefährlichen Weise. Die Spaltung zwischen Materie und Geist oder zwischen Körper und Seele, die mit Platons Philosophie begonnen hatte, ist jetzt vollständig. Gott ist sowohl vom Ich als auch von der Welt getrennt. Gott wird tatsächlich so hoch über die Welt und die Menschen erhoben, daß er schließlich in der Philosophie des Descartes nur als ein gemeinsamer Bezugspunkt erscheint, der die Beziehung zwischen dem Ich und der Welt herstellt."

The Cartesian split affected the progress of the whole of classical physics. Heisenberg writes: „Andererseits war diese Spaltung in der Naturwissenschaft für einige Jahrhunderte außerordentlich erfolgreich. Die Newtonsche Mechanik und alle anderen Teile der klassischen Physik, die nach ihrem Vorbild aufgebaut waren, beruhten auf der Annahme, daß man die Welt beschreiben kann, ohne über Gott oder uns selbst zu sprechen. Diese Möglichkeit galt beinahe als eine notwendige Voraussetzung für alle Naturwissenschaft."

Quantum mechanics, however, is entirely different from classical physics: We have first of all to overcome the Cartesian split. Heisenberg comments on its consequences on the ways many physicists think, and that this prevent them to fully understand the Copenhagen interpretation: „Wenn man an die großen Schwierigkeiten denkt, die selbst so bedeutende Naturwissenschaftler wie Einstein bei dem Verständnis und der Anerkennung der Kopenhagener Deutung der Quantentheorie hatten, so kann man die Wurzeln dieser Schwierigkeit bis zur cartesianischen Spaltung verfolgen. Diese Spaltung ist in den drei Jahrhunderten, die auf Descartes gefolgt sind, sehr tief in das menschliche Denken eingedrungen, und es wird noch lange Zeit dauern, bis sie durch eine wirklich neue Auffassung vom Problem der Wirklichkeit verdrängt ist."

For the Copenhagen (and the von Neumann-Wigner) interpretation the conscious mind has a special relevance, but the Cartesian split also imparts consequences for the understanding of the unconscious mind. In Whyte's 1960 book "The Unconscious Before Freud" dealing with the history of ideas about the unconscious mind, he defines the words "conscious" and "unconscious" [154]: „Conscious will be used to mean directly present in awareness (or "immediately known to awareness"). This adjective will be applied only to discrete aspects or brief phases of mental processes. No distinction will be made between "conscious" and "aware". Except for a few doubtful extreme cases (e.g., processes of logical or mathematical deduction) there appear to be no causally self-contained processes of which all aspects directly enter awareness. "Conscious" is a subjective term without, as yet, (i) interpretation in terms of physiological structure, or (ii) explanation of its function."

And further: „Unconscious, in the term "unconscious mental processes" will be used to mean all mental processes except those discrete aspects or brief phases which enter awareness as they occur. Thus "unconscious mental processes" (or the "unconscious", for short) is here used as a comprehensive term including not only the "subconscious" and "preconscious", but all mental factors and processes of which we are not immediately aware, whatever they be: organic or personal tendencies or needs, memories, processes of mimicry, emotions, motives, intentions, policies, beliefs, assumptions, thoughts, or dishonesties."

Whyte describes Descartes' philosophical motivation: „Descartes recognized that the "union of soul with body" is intuitively known, and that their pervasive connections are a matter of everyday experience. However, in his doctrine these connections are not ones of true causation, but an illusion resulting from some kind of parallelism between the two modes of being. It is therefore, in Descartes' view, abortive to study these connections in detail."

The basic comparison of different attitudes towards the unconscious mind from the positions of materialism, idealism and the Cartesian school follows: „During the late seventeenth century three main attitudes dominated European philosophical thought, corresponding to three interpretations of the nature of existence. Materialism treated physical bodies and their motions as the primary reality; idealism took it to be spirit or mind; while Cartesian dualism postulated two independent realms: the mental res cogitans and the material res extensa. For the first two schools there was no difficulty in recognizing unconscious mentality, though under other names... But to the third, Cartesian, school the admission of the existence of unconscious mental processes presented an acute philosophical challenge, for it demanded the discarding of the original conception of the dualism as one of two independent realms, matter in motion and mind necessarily aware. For those who were loyal to Descartes, all that was not conscious in man was material and physiological, and therefore not mental."

Whyte continues with the post-Cartesian chronology of the conceptual development of the unconscious mind up to Freud: „As often happens in the history of thought, an idea may be fashionable and even transform society before it is properly understood... The unconscious mind, in the post-Cartesian sense, was discovered around 1700; it is now transforming Western thought; who would dare to claim that its basic laws are yet understood?... I have given sufficient evidence to show that the general conception of unconscious mental processes was conceivable (in post-Cartesian Europe) around 1700, topical around 1800, and fashionable around 1870-1880. ...when Freud was forty-two he was unaware that at least fifty writers (probably many more) had been developing similar assumptions for over two hundred years. Finally in 1925, at the age of sixty-nine, Freud wrote: "The overwhelming majority of philosophers regard as mental only the phenomena of consciousness. For them the world of consciousness coincides with the sphere of what is mental." This curious mistake shows how narrow his reading had been, and how wrong a conception he must then have had of his own originality. The inference is that not only Freud but most of us are largely unaware of what has made us what we are and led us to think as we do, and it is sometimes as well that we should be ignorant."

From the above quote it is apparent that Freud might have been overemphasized at the expense of other researchers. His famous psychoanalysis is claimed to be a method for the treatment for mental-health disorders, but does actually in many cases also become misinterpreted disrupting individuals, families and societies.

Unfortunately Whyte's book ends up with Freud leaving out Jung, who in his later life opposed Freud. His primary disagreement with Freud stemmed from their differing conceptions of the unconscious, but also Freud's anti-religious attitudes mattered. Whereas Freud conceived the unconscious solely as a repository of repressed emotions and desires, Jung called this limited model "personal unconscious", and proposed the existence of a second, far deeper form of the unconscious underlying the personal one. In his work "The Archetypes and the Collective Unconscious" Jung explains the difference between personal and collective unconscious [155]: „A more or less superficial layer of the unconscious is undoubtedly personal. I call it the "personal unconscious". But this personal layer rests upon a deeper layer, which does not derive from personal experience and is not a personal acquisition but is inborn. This deeper layer I call the "collective unconscious". I have chosen the term "collective" because this part of the unconscious is not individual but universal; in contrast to the personal psyche, it has contents and modes of behaviour that are more or less the same everywhere and in all individuals."

The final definition of the collective unconscious acquiesces from the same work [155]: „My thesis then, is as follows: in addition to our immediate consciousness, which is of a thoroughly personal nature and which we believe to be the only empirical psyche (even if we tack on the personal unconscious as an appendix), there exists a second psychic system of a collective, universal, and impersonal nature which is identical in all individuals. This collective unconscious does not develop individually but is inherited. It consists of pre-existent forms, the archetypes, which can only become conscious secondarily and which give definite form to certain psychic contents."

On one hand the collective unconscious dictates how the structure of the psyche autonomously organizes experience, and on the other hand the collective unconscious collects and organizes those personal experiences in a similar way for all members of a particular species. This finding, due to Jung, appears particularly important for us as physicists. In addition to the Schrödinger cat paradox, resolved by the introduction of megascopic quantum jumps, the Copenhagen interpretation suffers from yet another conundrum - the paradox of Wigner's friend, see section 6. The Copenhagen view that the quantum state represents available knowledge to a community of communicating observers was set *ad hoc* and does not follow from any of the basic axioms of the Copenhagen interpretation. We live today in the time of Internet and mobile phones, and we can be proud of how ingeniously we are all connected. But this is a special type of connection mediated by the material world. If we accept that we are parts of the material world and communicate via this world, then Wigner's conclusion that each conscious being is able to collapse one single objective quantum state, regardless of whether the information is actually physically shared, is correct. Therefore Wigner looked for a way out of the crisis with his "desire for a less solipsistic theory". Only if we accept Jung's concept, we will understand that our conscious minds are all primarily connected in this Jung-like collective unconscious entirely outside the external material world, and this explains and provides the legitimacy of the shared knowledge of a community of communicating observers.

Incidentally sharing the knowledge of a community of communicating observers is not just one single form of sharing that descends from the collective unconscious. The communication and sharing of our discoveries is a similar case. Whyte presents a motto due to Goethe in his book [154]: „No one can take from us the joy of the first becoming aware of something, the so-called discovery. But if we also demand the honour, it can be utterly spoiled for us, for we are usually not the first. What does discovery mean, and who can say that he has discovered this or that? After all it's pure idiocy to brag about priority; for it's simply unconscious conceit, not to admit frankly that one is a plagiarist."

However, many of Jung's works relates to alchemy, though his interests were focused only on its arcane aspects. According to Linden [156], esoteric aspects of alchemy prevailed among psychologists, spiritual and new age communities, hermetic philosophers, and historians of esotericism, while exoteric aspects occurred among historians of the physical sciences, who did examine subjects in terms of protochemistry, medicine, and charlatanism. In this sense Jung's ideas and inspirations are full of occultism, gnosticism and spiritism. On the other hand, if we extract the scientific truth from the Jung's considerations, we immediately recognize that Jung was correctly aware of the collective unconscious as a mind component of alchemy, after the Cartesian split, whereas classical chemistry became a matter component. Wikipedia, in the year 2018, speaks about the origin of chemistry in this way: „Chemistry was preceded by its protoscience, alchemy, which is an intuitive but non-scientific approach to understanding the constituents of matter and their interactions... Chemistry as a body of knowledge distinct from alchemy began to emerge when a clear differentiation was made between them by Robert Boyle... Chemistry is considered to have become an established science with the work of Antoine Lavoisier..."

The comments above raises one of the most important questions of this article: is chemistry, as we know it today, the only true matter branch of alchemy after the Cartesian split? Looking first at the customary definition of chemistry, see e.g. Wikipedia, 2018: „Chemistry is the scientific discipline involved with compounds composed of atoms, i.e. elements, and molecules, i.e. combinations of atoms: their composition, structure, properties, behaviour and the changes they undergo during a reaction with other compounds. Chemistry addresses topics such as how atoms and molecules interact via chemical bonds to form new chemical compounds." By tradition, science dealing with material objects from elementary particles up to atoms is called physics, and from atoms up to molecules it is chemistry, however, from molecules up to the macroscopic level of fluids and solids it is shared with (condensed matter) physics again. At the quantum level the first and third parts essentially form quantum physics, and the second part mostly quantum chemistry. In this way quantum chemistry is a result of an embedding of post-Cartesian classical chemistry ideas into the pre-Cartesian quantum physics.

As Boyle and Lavoisier were fathers of classical chemistry, Löwdin is often regarded as father of quantum chemistry. A recent biography of Per-Olov Löwdin, see Brändas [157], describes his significant achievements for the development of the new sub-discipline of quantum chemistry. Assessing the new field and its specific goals it was stated: „He [Löwdin] referred to a so-called pure theory in a restrictive sense, 'if it derives for instance chemical data from the knowledge of only the physical values of the electronic mass and charge, Planck's constant, the atomic numbers and the form of the Schrödinger Equation, which itself represents the quintessence of a great deal of physical experience'…". Löwdin was of course well aware of the impossibility to derive the Coulomb Hamiltonian from first principles as well as of other limitations and possibilities offered by a structural approach. Even if the concepts of quantum chemistry is very expedient and practical, it does not, as already stated, give any answers to the Sutcliffe-Woolley clamped-nuclei paradox, see section 2, and to the Santilli's "no reduction theorem" paradox, see section 3. Authentic pre-Cartesian quantum chemistry differs from the pre-/post-Cartesian mix, i.e. standard quantum chemistry. The former deals with teleological megascopic phenomena, and in addition to chemistry and chemical reactions, it also comprises such phenomena that have no quantum physical explanation, such as the Einstein-de Haas effect and the brittle fracture phenomenon and those with false microscopic interpretation, like superconductivity and superfluidity.

Standard quantum chemistry has encouraged the idea that after centuries of progress physics and chemistry can finally 'sit under one roof' in the sense that these two disciplines of natural sciences refer to the same microscopic quantum laws. In contrast, however, authentic pre-Cartesian quantum chemistry forms a megascopic complement to microscopic quantum physics. Metaphorically, one might aver: what elementary particles and the conscious mind mean for pre-Cartesian quantum physics, is analogous to what the whole Universe and the collective unconscious mean for pre-Cartesian quantum chemistry. In the material world, elementary particles represent the smallest one and the Universe the Greatest One. There is a parallelism in the mind world: Observers' conscious minds represent the smallest one and the collective unconscious the Greatest One.

In microscopic quantum physics the von Neumann - Wigner rule binds together elementary particles with the conscious mind, and the megascopic rule binds together the whole Universe with the collective unconscious. An observer's activity, i.e. a measurement, causes microscopic quantum jumps and hence a discontinuity of the deterministic Schrödinger equation. All events in our world must have a reason. Therefore it is reasonable to postulate a similar megascopic analogy: the existence of megascopic "measurements" as an activity of the collective unconscious must cause megascopic quantum jumps. Due to Jung's celebrated formulation of the concept of the collective unconscious, we might venture to call the megascopic mirror of the von Neumann - Wigner rule as the Jung rule. Having finally found the last requested megascopic mirror, the Jung rule, we can summarize all mirrors in Table 2.

**Table 2**: Microscopic-megascopic mirrors

| First quantum floor: microscopic quantum physics | Second quantum floor: megascopic quantum chemistry |
|---|---|
| first complementarity | second complementarity |
| causality | teleology |
| projections | injections |
| Born rule | einselection rule |
| von Neumann - Wigner rule | Jung rule |

From our previous thoughts about the free will, formulated in section 8, two additional mirrors can be revealed, namely the reflections of microscopic quantum physical laws in the collective unconscious and the reflections of megascopic quantum chemical laws in the observer's conscious mind constituting a form of free will decisions. They will be called the crosswise-law mirrors. The first mirror, mentioned above, is supported by physicists such as Poincaré, Penrose and Compton, i.e. Poincaré's unconscious disorder born on chance, Penrose's largely unconscious 'putting-up', first phase of the two-stage model, and Compton's relationship between free will and quantum indeterminism and the randomness on the first phase of the two-stage model.

The second mirror is not apparent until we have introduced megascopic quantum chemistry. For instance, Doyle's concept of "adequate determinism" in his Cogito model, attempting to explain free will only on the basis of quantum physics, cannot be right. The physicist Compton and psychiatrist Flanagan were right in claiming that quantum physics is insufficient for the justification of free will. Compton argued that freedom involves an additional determining factor of choice, about which science tells us nothing. Moreover Flanagan stated that free actions need to be caused by me, in an undetermined and non-random manner. The combination of non-determinism and non-randomness follows only from the megascopic einselection processes of quantum chemistry, and not from microscopic physical processes that are indeterministic and random.

Moreover, the crosswise-law mirrors follow directly from the two-stage model formulated by Eccles and Popper, see section 8, which unifies nature's laws of biology (the matter world) and psychology (the mind world). All six mirrors, two microscopic-megascopic, two matter-mind, and two crosswise-law mirrors are pictured in Fig 2.

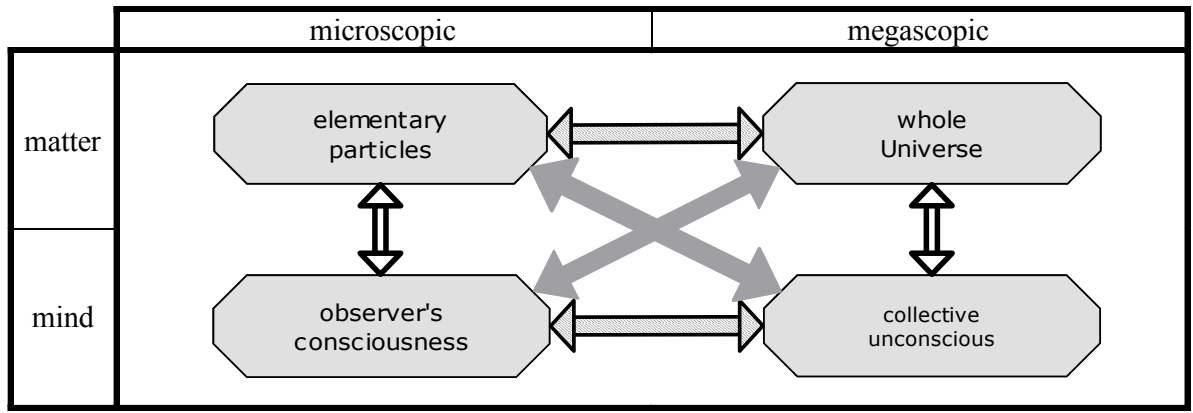

**FIG 2:** **Matter-mind relationship**

matter-mind mirror
microscopic-megascopic mirror
crosswise-law mirror

The crosswise-law mirrors have tremendous impact on our choice of the correct interpretation of quantum theory. They rule out all "objective" interpretations. The matter-mind dualism can never be removed and replaced by materialistic monism. Megascopic quantum chemistry adds "unconscious" measurements to "conscious" measurements as is known from microscopic quantum physics and this fact disqualifies the MWI interpretation, see section 7, and the arguments of its advocates, such as those of Zeh [158]: „He who considers this conclusion of an indeterminism or splitting of the observer's identity, derived from the Schrödinger equation in the form of dynamically decoupling ('branching') wave packets on a fundamental global configuration space, as unacceptable or 'extravagant' [159] may instead dynamically formalize the superfluous hypothesis of a disappearance of the 'other' components by whatever method he prefers, but he should be aware that he may thereby also create his own problems: Any deviation from the global Schrödinger equation must in principle lead to observable effects, and it should be recalled that none have ever been discovered [160]. The conclusion would of course have to be revised if such effects were some day to be found."

One can also find a nice response to this problem, i.e. of our false impression that all processes in the material world must be measurable, Norton [52]: „It has long been recognized that human action is the prototype of cause and effect. At its simplest, we identify processes as causal if they are sufficiently analogous... Using human action as a prototype, we identify terms in the cause and effect relation whenever we have one that brings about or produces the other; and we identify the process of production as the causal process... We restrict a science to some hospitable domain."

Norton's view is correct. If we confine our knowledge of the material world to causal quantum physics and fully ignore teleological phenomena, then we have restricted science to 'some hospitable domain'. From all known interpretations only the Copenhagen view has the ability to incorporate teleological phenomena via a megascopic mirroring of each of its basic axioms.

## 16. The Emerald Tablet

It is worth pointing out that important contributions to the concept of the collective unconscious of Jung and those of Darwin to the teleological concept of biological adaptation were not motivated by some true authentic science, but by esoteric occult ideas preserved since ancient times in various clandestine masonic, theosophical and new age societies. Predominantly in these societies, the ancient Emerald Tablet of Hermes Trismegistus has always drawn attention in special ways. Darwin did fulfil his grandfather's wish to apply the content of the Emerald Table to biological processes. From Mackey's Encyclopedia of Freemasonry we can cite this information [161]: „Before coming to Derby in 1788, Dr. [Erasmus] Darwin had been made a Mason in the famous Time Immemorial Lodge of Cannongate Kilwinning, No. 2, of Scotland. Sir Francis Darwin, one of the Doctor's sons, was made a Mason in Tyrian Lodge, No. 253, at Derby, in 1807 or 1808. His son Reginald was made a Mason in Tyrian Lodge in 1804. The name of Charles Darwin does not appear on the rolls of the Lodge but it is very possible that he, like Francis, was a Mason."

Most biologists believing that Darwinian evolution is a scientific theory, do not know that it originated in occultism in the disguise of a scientific theory. Applications of occultism, without a firm scientific background, is of course vulnerable to misunderstandings and misinterpretations of all kinds. There was no quantum physics at the time of Darwin, to e.g. vindicate random mutations, and even if Darwin was right with his teleological interpretation of the process of adaptation, the problem of teleology has not been resolved to this day. One might study how the central characters of Darwinian evolution, such as Mayr or Madrell, see section 7, have advanced the meaning of teleology, not in full accord with Darwin. The same applies to protagonists of quantum decoherence who attempt to replace the causal Born rule by a teleological einselection rule. They were all inspired with Universal Darwinism, but they must also realize that animate nature obeys the laws of inanimate nature and not conversely. It means that the Darwinian adaptation process must be first revealed in inanimate nature with the possibility of a consequent application to living organisms.

Philosophers like Popper regarded Darwinism as a tautology. Popper changed his mind after he had noticed similar patterns, see section 8, in the two-stage model, applicable both to biology and the psychology of free will. Maybe such arguments are convincing for philosophers, but not for physicists, since the latter will always look for primary patterns in inanimate nature. We have nevertheless shown that microscopic quantum physics and megascopic quantum chemistry form together the same two-stage model as was applied to biology and psychology by Eccles and Popper.

Note that once we have incorporated teleology into megascopic quantum chemistry, the Darwinian adaptation process becomes identical to megascopic quantum jumps that represent transitions between different symmetry broken states, i.e. between different Platonic Forms, see section 13. This view is synoptic with the prevailing view of Western biologists and philosophers until the 19th century, i.e. that the differences between individuals of a species are uninteresting departures from their Platonic idealism of created kinds. In this way the true Darwinian evolution can be applicable only to questions like why we have so many races of cats and dogs, but the cat will be forever a cat and the dog will be a dog. Darwin himself was aware of the problem of missing transitional forms [162]: „Why, if species have descended from other species by insensibly fine gradations, do we not everywhere see innumerable transitional forms? Why is not all nature in confusion instead of the species being, as we see them, well defined?... But, as by this theory innumerable transitional forms must have existed, why do we not find them embedded in countless numbers in the crust of the earth?... Why then is not every geological formation and every stratum full of such intermediate links? Geology assuredly does not reveal any such finely graduated organic chain; and this, perhaps, is the most obvious and serious objection which can be urged against my theory."

A few decades later Darwin was brainwashed by the illusory "success" of his own theory when he said [163]: „At some future period, not very distant as measured by centuries, the civilized races of man will almost certainly exterminate, and replace, the savage races throughout the world. At the same time the anthropomorphous apes... will no doubt be exterminated. The break between man and his nearest allies will then be wider, for it will intervene between man in a more civilized state, as we may hope, even than the Caucasian, and some ape as low as a baboon, instead of as now between the Negro or Australian and the gorilla."

Today these lines are far from being politically correct particularly since history has demonstrated for us the danger of disgusting "ape theories" and sickening pagan concepts. Returning instead to a closer look at the Emerald Tablet, Darwin's inspiration. It contains 14 verses but only two of them are relevant for an alignment with our current knowledge of natural sciences, i.e. the second and eighth ones. Citing the English translation of the standard 1613 Latin Beato's Aurelium Occultae Philosophorum [164].

Verse 2: *Whatever is below is similar to that which is above. Through this the marvels of the work of one thing are procured and perfected.*

Verse 8: *This ascends from the earth into the sky and again descends from the sky to the earth, and receives the power and efficacy of things above and of things below.*

Verse 2 seems to be intelligible and conceivable for us as physicists. Maruani [147], see also papers quoted therein, compares the microcosm and the macrocosm, and excerpts the same two verses from a similar 1680 Newton's translation and discusses some interesting relations between microphysics and astrophysics, electromagnetism and gravitation, and the "magical" relations between the fundamental physical constants.

Verse 8, on the other hand, has no support in contemporary quantum physics. Yet, based on this verse, Darwin proposed his two-stage model, and even if this proposition was successful, it is still occultism, since there is no scientific base, and hence we do not know the range, the limitations and the critical points of its applicability. But, if comparing verse 8 with Table 2, depicting microscopic-megascopic mirrors of all basic axioms, it is obvious that we may identify a decrypted version of the Emerald Tablet, expressed in terms of a clear and unambiguous language of quantum theory.

One might ask where such a magnificent wisdom of the Emerald Tablet comes from. We do not even know whether Hermes Trismegistus was a real person or only the name of the book dealing with the 'Emerald Tablet'. Most probably this is only a torso of some higher knowledge of antediluvian origin. There is a Jewish testimony of Flavius Josephus who in his Antiquities, Book I, Ch. 2: Concerning the Posterity of Adam, and the Ten Generations from Him to the Deluge [165], wrote about the significant discoveries of the descendants of Adam's son Seth that had survived the flood: „And that their inventions might not be lost before they were sufficiently known, upon Adam's prediction that the world was to be destroyed at one time by the force of fire, and at another time by the violence and quantity of water, they made two pillars, the one of brick, the other of stone: they inscribed their discoveries on them both, that in case the pillar of brick should be destroyed by the flood, the pillar of stone might remain, and exhibit those discoveries to mankind; and also inform them that there was another pillar of brick erected by them."

The philologist and historian van Bladel [166] also mentions an Arabic testimony in which Hermes Trismegistus is identified with Idris, which is the Arabic name of Enoch from the bloodline of Adam's son Seth, mentioned in the Bible for first time in Gen 5:18–24, and then in 1Ch 1:3, Luk 3:37, Heb 11:5, Jud 1:14-15, and Sir 44:16,49:14. We can assume that Enoch indeed fully understood the content of his teachings, which was part of the true original religion given to the first man by God. After the confusion of tongues in Babel, the Emerald Tablet became a source of occultism in many secret societies, since the key to its understanding has been lost.

It is interesting that this occultism has influenced not only secret societies, but also Folk sayings. One simple example: from all the megascopic phenomena summarized in Table 1, one of them, the brittle fracture, has been familiar with the whole population throughout the history of mankind. And each of us has been subject to attendant negative experiences in everyday life, such as broken dishes, china, porcelain, window glass, etc. However, there is a Folk saying, expressed by the same words in all Central European languages, entirely in opposition to impairments due to brittle fracture. In Czech this saying sounds: *Střepy přinášejí štěstí*, in German: *Scherben bringen Glück*, and the English literal translation is: *Shards bring luck*. Actually, brittle fractures are nightmares for mechanical and building engineers. It caused the loss of Titanic in 1912, the 1962 King Street Bridge collapse in Melbourne, and the 1967 Silver Bridge collapse in West Virginia, all resulting from the brittle fracture effect. There is a riddle: how can shards bring any luck?

No one knows the origin of this Folk saying, except that our pagan ancestors threw shards to drive off evil spirits, or that it was a heathen ritual of shattering the holy altars of clay after completing a sacrifice to the gods. All traces end in ancient Greek and Roman coffin tombs where often find shards donated to the deceased. In verse 8 of the Emerald Tablet, we have used the word "sky" as the English translation of the Latin "coelum". It could also be translated as "heaven", used e.g. in Newton's version. One should here be aware of the danger of capitalizing of the letter "h". Originally brittle fracture was understood as a "consecration of heaven", i.e. a megascopic effect. But after the confusion of tongues in Babel, when the key to the Emerald Tablet was lost, future generations explained it as a "consecration of Heaven", i.e. a supernatural or at least a preternatural phenomenon that brings the blessings of gods. In this way occultism infiltrates Folk sayings.

## 17. Conclusion

The core of this work originates in an independent proof of Goldstone's theorem [91] for molecules and solids. To better explain the significance of our message, assume at the outset that we have no prior knowledge of this theorem, simply rewriting the total molecular Hamiltonian of interacting nuclei and electrons into quantum field form. Instead of the usual formulation of the original electrons and the associated nuclear vibrational modes/phonons, we will introduce a new set of fermionic and bosonic operators in the most general way, and then optimizing the total transformed Hamiltonian. We have put forward five sets of well-known equations as limiting cases of the resulting equations of our procedure; a test that our solutions must pass and exactly agree with each other

1) Pople's equations for ab-initio calculations of vibrational frequencies [103]
2) Fröhlich's expression for the correction of the ground state energy [104,105]
3) Fröhlich's effective two-electron interaction [105]
4) Lee-Low-Pines' polarons and their self-energy [106]
5) Born-Huang's ansatz [7]

Starting from the field electron-vibrational (electron-phonon) Hamiltonian [101,102], only the first four sets pass the test successfully. For instance, in small molecules such as $H_2$ one obtains only a fragment of the whole contribution from the Born-Huang ansatz, and this is a major failure that unfortunately is not well-known in quantum chemistry and solid state physics. In quantum chemistry field methods are not often used, and usually only on the electronic level [98-100]. However, in solid state physics electron-phonon field mechanisms are frequently employed, and in most cases the contribution of the Born-Huang ansatz is negligible to such an extent that many physicists do not even realize any differences between the Born-Oppenheimer (B-O) and the adiabatic approximations and mistakenly even regard them as synonyms.

Hence the concept of the electron-vibrational field theory is clearly insufficient, and it must be replaced by a more rigorous field theory comprising two major upgrades [107,108]:

a) Together with the phonons two new types of quasiparticles appear: rotons and translons descending from rotational and translational degrees of freedom. The contribution from the roton and the translon quanta occurs even if the molecule as whole does not rotate nor move.
b) In analogy with Lorentz covariance, binding together space and time coordinates, a new covariance is necessary, binding together internal and external degrees of freedom, without the centre-of-mass (COM) separation, which normally applies in both classical and quantum mechanics.

Let us compare these results with the content of Goldstone's theorem [91]: „The spontaneous breaking of a continuous symmetry can be associated with a massless and spinless particle". According to this theorem there are three types of broken symmetries in molecules and crystals, composed of $N$ nuclei ($3N$ degrees of freedom): the Galilean associated with $3N$-6 phonons (or $3N$-5 for diatomic molecules), translational associated with 3 translons, and rotational associated with 3 (or 2) rotons. Unfortunately the scientific community does not seem to be aware of this fundamental fact with its origin in the Goldstone theorem, cf. the picture reflected in the (2018) Wikipedia phrase "Goldstone boson" that states the following incongruity: „In solids, the situation is more complicated; the Goldstone bosons are the longitudinal and transverse phonons and they happen to be the Goldstone bosons of spontaneously broken Galilean, translational, and rotational symmetry with no simple one-to-one correspondence between the Goldstone modes and the broken symmetries."

First of all the Goldstone theorem is a complement to the first Noether theorem where a one-to-one correspondence between the conservation laws and associated symmetries always exists. Therefore the unique and unambiguous one-to-one correspondence between the Goldstone modes and spontaneously broken symmetries in solids must necessarily exist as well. It means that vibrational modes/phonons in molecules/solids do not represent the full set of Goldstone bosons. The errors caused by disregarding the rest of the Goldstone modes are usually negligible in adiabatic cases, but in non-adiabatic situations they might lead to fatal consequences, i.e. an deficient view of spontaneously broken systems on the quantum level.

We have demonstrated that a field theory based only on phonons is not sufficient for the description of Born-Oppenheimer (B-O) degenerate states, such as Jahn-Teller (J-T) systems and superconductors, where a general field COM covariant theory incorporating all Goldstone bosons, i.e. phonons, rotons and translons is unavoidable. We have derived the formulae for the ground state energy and the excitation spectra of B-O degenerate quantum chemical and solid state systems together with a complete set of Goldstone bosons.

Within this background the Born-Huang ansatz plays the same role for the field COM covariance as Maxwell's equations for Lorentz covariance. Even though we have here only presented a special proof of the Goldstone theorem for interacting systems of electrons and nuclei, based on the field COM covariance resulting from the Born-Huang ansatz, we have nonetheless arrived at the correct set of Goldstone bosons. However, the general Goldstone-Salam-Weinberg proof [92] of Goldstone's theorem, based on the conserved currents of the first Noether theorem, appears not transparent enough in the case of a correct identification of the Goldstone bosons. The physicists inaccurately identify them only with phonons.

In fact this mistake has led to unfortunate consequences, i.e. an inadequate BCS theory [78] using only the electron-phonon interaction which has never been revisited. Moreover, Anderson in his article [95] added to this misunderstanding claiming that "the solid crystal violates translational and rotational invariance, and possesses phonons". This paper soon became a fateful marker, leading to the 'celebrated' Higgs mechanism, that was incorporated in the Standard model attempting to explain, how particles in the electro-weak interaction gain the mass. Higgs named his Nobel lecture "Evading the Goldstone theorem" [97]. This was alas no 'evading', but rather a misrepresentation of the theorem, since Goldstone bosons cannot after all be suppressed. We have summarized Comay's work [88], where he, after a careful analysis of Higgs' equations, did come to the conclusion that they do violate Bohr's correspondence principle, and therefore cannot be approved as valid equations of quantum physics.

Goldstone bosons, arising as the result of symmetry violations, i.e. from rotons and translons, produce singular behaviour in the original symmetrical positions. The system is forced to avoid them by removing the degeneracy, and after apt one-particle transformations, finding itself in new asymmetric positions. This principle unifies the formation of ground states of J-T molecules and superconductors. It should be noted how the quantum field formulation deals with the virtual degeneracies originating from approximative B-O solutions. They are profoundly different from the real degeneracies, where the quantum principle of superposition takes place, after being removed by some external perturbation, such as the Stark or the Zeeman effect. The quantum field simply does not share the centre of mass, defined in standard quantum mechanics, rather it solves the problem of the centre of gravity "in its own way".

We have also offered a reformulated version of the J-T theorem as it follows directly from the Goldstone theorem. Molecular and crystalline entities in a particular geometry of an electronically degenerate ground state are unstable except for the case when all matrix elements of electron-rotational and electron-translational interactions equal zero. This is a striking result, as it shows how a true quantum field, respecting fully the Goldstone theorem, copes with B-O degenerate systems, such as J-T molecules and superconductors. The superposition principle for the removal of the degeneracy is simply bypassed, since the rotons and the translons are actually responsible for the violation of symmetry. For instance in a superconductor, the symmetry breaking produces several geometrically different symmetry broken states, splitting the original half-occupied conducting band of the conductor into two bands, one fully occupied valence band and one empty conducting band, in such a way that the original conductor under the critical temperature becomes a multi-ground-state insulator.

The results discussed above clearly disqualify all theories based on B-O degeneracies, such as the usual solutions of the J-T effect [11,12] and the BCS theory of superconductivity [78]. Their validity was never justified from simple reasons, i.e. one query of extraordinary importance was never answered, cf. the question of Sutcliffe & Woolley [4]: „The interesting question is how to get from the quantum theory of an Isolated Molecule to a quantum theory of an individual molecule by rational mathematics." We have further concluded that quantum mechanics describing a system of nuclei and electrons can never resolve this task. The quantum theory of an Isolated Molecule is quantum mechanics with a solution avoiding the B-O concept, and this is the method of Monkhorst [5,6] that "takes a very atomic view of a molecule: instead of fixing the nuclei as in the BO approximation, the electrons and nuclei are both described quantum-dynamically within a centrosymmetric shell structure." Unfortunately this approach leads neither to any B-O degenerate states nor to any symmetry breaking.

A valid quantum theory of an individual molecule should be a true quantum field theory of fermions represented by renormalized electrons and Goldstone bosons standing for vibrational, rotational and translational modes, i.e. a system where nuclei from the mechanical description are replaced by the Goldstone bosons in the field description. This means, that true quantum mechanics deals solely with isolated systems, while true quantum field theory works with individual systems. Only a quantum field formulation can produce symmetry broken states and, unlike the B-O model, removing degeneracies bypassing the superposition principle.

Returning to the question of teleological principles in science, one might first deliberate over how classical physics copes with the spontaneous symmetry breaking (SSB). In the classical phenomenological picture of superconductors, the Ginzburg-Landau equations [77] exhibits quadratic and biquadratic functions of the order parameter, forming the well-known Mexican hat. We have also discussed similar Mexican hat-like situations, i.e. the Norton dome [52] and the hydrodynamic problem of Boussinesq [124]. These three cases have one thing in common, *viz.* the solutions of the corresponding equations are not unique. However, in addition to the standard causal solution also a teleological one is permitted. Maxwell was impressed by the third case of Boussinesq to such an extent that he for the first time spoke about teleological principles that exist outside the known laws of physics and called it [51] "some determining principle which is extra physical (but not extra natural)".

Physicists, contrary to biologists, usually abandon the teleological principle as unscientific, and they do therefore not agree with the problem raised regarding the classical description of the Meissner effect and its relation to the phenomenon of superconductivity. One discerns two camps of physicists. The first camp, following Meissner [64] and London [66], claims that superconductivity unlike ideal conductivity has no classical explanation, and that quantum physics is necessary in order to explain it. The second camp following Cullwick [71] identifies superconductors with perfect conductors. But they cannot both be right. Superconductors are not identical to perfect conductors and quantum physics (BCS theory) does not explain the Meissner effect, without properly explaining superconductivity first. Electromotive forces, described by Faraday's law of induction, are equal to zero in the stationary conditions of the Meissner effect, whereas existing theories do not suggest any other electric forces needed to accelerate the electrons until the steady state supercurrent is achieved. From the view of telicity, one concludes that classical physics provides two solutions, one for perfect conductors and one for superconductors. Undeniably classical physics can formulate, but not explain superconductivity. According to Peirce's philosophical analysis [125], that no physical theory based on corpuscular philosophy can solve the problem of the SSB bootstrap. Even if this analysis was written before the era of quantum physics, Peirce's argument is also applicable today. It builds on Democritus' atomic ideas, which is essentially the same corpuscular idea as in classical physics. It means that quantum physics in the Copenhagen interpretation is incapable of solving the problem of the SSB trigger. As we have seen, the teleological problem, which relates to Leibniz' principle of sufficient reason [114], to the second form of van Fraassen's argument [119], and Peirce's corpuscular philosophy argument [125], can be discounted in classical physics. Unfortunately quantum physicists usually ignore this problem. But this attitude is uncalled for, since it is motivated by the conviction, that quantum physics uses the same SSB archetype as the classical one.

In comparing classical and quantum physics, we conclude: classical physics does not distinguish superconductors from ideal conductors and sees them as one system with two possible solutions, one teleological and one causal, while quantum physics does recognize the difference, but, on the other hand does not separate them from multi-ground-state insulators, i.e. is only able to describe the ground states of superconductors, but never superconductivity itself, since the Bohr's correspondence principle does not account for teleological phenomena. Hence the application of the classical SSB, i.e. the Mexican hat, to quantum equations, the J-T effect, superconductivity or the Higgs mechanism becomes a mismatch. Quantum equations, based on a classical SSB, lead to paradoxical situations. With reference to SSB, the scientific community is again divided, i.e. whether SSB here is really present or not. We have, in this work, discussed the arguments both for and against in connection with the J-T effect, superconductivity, the Higgs mechanism and ferromagnetism.

From the addition of the lost Goldstone bosons, rotons and translons, found to be fully responsible for the mechanism of forming quantum states with spontaneously broken symmetries, it follows that the spontaneous symmetry breaking (SSB), only described phenomenologically at the level of classical physics, becomes misinterpreted on the quantum level, where every physical system can be described dually, either as a mechanical system or as a field theoretical system, in accordance with Weinberg's statement (repeated again) [126]: „If it turned out that some physical system could not be described by a quantum field theory, it would be a sensation; if it turned out that the system did not obey the rules of quantum mechanics and relativity, it would be a cataclysm." For molecules and solids treated equivalently, we have presented true quantum field solutions fully respecting the Goldstone theorem.

In a previous paper [108] I introduced the notion of property-object dualism: in quantum mechanics, nuclei and electrons represent objects, and vibrational modes are their common property. In quantum field objects, however, they are represented by electrons and Goldstone bosons, and the "nuclear" positions or "clamped nuclei" are properties of the pertinent field equations. We are now in a similar situation, with the description of the whole many-body system, as one was a century ago regarding the different aspects of a mechanical formulation and a field theory description of elementary entities that finally led to Bohr's complementarity. Obviously one needs a second type of Bohr complementarity on the many-body level as was already requested a long time ago by one of the co-founders of quantum mechanics, Pascual Jordan [134], even though, from a different reason, the Copenhagen interpretation based on the first Bohr complementarity, cannot explain classicality as an emergent property.

Linking Sutcliffe's and Woolley's [4] problem of transitions between an isolated and the individual systems with Bohm's [15] question of fragmentation and wholeness, implies that the second complementarity, embracing the whole Universe, is the megascopic mirror of the first Bohr microscopic complementarity. We have just touched upon an ancient knowledge, presenting the Universe as a mosaic where the smallest one resembles the Greatest One. We need to postulate a new axiom that is necessary for the justification of the quantum jumps of the Universe from its state of total wholeness to its total fragmented states and vice versa. This request certainly goes beyond the scope of the Copenhagen interpretation. It neither contradicts nor denies this interpretation, but rather appears to be its natural extension.

While microscopic processes are causal, megascopic phenomena are teleological. Teleology as the final cause can then be comprehended as downward causality, descending from the Greatest One - the whole Universe, in contrast to upward causality, ascending from the smallest entities, the elementary particles. The first complementarity represents the causal microscopic relationship between the mechanical and the field descriptions of elementary entities (particle-wave dualism), and the second complementarity represents the teleological megascopic relationship between the mechanical and the field descriptions of the whole Universe, where these two descriptions in particular deal with the centre of gravity in different ways.

Whereas microscopic quantum physics, dealing with elementary particles as 'basic building blocks' out of which everything is made, reflects the ancient materialistic ideas of Democritus, megascopic quantum phenomena can be seen as a reflection of Plato's Forms. In the latter one can finally find the true relationship between megascopic quantum jumps and time: these jumps do not proceed in time, but time is created by them. At the megascopic level this gives rise to the concept of quantum of time, which otherwise is unreachable at the microscopic level. In this sense megascopic quantum jumps are a true clock of the Universe.

Restating that on one hand we have time evolution of the wave function, representing only Aristotelian potentiality, and on the other reality created during the measurement process by timeless events, i.e. microscopic quantum jumps. This separation of events from time, as a classical entity, is the source of almost all philosophical problems of contemporary quantum physics, and since the origin of time is rooted in megascopic events, we are able to finally reunite time and quantum events, but in contrast to processes on a classical level these case events will be foundational and time subordinate. This idea goes back to Aristotle who defined time as, we repeat, "the number of movement in respect of before and after". Since it cannot exist without succession, it does not exist on its own but only relative to the motions of things.

To complete the formulation of megascopic quantum theory one needs to identify megascopic mirrors of all basic axioms of the microscopic quantum theory in its Copenhagen interpretation. There are five fundamental micro-megascopic mirrors:

1. First complementarity versus the second: While microscopic quantum theory is founded on Bohr's complementarity between mechanical (particle) and field (wave) descriptions of elementary entities, megascopic quantum theory is based on two megascopic complementary descriptions of the whole Universe, namely a mechanical one representing its total wholeness, and the field one representing its total fragmentation. The implementation of the second complementarity was for the first time mentioned by Jordan [134].

2. Causality versus teleology: Microscopic quantum physics inherits causality (or upward causality) from classical physics due to the Bohr principle of correspondence. Its megascopic mirror is teleology (or downward causality or final cause), mentioned for the first time by Maxwell [51] as resulting from a principle which, quote, unquote "is extra physical but not extra natural, and is acting without exerting any force or spending any energy".

3. Projections versus injections: Transitions between different states in microscopic quantum physics are expressed as vector projections in Hilbert space. The megascopic mirror of these projections is represented by injections that are responsible for the transitions from the fragmented Universe into its wholeness and vice versa, a statement given for the first time by Bohm [16].

4. Born rule versus einselection: The probability of transitions based on microscopic projections is calculated from the Born rule. Its megascopic mirror is einselection, taking into account influences from all environments. The concept of einselection was originally coined by Zurek [45] in his quantum decoherence program aimed to replace the Born rule. As we have shown, the Born law cannot be replaced, and we have revealed the true meaning of einselection as the megascopic counterpart of the microscopic Born rule. It means that the latter remains tied with the microscopic projections, while einselection determines the probabilities of the megascopic injection transitions.

5. von Neumann - Wigner rule versus Jung's rule: In microscopic quantum physics the von Neumann - Wigner rule binds together elementary particles with the conscious mind, and its perfect megascopic analogy must exist as well, binding together the whole Universe with the collective unconscious. An observer's activity, i.e. a measurement, causes microscopic quantum jumps and hence a discontinuity of the deterministic Schrödinger equation. Since all events in our world must have their reason, it is reasonable to postulate a similar megascopic analogy: the existence of megascopic "measurements" as the activity of the collective unconscious causing megascopic quantum jumps. Due to Jung's success formulating the concept of the collective unconscious [155], we have suggested to call the megascopic mirror of the von Neumann - Wigner rule as the Jung rule.

We have also applied the abovementioned megascopic rules to superconductivity, and we will give three main reasons why superconductivity (BCS) will not provide a microscopic explanation:

1) Every microscopic theory must result in the Born rule for estimating the probabilities of the density and the velocity of superconducting carriers, i.e. demonstrate that these quantities have to be experimentally measurable. But as we know, they are in principle non-measurable.

2) Microscopic theories do not allow us to avoid the universal concept of Bloch states for the description of superconducting carriers. It means that the mass of the carriers must take the effective mass of electrons into account. But this is in direct contradiction with measurements of the London moment, where only the bare electronic mass and not the effective one is reported.

3) According to the second form of van Fraassen's argument, asymmetry cannot arise ex nihilo. We know that the original asymmetry can be present in the form of an external magnetic field around the superconductor, but no microscopic theory is able to explain the mechanism of the Meissner effect, when a superconductor is cooled below the critical temperature. Moreover, a constant magnetic field cannot cause any acceleration of the superconducting carriers. It means that no microscopic theory is capable to implement the original asymmetry, contradicting the second form of van Fraassen's argument.

Nevertheless, superconductors can be, and have been, described macroscopically on the quantum level. Such a wave function representing macroscopically-occupied quantum states was proposed already by London [66], from which one could derive the Josephson equations [143] without the need for any microscopic concept of superconductivity, as it was also shown by Feynman [144]. Only the ground state and the excitation states of the superconductors have a microscopic explanation, in the same way as all solids must be microscopically described. However, microscopic theory yields merely the symmetry broken states but not transitions between them. We therefore need the concept of megascopic quantum jumps appearing either on the microscopic or the macroscopic level, standing in the same relation to the microscopic tunnelling as the concept of microscopic quantum jumps in the Bohr model stands to the classical electronic movement in the Rutherford model.

During these megascopic quantum jumps the pairs of electrons (valence orbitals forming some kind of "chemical bonds" occupied by two electrons with opposite spins) have a chance either to be relocated in diverse directions, or to return into their original position, depending on the external magnetic field that determines the einselection factor. The superconducting carriers do not recognize Cooper pairing; they are identical with the optimal transformed doubly occupied valence-like orbitals. Unlike BCS theory, which associates this phase with the motion of carriers in Bloch's $\mathbf{k}$ space, our megascopic interpretation associates the phase with the macroscopic $\mathbf{l}$ space, orthogonal to Bloch's $\mathbf{k}$ space.

Hence it is evident that a microscopic theory (BCS), based on the microscopic dynamics of  superconducting carriers, whatever these may be, and whether unpaired or Cooper paired electrons, polarons, bipolarons etc. – is, in view of our megascopic interpretation, wrong. To overthrow the corpuscular philosophy in the Peirce's argument, and to explain the Meissner effect and superconductivity a holistic megascopic quantum theory is necessary.

In addition to superconductivity and superfluidity, we have also analysed other megascopic phenomena on both the microscopic and the macroscopic levels, taking into account equilibrium processes of both adiabatic and non-adiabatic systems as well as non-equilibrium processes, see e.g. the complete list in the Table 1: isomeric transitions, the Jahn-Teller effect, chemical reactions, the Einstein-de Haas effect, superconductivity and superfluidity, and the brittle fracture.

Note that post-Cartesian classical chemistry deals with material objects from atoms up to molecules, while standard quantum chemistry is a result of an implantation of post-Cartesian classical chemistry ideas into pre-Cartesian quantum physics. On the other hand, one can alternatively build an authentic pre-Cartesian quantum chemistry on quantum megascopic axioms, embracing in addition to chemistry and chemical reactions also all the above mentioned megascopic phenomena.

Furthermore, the concept of megascopic quantum jumps resolves many of the paradoxes of quantum physics: accepting that all processes in the Universe consist of only microscopic and megascopic irreversible events and that the origin of time is rooted in megascopic events, then the problem of the arrow of time is automatically solved. Emergent reversibility can only appear in subsystems with no SSB (most adiabatic systems) where a one-to-one correspondence between the mechanical and the field states holds, with the result that these subsystems can be fully described by the time-reversible Schrödinger equation. This is commensurate with the answer to von Weizsäcker's problem [26], where he correctly recognized irreversibility as a necessary condition for classicality, but he left open the question if this condition is also sufficient. We know now, that this condition is not sufficient in the framework of the microscopic Copenhagen interpretation, but is fully sufficient after the inclusion of megascopic irreversible processes.

This establishes also a fulfilment of Jordan's request [134] to introduce a second complementarity to explain classicality. Megascopic chemical processes are responsible for the decoherence of the wave function, and the classical appearance of the silver grain on the photographic plate, discussed in Jordan's article, ensues. On the same basis Santilli's "no reduction theorem" [13] paradox is explained and Schrödinger's cat can never be in some state of superposition: Either the irreversible megascopic event representing the chemical reaction of the poison with the cat's body happens or does not, and nothing in between.

We also know that our conscious minds are all primarily connected in Jung's collective unconscious entirely outside the external material world, explaining the legitimacy of shared knowledge of a population of communicating observers resolving the paradox of Wigner's friend [37].

Finally we have also discussed the paradox of the residual entropy arising in chemistry, if physical entropy is assigned to the multitude of equivalent isomeric molecular ground states. We have quoted two examples of residual entropy: the first one pointed out by Pauling [148] to describe water-ice, and the second one, the Kauzmann paradox [149], related to the liquid-glass transition with actually negative residual entropy. One needs here to introduce a separation of the physical microscopic entropy from the megascopic chemical one, in order to avoid violating the third law of thermodynamics.

A microscopic, ductile to megascopic, brittle transition is analogous to a transition between microscopic normal conductivity and megascopic superconductivity under critical temperatures. Materials that usually fracture in a brittle manner are glasses, ceramics, and some polymers and metals. This also explains why good superconductors are usually formed with brittle ceramics on the verge of rupture.

The most interesting result of the present work is the aspect that microscopic quantum physics and megascopic quantum chemistry together form a two-stage model, ontologically compatible with the two-stage model formulated by Eccles and Popper [60], which unifies nature's laws of the material world of biology with the mind world of psychology. One can now perform a unification of this two-stage model that embraces the realm of inanimate nature, the realm of animate nature, and the realm of psychology.

Moreover, we have revealed the crosswise-law mirrors that follow directly from the two-stage model formulated by Eccles and Popper, namely the reflections of microscopic quantum physical laws in the collective unconscious and the reflections of megascopic quantum chemical laws in the observer's conscious mind, the latter in a form sustaining his free will decisions. The crosswise-law mirrors have a tremendous impact on our choice of a correct consistent interpretation of quantum theory, disqualifying all "objective" interpretations. Basically, matter-mind dualism cannot be removed and replaced by materialistic monism. Megascopic quantum chemistry adds "unconscious" measurements to the "conscious" ones known from microscopic quantum physics, eliminating also Everett's interpretation [41], since its basic stipulation presupposes a deterministic quantum mechanical wave function of the whole Universe. The Bohm - de Broglie interpretation [14] was already outdated by Bohm himself, when he formulated several additional requirements [15,16] that a true quantum interpretation must fulfil.

So the final verdict reads: From all the known conceptions of microscopic theories of nature only the Copenhagen interpretation has the ability to incorporate teleological phenomena via the megascopic mirroring of each of its basic axioms, and in that way squaring up with all the universal paradoxes discussed here.

## Acknowledgements

The author would like to express his greatest appreciation to E. Brändas for his careful reading of the manuscript, offering many linguistic improvements, as well as extensive constructive suggestions for valuable reformulations.

## R e f e r e n c e s